\documentclass[11pt]{article} 

\usepackage[utf8]{inputenc} 
\usepackage{amsmath,amssymb,amsfonts,multicol}
\usepackage[numbers,sort&compress]{natbib}
\usepackage[colorlinks,citecolor=blue,urlcolor=blue,linkcolor=blue]{hyperref}
\usepackage{amssymb}
\usepackage{geometry} 
\usepackage{graphicx} 

\usepackage{booktabs} 
\usepackage{array} 
\usepackage{paralist} 
\usepackage{verbatim} 
\usepackage{subfig} 

\usepackage{fancyhdr} 
\usepackage{sectsty}
\allsectionsfont{\sffamily\mdseries\upshape} 

\usepackage[nottoc,notlof,notlot]{tocbibind} 
\usepackage[titles,subfigure]{tocloft} 

\usepackage{authblk}

\title{Images of the accretion disk in Hybrid metric-Palatini gravity}
\author[1 a]{P.I. Dyadina\footnote{Corresponding author. E-mail: guldur.anwo@gmail.com}}
\author[2 b]{N.A. Avdeev}
\affil[1]{Sternberg Astronomical Institute, Lomonosov Moscow State University, Universitetsky Prospekt, 13, Moscow, 119234, Russia}
\affil[2]{Istituto Nazionale di Fisica Nucleare, via Pietro Giuria 1, Torino, Italy}
\affil[a]{\textit{e-mail: guldur.anwo@gmail.com}}
\affil[b]{\textit{e-mail: naavdeev1995@mail.ru}}
\date{} 

\begin{document}
\maketitle
\begin{abstract} 
In this paper we obtain and study images of accretion disks around static spherically symmetric black holes in hybrid metric-Palatini gravity.  We use Novikov-Thorne thin-disk model. Semi-analytic ray-tracing methods in curved spacetime are employed to generate the images of the disk for different scalar field configurations, including both Higgs-type potentials and cases without a potential. The resulting images, including both redshift and intensity maps, are analyzed. The results show that the scalar field parameters play a significant role in shaping both the direct and secondary images of the disk, while the inclination angle primarily affects the asymmetry and brightness distribution. In particular, configurations with extreme scalar field values lead to cooler and dimmer disk compared to General Relativity. Furthermore, the structure and angular size of the secondary ring exhibit noticeable deviations from General Relativity, offering a potential observational signature. This work should be regarded as a first step toward modeling realistic accretion disks within the hybrid metric-Palatini gravity framework, and toward assessing their potential observational distinguishability from General Relativity predictions.
\end{abstract}

\section{Introduction}\label{sec1}
Black holes (BHs) are among the most fascinating and enigmatic objects in the Universe. Due to their extremely strong gravitational fields, they are capable of capturing not only matter but also electromagnetic radiation. At the same time, this  feature makes the vicinity of BHs an ideal laboratory for testing gravitational theories in the strong-field regime. Although BHs were long considered purely theoretical constructs lacking direct observational support, this view has significantly changed over the past decades. The detection of gravitational wave signals from binary BH mergers by the LIGO and Virgo collaborations \cite{Abbott_2016}, along with the imaging of the shadows of supermassive BHs at the centers of the Milky Way \cite{Event_Horizon_Telescope_Collaboration_2022a, Event_Horizon_Telescope_Collaboration_2022b, Event_Horizon_Telescope_Collaboration_2022c, Event_Horizon_Telescope_Collaboration_2022d} and M87* \cite{2019a, 2019b, 2019c, 2019d, 2019e, 2019f} by the Event Horizon Telescope (EHT), have made it possible to directly probe the extreme gravitational environment near these objects. These breakthroughs offer powerful new tools for testing the predictions of General Relativity (GR) and exploring possible signatures of modified theories of gravity in the strong-field regime. One of the most promising approaches for testing gravitational theories in this context is the construction of BH images and their comparison with observational data.

The observed image of a BH is largely determined by the radiation emitted from its accretion disk. Since the 1970s, the imaging of accretion disks around BHs has been an important subject in astrophysics. The foundation for this line of research was laid with the development of the standard model of geometrically thin and optically thick accretion disks, originally proposed by Shakura and Sunyaev \cite{Shakura1973} and later expanded by Novikov and Thorne \cite{Novikov1973}. Early studies by Cunningham and Bardeen \cite{1972Cunningham, 1975Cunningham} computed the optical appearance of a star orbiting a Kerr BH, demonstrating how relativistic effects affect the observed image. Two primary techniques have since emerged for modeling the appearance of thin accretion disks: semi-analytic methods and ray-tracing combined with radiative transfer. A classic work by Luminet \cite{luminet} employed a semi-analytic approach to derive both primary and secondary images of an accretion disk around a Schwarzschild BH. A distinctive feature of this method is that observable quantities such as radiation flux and redshift can be expressed analytically in terms of the impact parameter and elliptic integrals.

Later, Fukue and Yokoyama \cite{Fukue} analyzed color images of accretion disks surrounding stellar-mass Schwarzschild BH, examining in detail the origins of both horizontal and vertical asymmetries in the observed structure. Marck introduced a computational shortcut for solving geodesic equations in Schwarzschild spacetime, which enabled the efficient generation of disk images as viewed from various observer positions \cite{Marck_1996}. In the work \cite{Muller} M{\"u}ller analytically explores how light bending and time delay affect the apparent position of a star orbiting a Schwarzschild black hole, illustrating key optical effects related to black hole shadow formation. In the article \cite{Tian_2019}, authors proposed a parametrized Schwarzschild metric and studied the resulting accretion disk images under different viewing angles. Notably, their models accounted for projection effects, acknowledging that light rays do not necessarily escape vertically from the disk surface. In addition, the properties of shadows of rotating black holes in the Kerr metric were studied in the works \cite{Zakharov_1986, Zakharov_1989, Zakharov_2005}. 

More recently, ray-tracing studies have systematically extended accretion-disk imaging and related radiative diagnostics to a broad class of non-Kerr spacetimes and beyond-GR black hole models. For instance, the radiative properties and bolometric appearance of thin disks were explored for rotating hairy black holes generated via gravitational decoupling \cite{Li_2025}, as well as for charged string-inspired solutions such as Kerr--Sen black holes \cite{Wang_2025} and Einstein--Maxwell--dilaton black holes, including the role of accretion-flow prescriptions and dilaton charge \cite{Chen2025_SC}. Additionally, complementary analyses in Einstein gravity with minimally coupled scalar fields and in models with synchronized scalar hair demonstrated that scalar degrees of freedom can lead to qualitatively new optical features, including strongly deformed or even chaotic disk images with multiple disconnected components in certain regions of parameter space \cite{Bogush2022_PRD, Gyulchev2025_JPCS}. Comprehensive overview of the development of BH imaging techniques is presented in \cite{Falcke_2017, luminet2019illustratedhistoryblackhole, falcke2022roadimagingblackhole}.

	In this context, motivated by cosmological and astrophysical observations that challenge GR at large and small scales, numerous studies have investigated how modified theories of gravity, such as scalar-tensor theories \cite{Heydari-Fard2024},  Einstein-Gauss-Bonnet gravity \cite{Gyulchev_2021},  Moffat gravity (MOG) \cite{2022hu, 2022hua}, affect the appearance of accretion disks. These works typically aim to identify qualitative or quantitative deviations in disk morphology, brightness asymmetry, or polarization patterns that could serve as observational signatures of deviations from GR. In this work, we seek to extend the list of gravitational theories tested via accretion disk imaging by investigating one of the promising frameworks proposed in recent years, the hybrid metric-Palatini gravity \cite{Harko2012, Capozziello2015, Harko2020}.
	
	Hybrid metric-Palatini gravity can be viewed as a natural extension of the ideas behind $f(R)$-gravity. It combines the metric and Palatini formalisms in a single framework, retaining the advantages of both while avoiding their respective shortcomings. One of the key strengths of this theory is its ability to explain the late-time accelerated expansion of the Universe without conflicting with observational data obtained from local astrophysical systems \cite{Capozziello2015}. Notably, it achieves this without the need for screening mechanisms. In addition, HMPG can account for certain phenomena usually attributed to dark matter, while requiring significantly less of it \cite{Capozziello2013a}. Altogether, these features make hybrid metric-Palatini gravity one of the most promising modern alternatives to GR.
	
	HMPG has been actively studied across a wide range of physical regimes, from cosmological scales to compact astrophysical objects. On cosmological scales, it has been shown to successfully reproduce the observed  late-time acceleration \cite{Boehmer2013, Lima2016}. The viability of the model is assessed by combining constraints from Supernovae Ia and Baryon Acoustic Oscillation data \cite{Leanizbarrutia2017}. At galactic scales, the theory allows for the description of rotation curves and cluster dynamics without invoking large amounts of dark matter \cite{Capozziello2013a, Capozziello13}. In the solar system, it has passed observational constraints without the need for screening mechanisms, as confirmed through parametrized post-Newtonian (PPN) analyses \cite{Leanizbarrutia2017, Dyadina2019, Dyadina_2022}. The theory has also been tested in strong-field scenarios, such as binary pulsar systems \cite{Dyadina2018, Avdeev2020} and compact stars, including neutron stars, quark stars, and Bose-Einstein condensate stars \cite{Danil2017}. Furthermore, gravitational wave propagation in this framework has been investigated \cite{Dyadina2022, kausar2018, Dyadina2024}, and static, spherically symmetric BH solutions have been obtained numerically \cite{Danila2019}. More exotic configurations, such as  exact electrically charged solutions, including BHs, traversable wormholes,  and black bounces have also been explored \cite{rois2024novelelectricallychargedwormhole, rois2025horizonsthroatsbounceshybrid}. Different scenarios of inflation were considered in number of works \cite{Asfour_2024, bombacigno2024inflationnonlocalhybridmetricpalatini}. Also,  a generalized version of the HMPG was studied \cite{Rosa20, Rosa2021, Rosa_2024}. These results demonstrate the versatility and robustness of HMPG in both weak and strong gravity regimes. For the most complete review of studies, see \cite{Capozziello2015, Harko2020}.
	
	The present study builds upon our previous work on accretion onto static, spherically symmetric BHs in the framework of HMPG, conducted within the Novikov-Thorne thin-disk model \cite{meacc}. In that study, the energy flux, temperature distribution, emission spectrum, and energy conversion efficiency of the accretion disk were computed numerically. It was found that, compared to GR, BHs in HMPG produce accretion disks that are cooler and less luminous under the same conditions.

In this work, we aim to extend that analysis by constructing images of such accretion disks and investigating their observable features. Our goal is to identify potential signatures in the optical appearance of the disk that may distinguish HMPG from GR using current or future high-resolution observations. The BH solutions employed here are based on the numerical results obtained in \cite{Danila2019}, where static spherically symmetric BHs were derived within HMPG. As in our previous study, we consider two cases: one without a scalar field potential, and one with a Higgs-like potential. In constructing the images of accretion disks, we follow the method originally proposed by Luminet and later extended in Refs. \cite{Tian_2019, 2022hu, Liu_2022}. As in the previous work, we use the Novikov-Thorne thin-disk framework, which provides a standard, geometrically thin and radiatively efficient description of disk accretion in stationary spacetimes. While this approximation neglects several aspects of realistic accretion physics, such as magnetic fields, disk thickness, and time-dependent effects, it offers a well-defined and analytically tractable baseline for studying how HMPG affects observable disk properties. This choice allows us to isolate the impact of the underlying spacetime geometry from the complications introduced by more complex accretion dynamics. Consequently, the present work should be regarded as a first step toward constructing fully consistent, physically realistic accretion-disk models within hybrid metric-Palatini gravity.


The article is organized into six sections. The first and last are the introduction and conclusion respectively. The section \ref{sec2} provides a description of the HMPG and its scalar-tensor representation. The section \ref{sec3} outlines the numerical calculations for the static spherically symmetric BH metric. In the  section \ref{sec4} we analyze the geodesic equations for the time-like and light-like particles and describe the method for obtaining images of accretion disks. In the section \ref{sec5} we discuss obtained images of accretion disks in HMPG. The conclusion \ref{conc} summarizes our findings.

	Throughout this paper the Greek indices $(\mu, \nu,...)$ run over $0, 1, 2, 3$ and the signature is  $(-,+,+,+)$. In this paper, we use the CGS unit system, except for section \ref{sec4}, where we switch to the unit system with $G = c = 1$ for the sake of convenience in presenting the equations.

\section{Hybrid metric-Palatini gravity}\label{sec2}
	In this section,  we introduce the foundational aspects of the hybrid metric-Palatini gravity, a modification of GR that incorporates both metric and Palatini approaches. This theory extends the Einstein-Hilbert action by including a term that depends on the Palatini curvature scalar, thereby allowing for a richer geometrical structure.  The action for HMPG is formulated as follows \cite{Harko2012, Capozziello2015}: 
	\begin{equation}
		\label{action1}
		S=\frac{1}{2k^2}\int d^4x\sqrt{-g}[R+f(\Re)]+S_m,
	\end{equation}
where $k^2=\frac{8\pi G}{c^4}$, $G$ is the gravitational constant, $c$ is the speed of light, $g=det\{g_{\mu\nu}\}$~ is the determinant of the metric, $S_m$~ is the matter action, $R$ is the metric Ricci scalar and the Palatini curvature is defined as
	\begin{eqnarray}
		\label{R}
		\Re= g^{\mu\nu}\Re_{\mu\nu}\equiv g^{\mu\nu}\left(\hat{\Gamma}^\alpha_{\mu\nu,\alpha}-\hat{\Gamma}^\alpha_{\mu\alpha,\nu}+\hat{\Gamma}^\alpha_{\alpha\lambda}\hat{\Gamma}^\lambda_{\mu\nu}-\hat{\Gamma}^\alpha_{\mu\lambda}\hat{\Gamma}^\lambda_{\alpha\nu}\right),
	\end{eqnarray}
 where $\hat{\Gamma}^\alpha_{\mu\nu}$ is an independent connection.
 
To facilitate the analysis of HMPG, it is advantageous to reformulate the theory in its scalar-tensor representation. This approach simplifies the computational process and provides clearer insights into the underlying dynamics. Through a series of transformations (detailed in \cite{Harko2012, Capozziello2015}), the action of HMPG can be equivalently expressed in terms of a scalar field, leading to the following form:
	\begin{equation}
		\label{action4}
		S=\frac{1}{2k^2}\int d^4x\sqrt{-g}\left[(1+\phi) R+\frac{3}{2\phi}\partial_\mu\phi\partial^\mu\phi-V(\phi)\right]+S_m,
	\end{equation}
where $\phi$ is a scalar field and $V(\phi)$ is a scalar potential.

Upon varying the action in the scalar-tensor form \eqref{action4} with respect to both the metric and the scalar field, we derive the respective field equations \cite{Harko2012, Capozziello2015}:
	\begin{eqnarray}
		\label{feq_g}
		\frac{1}{1+\phi}\Bigg[k^2\left(T_{\mu\nu}-\frac{1}{2}g_{\mu\nu}T\right)+\frac{1}{2}g_{\mu\nu}\left(V+\nabla_\alpha\nabla^\alpha\phi\right)+\nabla_\mu\nabla_\nu\phi -\frac{3}{2\phi}\partial_\mu\phi\partial_\nu\phi\Bigg]=R_{\mu\nu},
\end{eqnarray}
\begin{equation}
		\label{feq_phi}
		-\nabla_\mu\nabla^\mu\phi+\frac{1}{2\phi}\partial_\mu\phi\partial^\mu\phi+\frac{\phi[2V-(1+\phi)V_\phi]}{3}=\frac{\phi k^2}{3}T.
	\end{equation}
 
 It is important to emphasize that scalar field is dynamical in HMPG unlike the Palatini case. 
 
 Further we consider the theory exclusively in scalar-tensor form.
 
  \section{Numerical spherically symmetric black hole solution}\label{sec3}
 The main objective of this paper is to obtain images of accretion disks around a static spherically symmetric BHs in HMPG. The corresponding BH solution was found in the article \cite{Danila2019}. In our work, we repeated this result and obtained the static spherically symmetric BH solution in numerical form using integration methods from Python's SciPy library. The details of obtaining the solution are given below.
  
The line element describing the spherically symmetric space-time around a massive body is given as follows:
 	\begin{equation} \label{metric}
 	ds^2=-e^{\nu(r)}c^2dt^2+e^{\lambda(r)}dr^2+r^2(d\theta^2+\sin^2\theta d\varphi^2).
 	\end{equation}
 	The metric functions $\nu(r)$ and $\lambda(r)$ are solely functions of the radial coordinate $r$, which ranges from 0 to $\infty$. Considering the metric defined in (\ref{metric}), the field equations (\ref{feq_g}) and (\ref{feq_phi}) can be expressed as follows \cite{Danila2019}:
\begin{equation}   
 \frac{d\phi}{d\xi}=-\frac{U}{\xi^2},
 \end{equation}
  \begin{eqnarray}
 \frac{dM_{eff}}{d\xi}=\frac{(1-M_{eff}\xi)[\xi^2dU/d\xi+3U^2/4\phi-2\xi U]+M_{eff}\xi^3(1+\phi)-v}{\xi^4(1+\phi+U/2\xi)} -\frac{M_{eff}}{\xi},
 \end{eqnarray}

 \begin{equation}
  \frac{d\nu}{d\xi}=-\frac{\xi-\bigl\{\frac{U(\xi)[8\phi+3U(\xi)/\xi]}{4\phi(1+\phi)}+\xi\bigr\}[1-\xi M_{eff}(\xi)]-\frac{v(\phi)}{\xi(1+\phi)}}{\xi^2[1-\xi M_{eff}(\xi)]\bigl[1+\frac{U(\xi)}{2\xi(1+\phi)}\bigr]},
 \end{equation}
\begin{align}
\frac{d^2\nu}{d\xi^2}=&\frac{(1-\frac{\xi}{2}\frac{d\nu}{d\xi})(-\xi\frac{dM_{eff}}{d\xi}-M_{eff})}{\xi(1-\xi M_{eff})}-\frac{5U(\xi)^2}{2\xi^4\phi(1+\phi)}+\frac{2U}{\xi^3(1+\phi)}-\frac{1}{2}\biggl(\frac{d\nu}{d\xi}\biggr)^2+\frac{1}{\xi}\frac{d\nu}{d\xi}\nonumber\\
 &-\frac{2}{\xi^4(1+\phi)(1-\xi M_{eff})}\biggl\{\frac{2\phi}{3}[2v-(1+\phi)v_\phi]+v\biggr\}, 
\end{align}
\begin{gather}
\begin{split}
  \frac{dU(\xi)}{d\xi}=&\frac{\frac{\xi^2 U(\xi)}{2}\bigl[\xi\frac{d M_{eff}(\xi)}{d\xi}+M_{eff}(\xi)\bigr]}{\xi^2(1-\xi M_{eff}(\xi))}+\frac{2U(\xi)}{\xi}-\frac{1}{\xi^2}\frac{U^2(\xi)}{2\phi}\\
&-\frac{\frac{2\phi}{3}\bigl[2v(\phi)-(1+\phi)v_\phi(\phi)\bigr]}{\xi^2(1-\xi M_{eff}(\xi))}-\frac{U(\xi)}{2}  \frac{d\nu}{d\xi}. 
\end{split}
\end{gather}

These equations were derived using the following dimensionless variables:
\begin{equation}
\xi=\frac{2GM_{BH}}{c^2r},\ \ \ \ \frac{d\phi}{d r}=\frac{c^2}{2GM_{BH}}U, \ \ \ \ 
V(\phi)=2\biggl(\frac{c^2}{2GM_{BH}}\biggr)^2v(\phi),
\end{equation}
where $M_{BH}$ is the BH mass.
The metric function $e^{-\lambda(r)}$ is redefined as
\begin{equation}
e^{-\lambda(r)}=1-\frac{2GM_{BH} M_{eff}(r)}{c^2r}.
\end{equation}

To obtain numerical static spherically symmetric solutions, we used the following fixed initial conditions
\begin{equation}
M_{eff} (0) = 1, \ \ \ \nu(0) = 0, \ \ \ \nu'(0) = 0
\end{equation}
and arbitrary numerical values for 
\begin{equation}
u(0) = u_0, \ \ \ \  \phi(0) =\phi_0.
\end{equation}

More details of the calculations can be found in the article \cite{Danila2019}.

\section{Thin accretion disk}\label{sec4}

An accretion disk is a structure that forms around a massive object, consisting of diffuse matter orbiting the central body. In this work, we focus on thin accretion disks, a model initially developed by Shakura and Sunyaev \cite{Shakura1973} and later extended by Novikov, Thorne, and Page \cite{Novikov1973, Page1974}. A distinguishing feature of such disks is their small vertical thickness \( h \) compared to their radial size (\( h \ll r \)). In these disks, particles follow Keplerian orbits, and the disk is located in the equatorial plane of the compact object. The thin accretion disk model assumes a steady-state configuration, where the mass accretion rate \( \dot{M}_0 \) remains constant over time, and the accreting matter is in thermodynamic equilibrium. This approximation is treated here as a first step toward constructing more realistic accretion-disk models. It allows us to isolate and identify the fundamental geometric differences between HMPG and GR before introducing the additional physical complexities of realistic accretion disks.

Further we consider the geometry of space-time around a compact object in the case of spherically symmetric BH in HMPG. We divide our narrative into two parts: an overview of the motion of massive and massless particles. In this section, we use geometric units in which $G=c=1$.

\subsection{Motion of massive particles}
The geodesic motion of test particles in the vicinity of a massive body can be described using Lagrangian formalism as follows:
\begin{equation}\label{lagrangian}
\mathcal{L}=\frac{1}{2}g_{\mu\nu}\dot x^\mu \dot x^\nu,
\end{equation}
where the dot means the derivative with respect to an affine parameter $\tau$ along the
geodesic $x^\mu(\tau)$; $g_{\mu\nu}$ is the metric of a static spherically symmetric space-time:
	\begin{equation}
	ds^2=g_{00}dt^2+g_{11}dr^2+g_{22}d\theta^2+g_{33}d\varphi^2,
	\end{equation}
	for which the elements $g_{00}, g_{11}, g_{22}, g_{33}$ only depend on the radial coordinate $r$. It is worth to noting that for massive particles, the Lagrangian $\mathcal{L}=1/2$, while for massless particles, such as photons, $\mathcal{L}=0$.  As we assume a motion along Keplerian orbits and use an equatorial approximation, we imply $|\theta - \pi/2| \ll 1$. Using the Euler-Lagrange equations, we derive  the specific energy $\tilde E$ and specific angular momentum $\tilde L$ of the accretion disk \cite{Page1974}:
  \begin{equation} \label{E}
		-g_{00}\dot t=\tilde E,
	\end{equation}	
		\begin{equation}\label{L}
		g_{33}\dot \varphi=\tilde L,
	\end{equation}
	\begin{equation}\label{V}
			-g_{00}g_{11}\dot r^2+V_{eff}(r)=\tilde E^2.
	\end{equation}
	In this context, $\tilde E=E/m_0c^2$ and $\tilde L=L/m_0c$, where $E$ denotes the total energy of the particle in its orbit, $m_0c^2$ is the rest mass energy of this particle and $L$ represents the angular momentum of the particle (here we have reinstated $c$ for clarity). In the equation \eqref{V} $V_{eff}(r)$ is the effective potential. This quantity is defined as
	\begin{equation}
		V_{eff}(r)=-g_{00}\biggl(1+\frac{\tilde L^2}{g_{33}}\biggr).
	\end{equation}

The effective potential plays a crucial role in determining the radial motion of particles and the stability of their orbits around massive objects. Stable circular orbits occur where the minimum of the effective potential is located, with the conditions $V_{eff}(r) = 0$ and $V_{eff,r}(r) = 0$ ensuring stability. If the potential has no minimum, the orbits are unstable. By applying these conditions, one can derive key parameters such as the specific energy $\tilde E$, specific angular momentum $\tilde L$, and angular velocity $\Omega$ for particles in stable circular orbits:
\begin{equation}\label{E1}
		\tilde E=-\frac{g_{00}}{\sqrt{-g_{00}-g_{33}\Omega^2}},
	\end{equation}	
		\begin{equation}\label{L1}
		\tilde L=\frac{g_{33}\Omega}{\sqrt{-g_{00}-g_{33}\Omega^2}},
	\end{equation}	
		\begin{equation}
		\Omega=\frac{d\varphi}{dt}=\sqrt{\frac{-g_{00,r}}{g_{33,r}}}.
	\end{equation}
The condition $V_{eff,rr}(r) = 0$ determines the radius of  the innermost stable circular orbit $r_{isco}$. This leads to the following equation:

		\begin{equation}
		\tilde E^2 g_{33,rr}+\tilde L^2 g_{00,rr}+(g_{00}g_{33})_{,rr}=0.
	\end{equation}
	Solving this equation with respect to $r$ and taking into account expressions (\ref{E1}) and (\ref{L1}), we obtain the $r_{isco}$.
	
	\subsection{Motion of massless particles}
   	Now, let us describe the trajectory of a light ray emitted by the accretion disk and deflected by the BH. In this case, the dynamics of the photon can be described by the Lagrangian \eqref{lagrangian} with the condition $L = 0$. Because photons move through a static, spherically symmetric spacetime, we can simplify the analysis by restricting their motion to any equatorial plane without losing generality. Then we can use the  set of metric functions from \eqref{metric}.
		
	Thus, similarly to the case of a massive particle, we can determine the key parameters for describing the motion of a photon:
  \begin{equation} \label{E2}
		-e^{\nu(r)}\dot t=\mathcal{E},
	\end{equation}	
		\begin{equation}\label{L2}
		r^2\dot \varphi=\mathcal{L},
	\end{equation}
	\begin{equation}\label{V2}
			e^{\nu(r)+\lambda(r)}\dot r^2+\tilde V_{eff}(r)=\mathcal{E}^2.
	\end{equation}
	However, in this case, the effective potential differs from that of a massive particle and is defined as follows:
		\begin{equation}
		\tilde V_{eff}(r)=\frac{e^{\nu(r)}\mathcal{L}^2}{r^2}.
	\end{equation}
	
	Next, let us introduce an important quantity for our purposes, known as the impact parameter: 
	\begin{equation}
	b=\mathcal{L}/\mathcal{E}.
	\end{equation}
This parameter can be defined for each individual particle, and its value determines the trajectory of a photon. In this context, it is essential to introduce the concept of the critical impact parameter $b_c$. If a particle's impact parameter exceeds this critical value $b>b_c$, it will be deflected by the gravitational field of the compact object and escape to infinity. If the impact parameter is approximately equal to the critical value $b\approx b_c$, the particle will move in a nearly unstable orbit around the BH, completing several revolutions before either escaping or falling into the BH due to small perturbations. However, if the impact parameter is smaller than the critical value $b<b_c$, the particle will inevitably cross the event horizon and be captured by the BH. The critical impact parameter is determined by photons located in the region of the photon sphere and can be found by satisfying the following two conditions:$\frac{d V_{\text{eff}}}{dr} = 0$ and $\dot{r} = 0$. Then we have the following expression:
\begin{equation}
b_c=\frac{r}{\sqrt{e^{\nu(r)}}}.
\end{equation}
It is important to note that $b_c$ corresponds to the radius of the photon ring as seen by a distant observer.

Our goal is to derive the trajectory of a photon $\varphi(r)$ in the equatorial plane. For this purpose, we use expressions \eqref{L2} and \eqref{V2}. By expressing $\dot\varphi$ and $\dot r$, we divide equation \eqref{V2} by equation \eqref{L2} and obtain:
\begin{equation}\label{rf}
\frac{dr}{d\varphi}=\frac{|\mathcal{L}|}{\mathcal{L}}\cfrac{r^2}{\sqrt{e^{\nu(r)+\lambda(r)}}}\sqrt{\frac{1}{b^2}-\frac{e^{\nu(r)}}{r^2}}.
\end{equation}
The sign of $\frac{|\mathcal{L}|}{\mathcal{L}}$ determines the direction of photon motion. Further we can introduce new variables $u=1/r$, $f(u)=e^{\nu(u)}$ and $g(u)=e^{\lambda(u)}$, than the expression \eqref{rf} transforms into
\begin{equation}\label{uf}
\frac{du}{d\varphi}=\frac{|\mathcal{L}|}{\mathcal{L}}\sqrt{\frac{1}{b^2f(u)g(u)}-\cfrac{u^2}{g(u)}}.
\end{equation}

After that we obtain:
\begin{equation}\label{intuf}
\gamma=\frac{|\mathcal{L}|}{\mathcal{L}}\int^{u_s}_{u_o}\cfrac{du}{\sqrt{\frac{1}{b^2f(u)g(u)}-\frac{u^2}{g(u)}}},
\end{equation}
where $u_s = 1/r_s$ and $u_o = 0$ denote the positions of the light source and distant observer, respectively; the quantity $\gamma$ arises from integrating over the azimuthal angle $\varphi$.

When calculating the trajectory of a photon, it is important to consider turning points \cite{2022hu}. Turning points are the locations where the radial velocity of the photon becomes zero, i.e., $\dot r=0$. At these points, the photon reverses direction and starts moving in the opposite radial direction. The value of the impact parameter $b$ for the turning points can be determined from the equation of the effective potential by satisfying the condition 
$\tilde V_{eff}=\mathcal{E}^2$ together with $\dot r=0$. This leads us to the following expression
\begin{equation}
\frac{1}{b^2}=u_p^2f(u_p),
\end{equation}
where $u_p$ is the turning point. 

Thus, in order to account for all possible photon trajectories, we need to split the integral into a sum and carefully consider the direction of the photon's motion, specifically the sign of the expression $|\mathcal{L}|/\mathcal{L}$ \cite{2022hu}:
\begin{equation}\label{intuf1}
\gamma=\frac{|\mathcal{L}|}{\mathcal{L}}\int^{u_p}_{u_o}\cfrac{du}{\sqrt{\frac{1}{b^2f(u)g(u)}-\frac{u^2}{g(u)}}}+\frac{|\mathcal{L}|}{\mathcal{L}}\int^{u_s}_{u_p}\cfrac{du}{\sqrt{\frac{1}{b^2f(u)g(u)}-\frac{u^2}{g(u)}}}.
\end{equation}
Each point on the accretion disk observed by a distant observer corresponds to a unique pair of impact parameter and polar angle \((b, \eta)\) in the photographic plate. Therefore, mapping the \((b, \eta)\) space is equivalent to covering the entire disk. Thus, we need to determine all possible pairs \((b, \eta)\) to construct a complete view of the disk. Expression \eqref{intuf1} includes the impact parameter $b$ but does not account for the polar angle $\eta$, which can be obtained from the following assumptions.

\begin{figure}
\begin{center}
		\begin{minipage}[h]{0.8\textwidth}
			\includegraphics[width=\columnwidth]{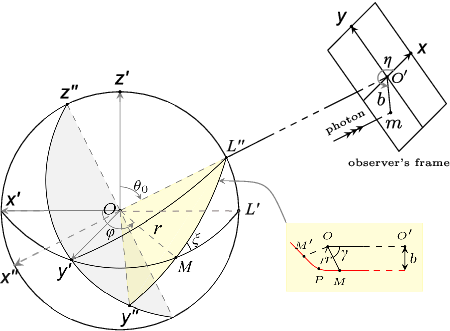}
					\end{minipage}

		\caption{The coordinate system of the disk}
		\label{fig:1}
	\end{center}
	\end{figure}

We selected a coordinate system (see Fig. \eqref{fig:1}) developed on the basis of the following publications \cite{2022hu, Tian_2019, luminet}. The accretion disk is located in the \( \overline{x'oy'} \)-plane, and the BH's singularity is positioned at point \( o \). The disk consists of elements, denoted as \( M \), which emit light that reaches point \( m \) on the screen of a distant observer. The photon trajectory lies in the \( \overline{y''oL''} \)-plane   (shown in yellow in the online version) and becomes parallel to the \( oo' \) line when the photon is far from the BH. The observer is assumed to be at infinity in the coordinate system \( \overline{xo'y} \). The polar distance from \( m \) to \( o' \) corresponds to the impact parameter of the trajectory, \( b \). It is important to note that we define the polar angle differently from most previous studies: we follow the classical convention, measuring the angle counterclockwise from the $x$-axis to the line connecting point $m$ with point $O'$. In this setup, the  angle $y'Oy''$ is equal to $\eta - \pi$. Thus, light emitted from point \( M \) with coordinates \( (r, \phi) \) reaches point \( m \) on the screen of the distant observer, where its coordinates are \( (b, \eta) \). 

Exploiting spherical symmetry, the deflection angle $\gamma$ can be expressed through the inclination angle $\theta_0$ and the polar angle $\eta$ as
\begin{equation}\label{cos}
\cos{\gamma}=-\cfrac{\sin{\eta}\tan{\theta_0}}{\sqrt{\sin^2{\eta}\tan^2{\theta_0}+1}}.
\end{equation}
It is important to note that this expression differs from those presented in \cite{Tian_2019, luminet}, but agrees with the results of \cite{2022hu}. This discrepancy arises from the fact that we adopt a different convention for measuring the polar angle $\eta$. The polar coordinate \( \eta \) lies within \( [0, 2\pi] \), the deflection angle \( \gamma \) satisfies \( \pi/2 - \theta_0 \leq \gamma \leq \pi/2 + \theta_0\) for direct images, and $\theta_0 \in [0, \pi/2]$. However, some light rays are deflected at larger angles or even complete multiple orbits before reaching the observer, forming secondary and higher-order images.  The formation of direct and secondary images around a BH results from gravitational lensing, where light rays follow different paths due to spacetime curvature. The direct image consists of photons traveling directly from the accretion disk to the observer with minimal deflection, forming the brightest and most distinct disk image. The secondary image arises from photons that complete additional orbits around the BH before reaching the observer, making it fainter due to the longer optical path and gravitational redshift.  To account for all these contributions, equation \eqref{cos} must be modified as
\begin{equation}\label{cos1}
\gamma=k\pi-\arccos{\cfrac{\sin{\eta}\tan{\theta_0}}{\sqrt{\sin^2{\eta}\tan^2{\theta_0}+1}}},
\end{equation}
 where \( k = 1 \) corresponds to direct  images, and \( k = 2 \) represents secondary  images. Higher-order images (\( k \geq 2 \)) appear increasingly closer to the photon ring \cite{Gyulchev_2019, Gyulchev_2020, Gyulchev_2021}.

Therefore, we have the relation between the impact parameter $b$ and the
polar coordinate $\eta$ as
\begin{equation}\label{int_1}
\frac{|\mathcal{L}|}{\mathcal{L}}\int^{u_s}_{u_o}\cfrac{du}{\sqrt{\frac{1}{b^2f(u)g(u)}-\frac{u^2}{g(u)}}}=k\pi-\arccos{\cfrac{\sin{\eta}\tan{\theta_0}}{\sqrt{\sin^2{\eta}\tan^2{\theta_0}+1}}},
\end{equation}
for rays without a turning point. For rays with a turning point this relation is
\begin{eqnarray}\label{int_2}
&\frac{|\mathcal{L}|}{\mathcal{L}}\int^{u_p}_{u_o}\cfrac{du}{\sqrt{\frac{1}{b^2f(u)g(u)}-\frac{u^2}{g(u)}}}+\frac{|\mathcal{L}|}{\mathcal{L}}\int^{u_s}_{u_p}\cfrac{du}{\sqrt{\frac{1}{b^2f(u)g(u)}-\frac{u^2}{g(u)}}}\nonumber\\
&=k\pi-\arccos{\cfrac{\sin{\eta}\tan{\theta_0}}{\sqrt{\sin^2{\eta}\tan^2{\theta_0}+1}}}.
\end{eqnarray}
 To construct the image of the accretion disk, it is necessary to determine the full set of impact parameters $b$ and polar angles $\eta$ corresponding to all observed disk points. The given integral equation \eqref{int_2} relates the impact parameter on the left-hand side to the polar angle on the right-hand side. By calculating the integral numerically on the left for various values of $b$, we can determine whether a corresponding $\eta$ in the range of $[0,2\pi]$ exists that satisfies the equation and identify the specific polar angles for which this holds. So we obtain a point in the picture plane in polar coordinates $(b,\eta)$. For constructing image we repeat this procedure for each point in the picture plane. For this reason we used numerical methods from SciPy package for Python. The corresponding implementation is available online (see Code Availability Statement).

To construct an image of a BH, it is essential to consider the time-averaged energy flux, which is a key characteristic of the accretion disk. This flux is derived from the conservation laws of energy and angular momentum. A detailed study of this quantity in the framework of HMPG is presented in \cite{meacc}; here, we summarize the main aspects. 

The avereged energy flux per unit area can be expressed as:
	\begin{equation}\label{page_thorne_flux}
		F_e(r)=-\frac{\dot M_0}{4\pi\sqrt{-g}}\frac{\Omega_{,r}}{(\tilde E-\Omega \tilde L)^2} \int^r_{r_{isco}}(\tilde E-\Omega \tilde L) \tilde L_{,r}rdr,
	\end{equation}
where  $r$ corresponds to stable circular orbits in the disk and the emitted energy flux is $F_e$. The observed energy flux $F_o$ is connected with the emitted quantity through redshift factor $\mathfrak{g}$ as
\begin{equation}
F_o=\cfrac{F_e}{\mathfrak{g}^4}.
\end{equation}

In deriving the expression for gravitational redshift factor $\mathfrak{g}$, we rely on the following studies \cite{luminet, Tian_2019, 2022hu}. Let us briefly review the derivation. By definition, gravitational redshift is the ratio of the photon’s energy at the point of emission $E_e$ to its energy at the point of observation $E_o$:
\begin{equation}
\mathfrak{g}=1+z=\cfrac{E_e}{E_o}=\cfrac{(p_\mu u^\mu)_e}{(p_\mu u^\mu)_o},
\end{equation}
where $p^\mu$ is photon's 4-momentum, $u_\mu$ is the 4-velocity of the emitting particle. For a particle moving in a circular orbit in the plane $\theta=\pi/2$, the 4-velocity has the form: $u_e^\mu=(u^t,0,0,\Omega u^t)$, where $\Omega=d\phi/dt$. From $u^\mu u_\mu=-1$ we obtain $u^t=1/\sqrt{-g_{00}-g_{33}\Omega^2}$. At the same time
 $p_\mu=(p_t,p_r,p_\theta,p_\phi)=(-\mathcal{E},p_r,p_\theta,\mathcal{L})$. Then we obtain the following expression:
\begin{equation}
E_e=p_tu^t+p_\phi u^\phi=p_tu^t\biggl(1+\frac{p_\phi}{p_t}\Omega\biggr).
\end{equation}
The only nonzero component of the observer's 4-velocity is $u^t_{o}=1$. Consequently, the observed energy of this photon is $E_o=p_tu^t_o=p_t$.

Further we consider the ratio $p_\phi/p_t$.  When the photon's plane of motion does not coincide with the equatorial plane of the disk, it is necessary to project the components of the 4-momentum onto the equatorial plane. The time component \( p_t \), associated with energy conservation, remains unchanged, while the azimuthal component \( p_\phi \), representing the angular momentum relative to the axis of symmetry, is projected as \( p_\phi = \mathcal{L} \cos(\xi) \), where \( \xi \) is the angle between the photon's plane of motion and the equatorial plane. As a result, the ratio of the 4-momentum components is expressed as \( \frac{p_\varphi}{p_t} = -b \cos(\xi) \), where \( b = \frac{\mathcal{L}}{\mathcal{E}} \) is the impact parameter.

Using the sine theorems of spherical triangles $\triangle My''y'$ and $\triangle ML''L'$ (see Fig. \ref{fig:1}) we obtain the following relationships \cite{Tian_2019, Liu_2022}:
\begin{equation}
\sin{\xi}=\cfrac{\cos{\theta_0}}{\sin{\gamma}}, \ \ \ \ \ \ \cfrac{\sin{\theta_0}}{\cos{\gamma}}=-\cfrac{\sin{\xi}}{\sin{\eta}}.
\end{equation}
Using these ratios and expression \eqref{cos} we obtain
\begin{equation}
\cos{\xi}=-\sin{\theta_0}\cos{\eta}.
\end{equation}

 Thus we can find the expression for the gravitational redshift:
\begin{equation}
\mathfrak{g}=1+z=\cfrac{1+\Omega b\sin{\theta_0}\cos{\eta}}{\sqrt{-g_{00}-g_{33}\Omega^2}}.
\end{equation}
 This definition of the redshift is in agreement with that presented in \cite{2022hu}.

We now have all the necessary components to construct images of accretion disks. The fundamental parameters, including the time-averaged energy flux and gravitational redshift, provide the basis for accurately modeling the disk’s appearance. In the next section, we will discuss these parameters in greater detail and examine the impact of the HMPG parameters on the resulting images. This will lead to a comprehensive analysis of the accretion disk visuals and their physical interpretation.

\section{Images of black hole}\label{sec5}
In this section, we proceed directly to the discussion of our obtained results, namely the images of accretion disks around static spherically symmetric BHs within HMPG. We study two cases: a case without potential $V=0$ and a case where potential takes the Higgs-type form $V=-\frac{\mu^2}{2}\phi^2+\frac{\zeta}{4}\phi^4$.

In our previous work \cite{meacc}, we studied accretion disks in HMPG. The energy flux, temperature distribution, emission spectrum and  efficiency of accretion disks around HMPG BHs were numerically calculated. In a case without potential we considered a range of values for \( \phi_0 \)  with a fixed \( u_0 \) and vice versa. Additionally, we examined the case where a connection between these two parameters originating from the post-Newtonian formalism was observed. The most significant deviations from the Schwarzschild BH were observed when we fixed the parameter \( u_0 \) and considered a range of \( \phi_0 \). In contrast, introducing a connection between the parameters \( u_0 \) and \( \phi_0 \) led to results that differed little from the predictions for a Schwarzschild BH, with the difference in the maximum of energy flux amounting to less than 0.1\%. Therefore, in this work, we  focus specifically on these two cases to avoid overloading the text with images. From the entire range of \( \phi_0 \), we consider only extreme case \( \phi_0 = 0.5 \),  keeping the parameter \( u_0 =5.12\times 10^{-7}\). In the case of a connection between $\phi_0$ and $u_0$, this relationship was obtained from the post-Newtonian analysis. This connection has the following form \cite{meacc}:
\begin{equation}\label{u0}
u_0=\frac{2GM\phi_0}{3c^2r^2}.
\end{equation} 
It can be obtained from the expression for scalar perturbation $\varphi=\frac{-2GM\phi_0e^{-m_\phi r}}{3c^2r}$ \cite{Dyadina2019}. In this case, we consider $|\phi_0|\lesssim4\times10^{-5}$ \cite{Leanizbarrutia2017}. This limit was imposed on the initial value of the scalar field using the data from the Cassini experiment \cite{Cassini2003}. In this case $u_0=1.3\times10^{-11}$.

Also we consider the  case with potential of Higgs-type form:
\begin{equation}
V=-\frac{\mu^2}{2}\phi^2+\frac{\zeta}{4}\phi^4,
\end{equation}
where $\mu^2$ and $\zeta$ are constants. Firstly, we redefine constants $\mu^2$ and $\zeta$ into a dimensionless form as \cite{Danila2019}
\begin{equation}\label{valpha}
v(\phi)=\alpha \phi^2+\beta \phi^4,
\end{equation}
where
\begin{equation}
\alpha=-\frac{1}{4}\biggl(\frac{ 2GM_{BH} }{ c^2}\biggr)^2\mu^2,\ \ \ \ \ \ \beta=\frac{1}{2} \biggl(\frac{ 2GM_{BH} }{ c^2}\biggr)^2\zeta.
\end{equation}
Thus the Higgs-type potential yields four-parameter $(\alpha, \beta,\phi_0, u_0)$ solutions of the static gravitational field equations in HMPG. In article \cite{meacc}, we studied accretion disks by varying all these parameters over a wide range of values. In this work, however, we focus on the case that deviates the most from the predictions of GR. This case corresponds to the set of parameters \(\alpha=-10^{-10}, \beta=10^{-11}, \phi_0=4\). Also we save the connection between $u_0$ and $\phi_0$, which is known from PPN analysis as derivative of $\varphi=\frac{-2GM\phi_0e^{-m_\phi r}}{3c^2r}$ with respect to $r$. As a result, we get the expression \cite{meacc}:
\begin{equation}\label{u0V}
u_0=-\frac{2GM\phi_0e^{-m_\phi r}m_\phi}{3c^2r} - \frac{2GM\phi_0e^{-m_\phi r}}{3c^2r^2}. 
\end{equation}
 Additionally, following \cite{Capozziello2015, Dyadina2019}, the mass of the scalar field is defined as
\begin{equation}\label{ppn}
m_\varphi^2=[2V_0-V_\phi-(1+\phi)\phi V_{\phi \phi}]/3|_{\phi=\phi_0},
\end{equation} 
where the subscript $\phi$ denotes differentiation with respect to the scalar field.
Accordingly, $m_\varphi^2$ can be related to the parameters $\alpha$ and $\beta$ via
\begin{equation}\label{hyggs}
m_\varphi^2=[- 4/3\alpha\phi_0 - 16/3\beta\phi_0^3 -10/3\beta\phi_0^4 ] \times 2\bigg(\frac{c^2}{2GM_{BH}}\bigg)^2.
\end{equation}

Besides, we present the case of a Schwarzschild BH for comparison. All the cases considered in this paper are summarized in Tab.~\ref{tab:models}.

			\begin{table}[t]
\centering
\caption{Model variants and corresponding parameter values}
\label{tab:models}
\begin{tabular}{ll}
\hline
Label & Description / parameters \\
\hline
(I)   & HMPG with $\phi_0=0.5$ and $u_0=5.12\times10^{-7}$ \\
\\
(II)  & HMPG with post-Newtonian relation between $\phi_0$ and $u_0$ \\
      & ($\phi_0=4\times10^{-5}$, $u_0=1.3\times10^{-11}$) \\
      \\
(III) & HMPG with a Higgs-type scalar-field potential \\
      & ($\alpha=-10^{-10}$, $\beta=10^{-11}$, $\phi_0=4$) together with the \\
      & post-Newtonian relation between $\phi_0$ and $u_0$ \\
      \\
(IV)  & Schwarzschild space-time (for comparison) \\
\hline
\end{tabular}
\end{table}

\begin{figure}
\begin{center}
		\begin{minipage}[h]{0.7\textwidth}
			\includegraphics[width=\columnwidth]{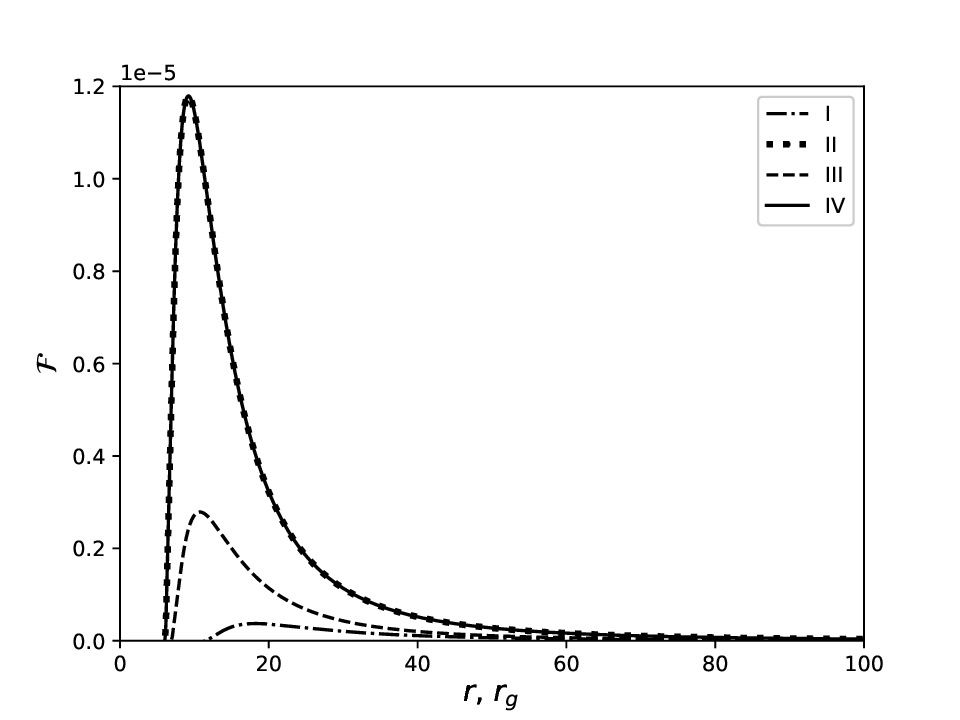}
					\end{minipage}

		\caption{Radial profiles of the normalized emitted flux $\tilde{\mathcal{F}}(r)$ for the considered black-hole configurations
}
		\label{fig:2}
	\end{center}
	\end{figure}

	Before turning to the ray-traced images, we first compare the radial profiles of the emitted flux for all considered configurations. The flux \eqref{page_thorne_flux} is presented in the normalized form
	\begin{equation}
\tilde{\mathcal{F}}=\frac{F_e(r)\, r_g^2}{c^2\dot M},
\end{equation}
 which removes the trivial scaling with the accretion rate and highlights purely geometrical differences between the models. Fig.~\ref{fig:2} shows $\tilde{\mathcal{F}}(r)$ for cases (I)--(III) and for the Schwarzschild metric. In all cases the flux vanishes at the inner edge, $r=r_{\rm isco}$, increases rapidly to a maximum at a few gravitational radii, and then gradually decreases with radius. The largest flux is obtained for the Schwarzschild metric (case~(IV)) and for the case~(II), while the smallest flux corresponds to the case ~(I); the Higgs-type potential case (case~(III)) yields intermediate flux values. Moreover, case~(II) is practically indistinguishable from the Schwarzschild case~(IV) over the entire disk.
 
 A more detailed discussion of closely related flux profiles, together with the corresponding temperature distributions, emission spectra, and energy conversion efficiencies for accretion disks around the same class of black holes, can be found in our previous work~\cite{meacc}. 
There we considered a broader set of scalar-field configurations and focused on stellar-mass black holes without using the above normalization. While we employ the normalized flux $\tilde{\mathcal{F}}$ in this paper, the qualitative conclusions remain unchanged: configurations that deviate more strongly from GR generally yield cooler disks and a lower emitted flux than the Schwarzschild case.


\begin{figure*}
		\begin{minipage}[h]{.37\textwidth}
		{\small (I)}\\
			\includegraphics[width=\columnwidth]{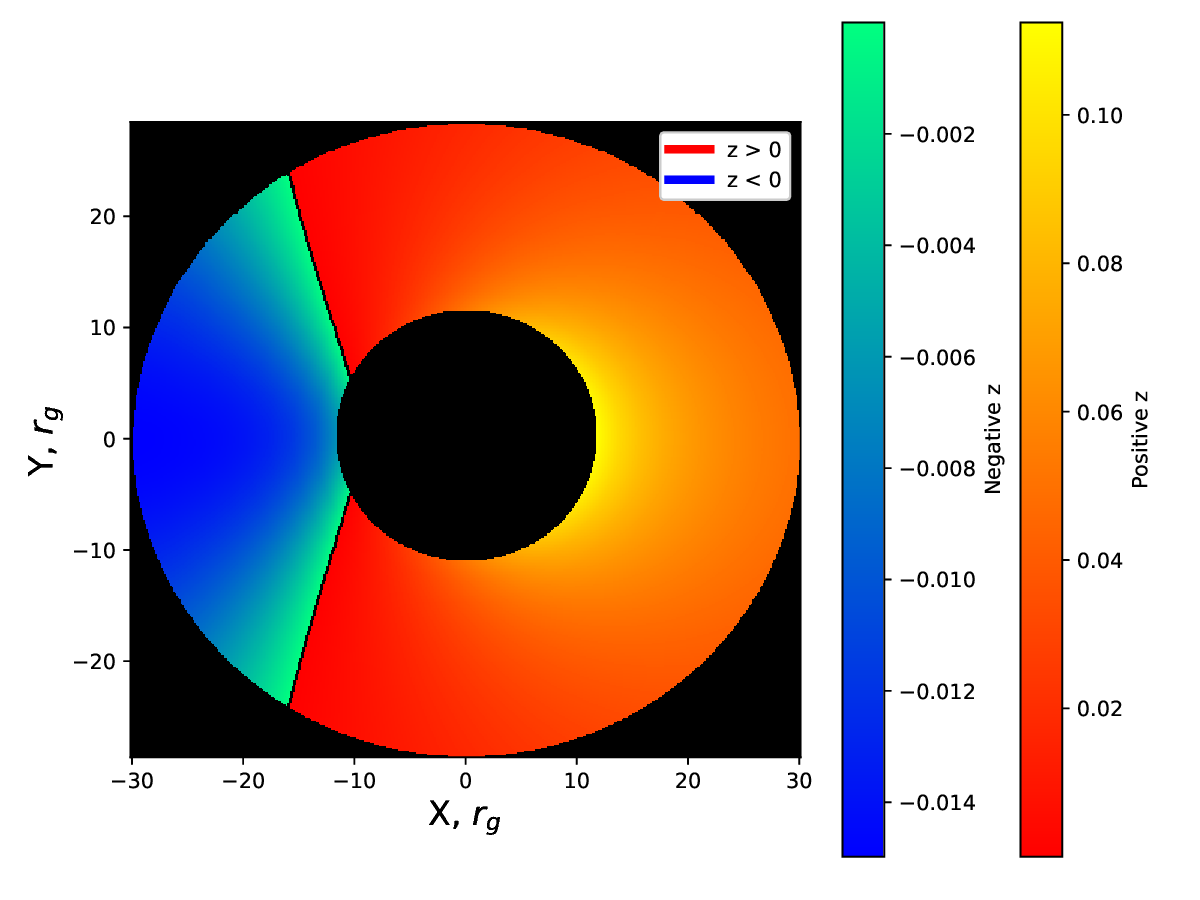} \\ 
					\end{minipage}
\hfill
		\begin{minipage}[h]{.37\textwidth}
	{\small (II)}\\
			\includegraphics[width=\columnwidth]{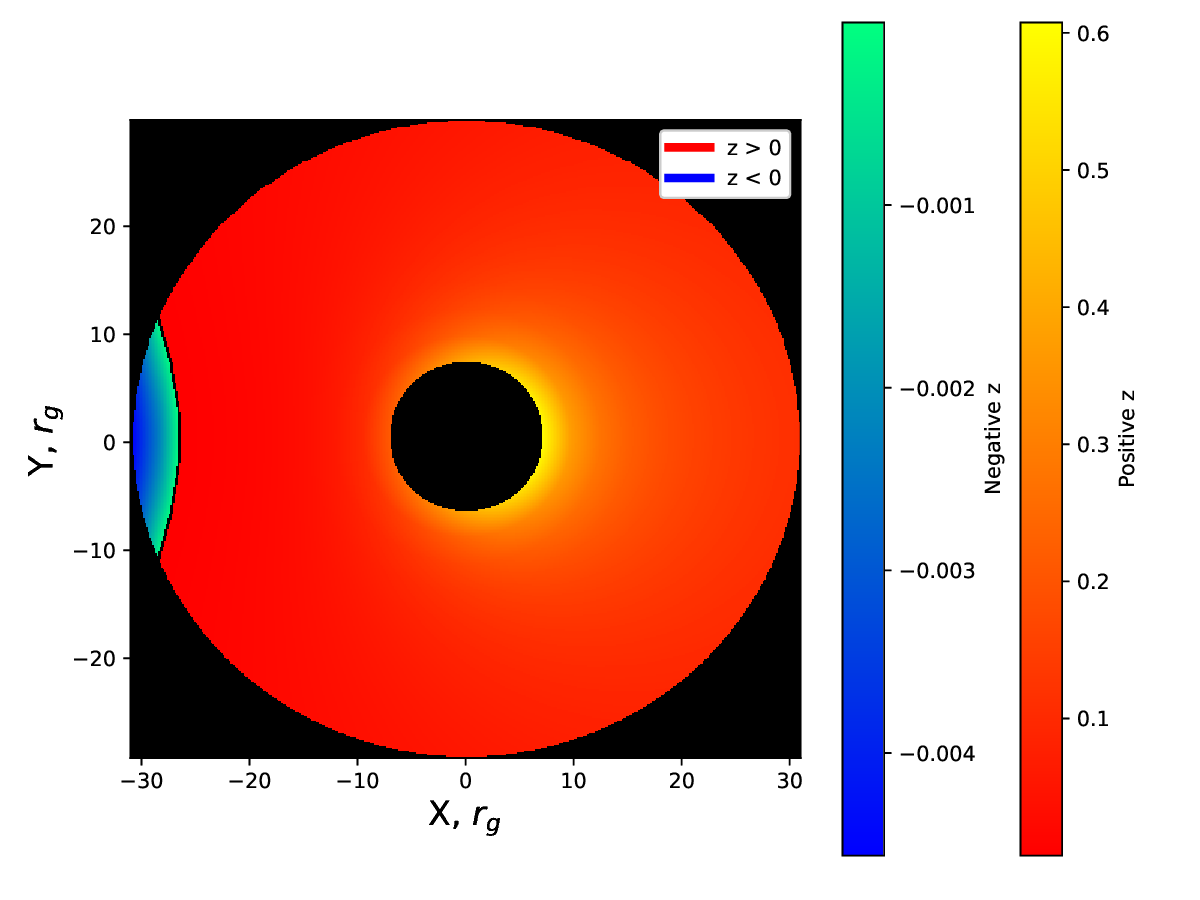} \\ 
					\end{minipage}
					\hfill
\begin{minipage}[h]{.37\textwidth}
			\includegraphics[width=\columnwidth]{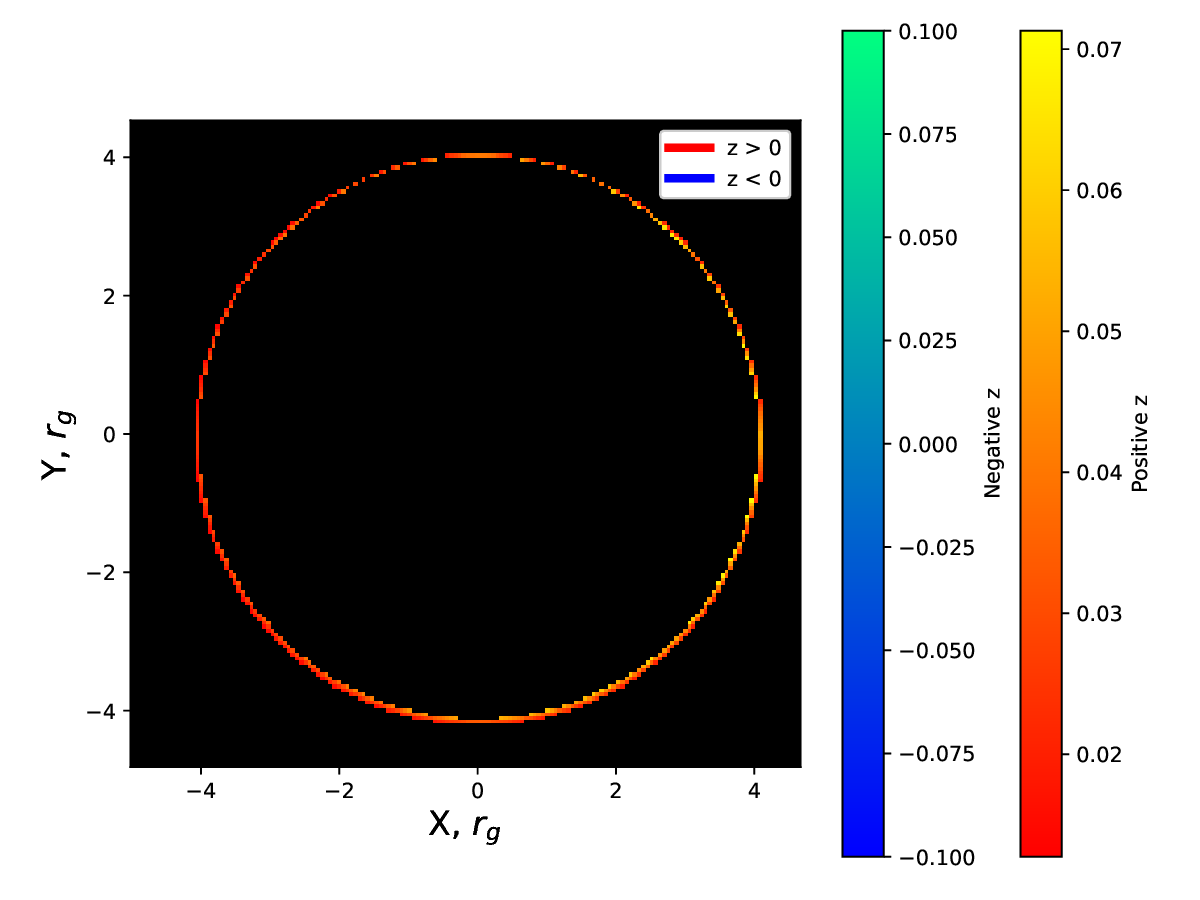} \\ 
					\end{minipage}
\hfill
		\begin{minipage}[h]{.37\textwidth}
			\includegraphics[width=\columnwidth]{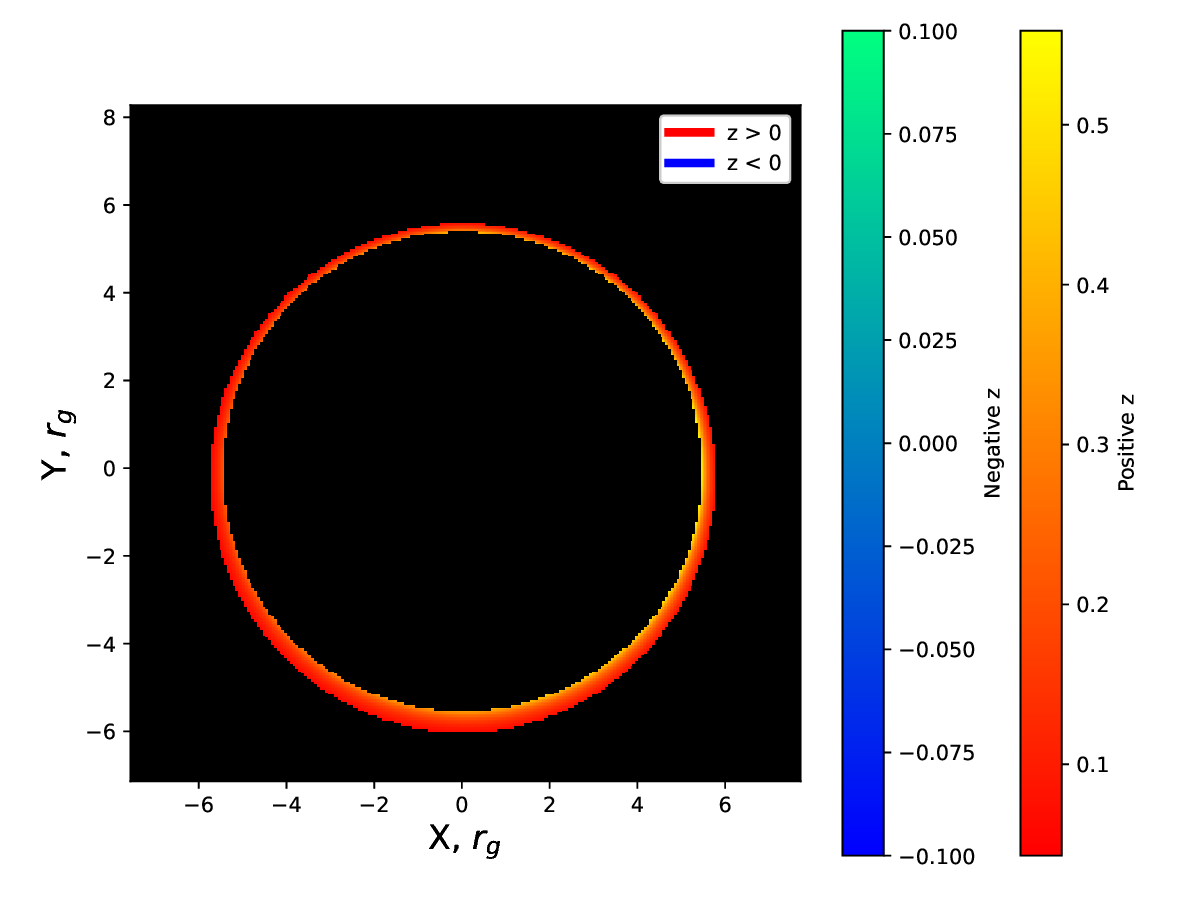} \\ 
					\end{minipage}
\hfill
		\begin{minipage}[h]{.37\textwidth}
		{\small (III)}\\
			\includegraphics[width=\columnwidth]{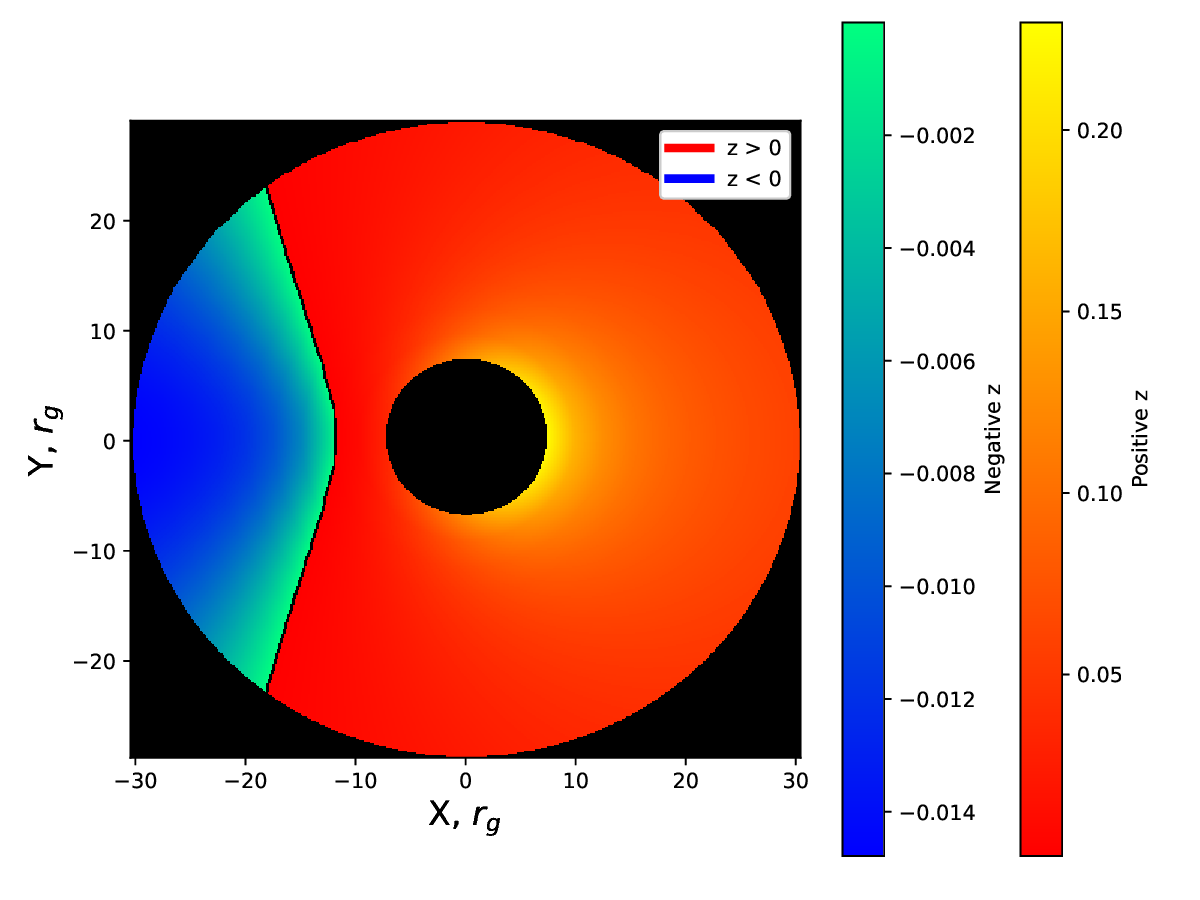} \\ 
					\end{minipage}
\hfill
		\begin{minipage}[h]{.37\textwidth}
		{\small (IV)}\\	\includegraphics[width=\columnwidth]{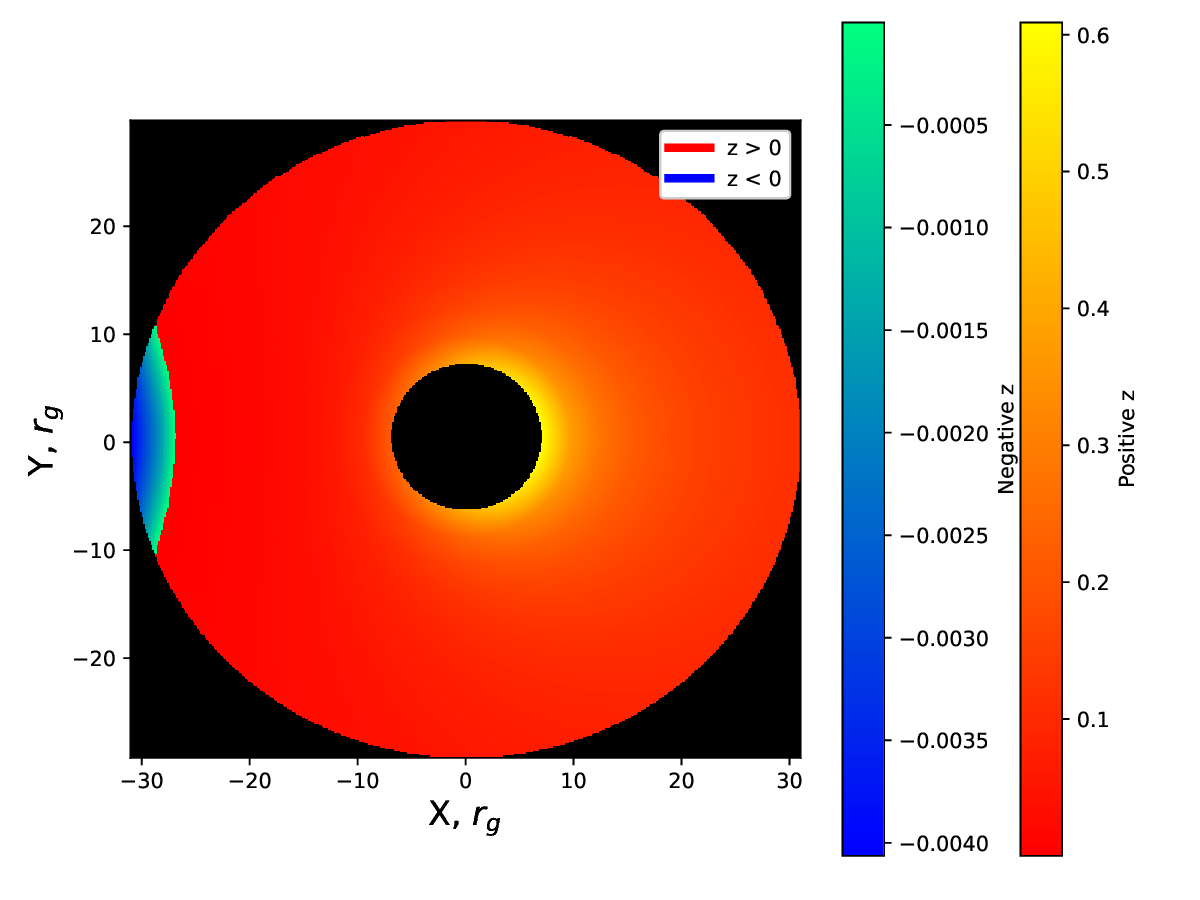} \\ 
					\end{minipage}
\hfill
		\begin{minipage}[h]{.37\textwidth}
			\includegraphics[width=\columnwidth]{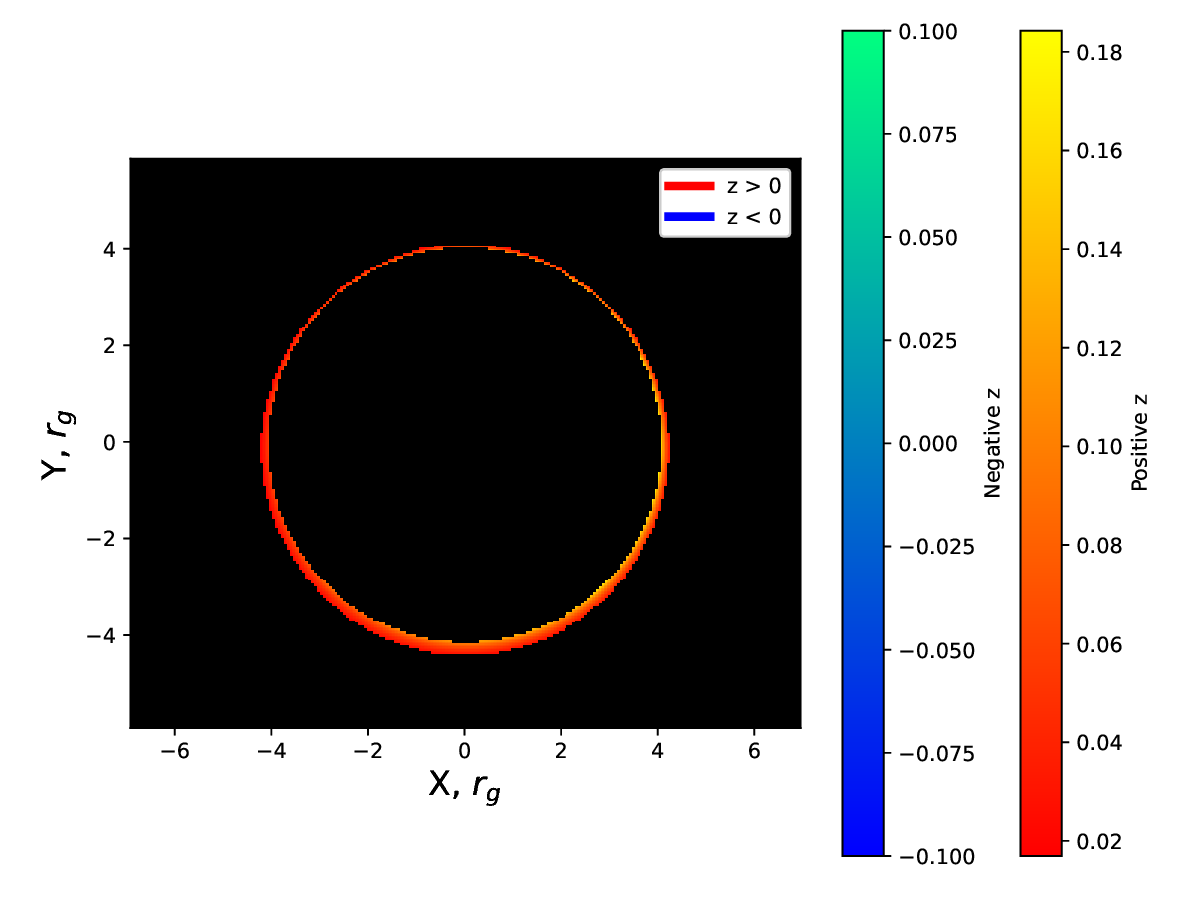} \\ 
					\end{minipage}
\hfill
		\begin{minipage}[h]{.37\textwidth}
			\includegraphics[width=\columnwidth]{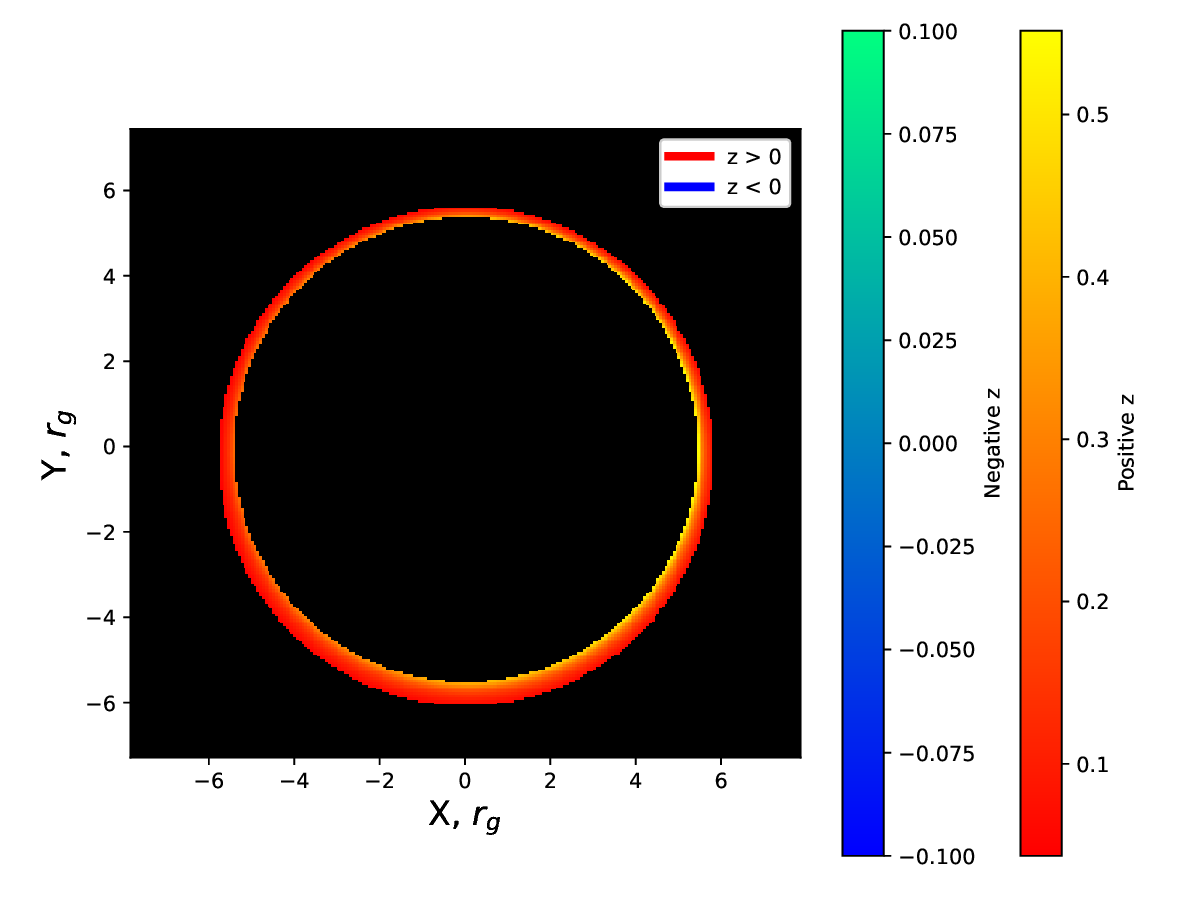} \\ 
					\end{minipage}

\caption{Continuous distributions of the redshift $z$ in the direct (odd rows) and secondary (even rows) images cast by the accretion disk for the configurations listed in Tab.\ref{tab:models}  at a fixed inclination angle of $i=17^\circ$. The inner and outer boundaries of the accretion disk correspond to stable circular orbits with radii $r=r_{\rm isco}$ and $r=30\,r_g$, respectively
}
\label{fig2a}	
\end{figure*}

\begin{figure*}
		\begin{minipage}[h]{.37\textwidth}
		{\small (I)}\\
			{\includegraphics[width=\columnwidth]{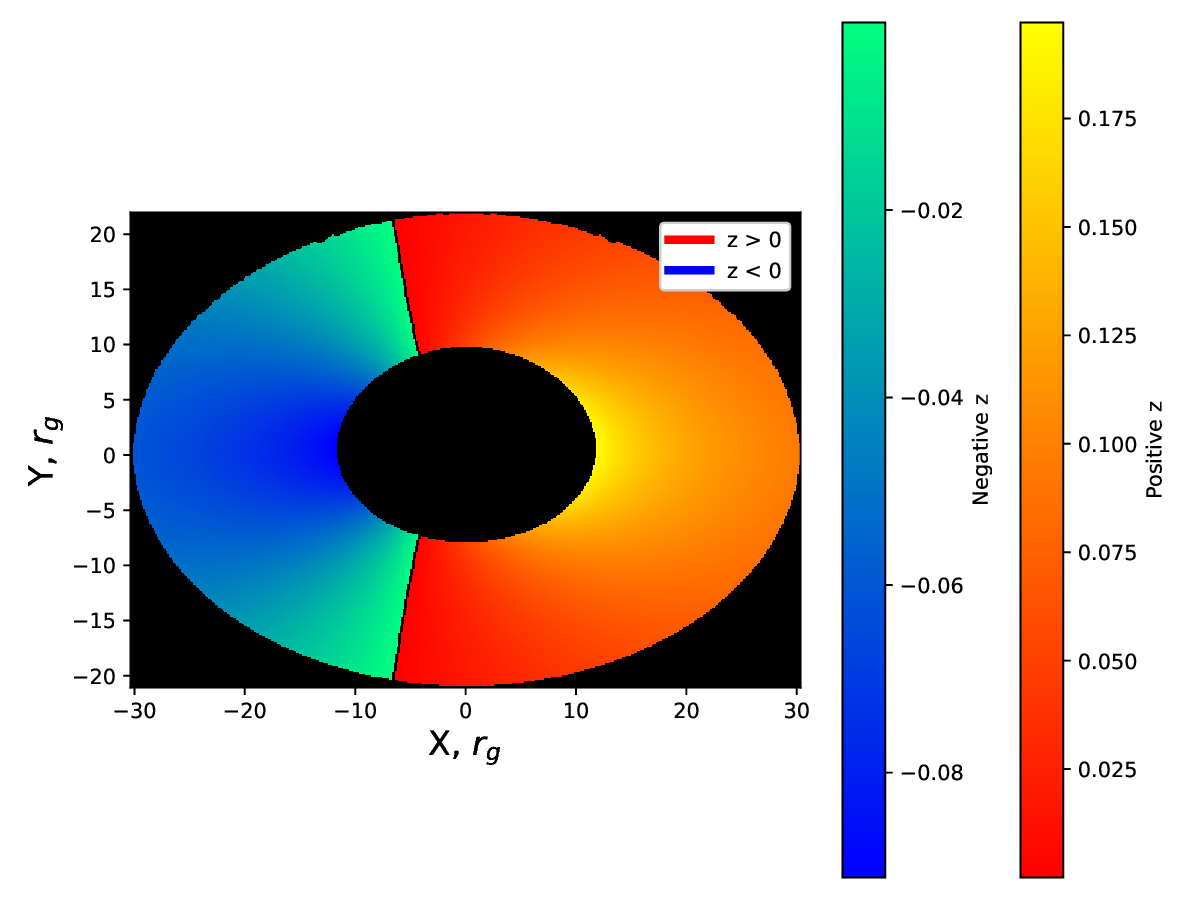}} \\ 
				
		\end{minipage}
\hfill
		\begin{minipage}[h]{.37\textwidth}
		{\small (II)}\\
			{\includegraphics[width=\columnwidth]{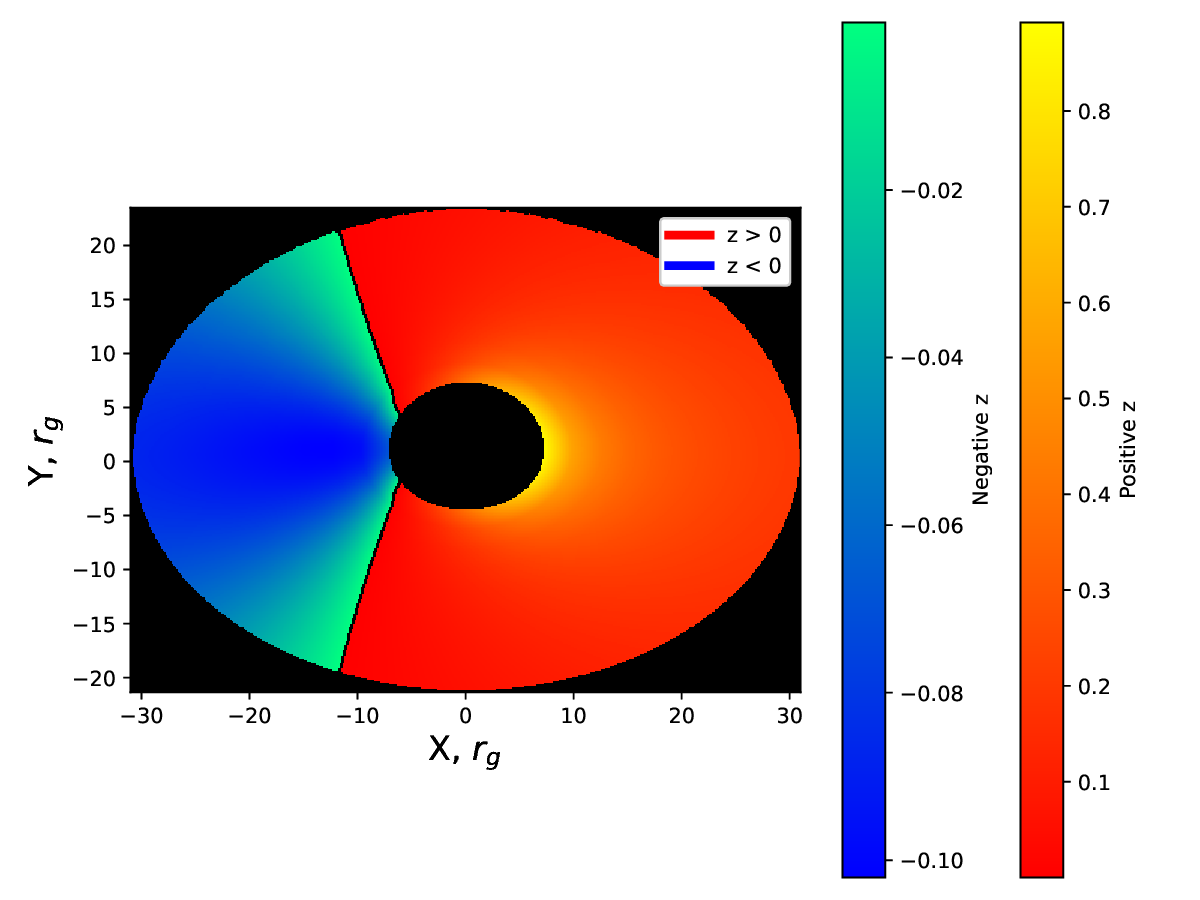}} \\ 
				
		\end{minipage}
		\hfill
		\begin{minipage}[h]{.37\textwidth}
			{\includegraphics[width=\columnwidth]{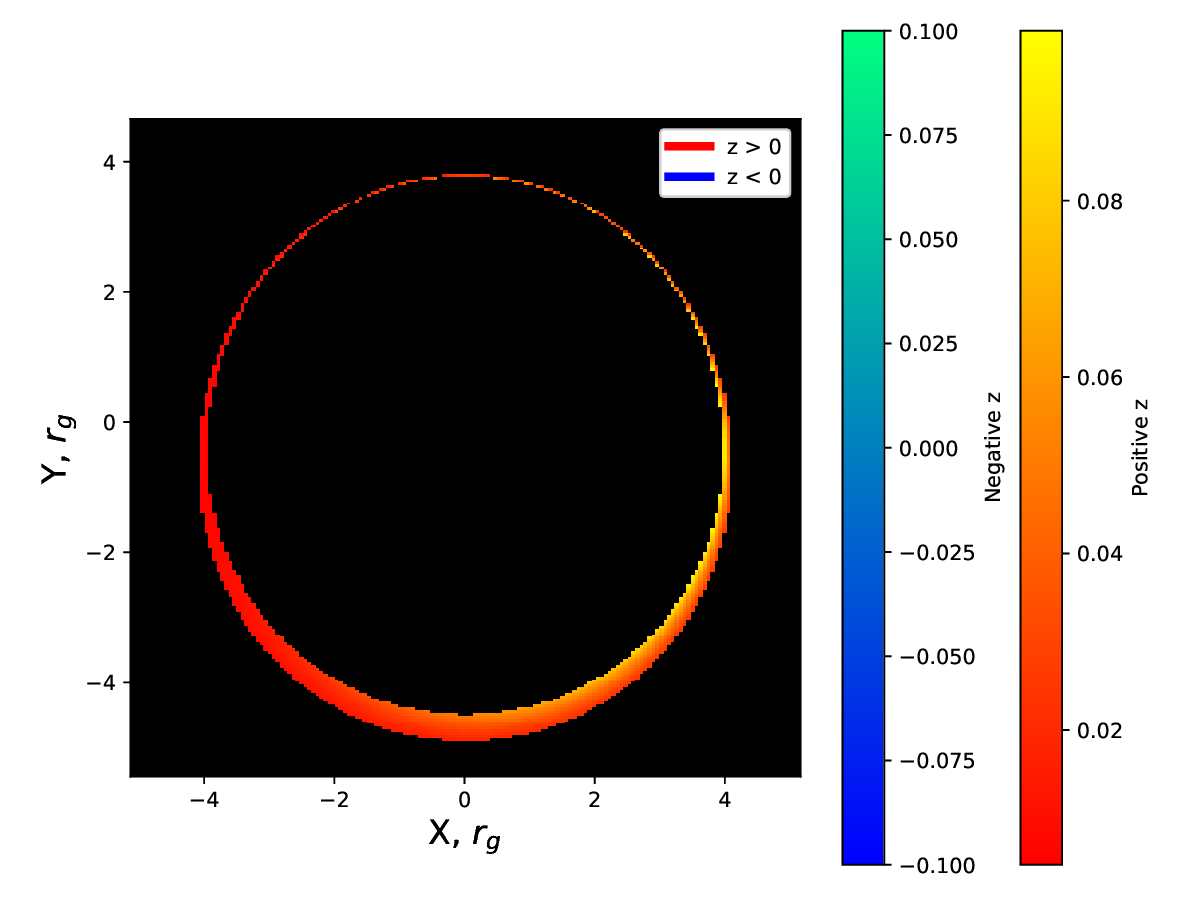}} \\

		\end{minipage}
\hfill
		\begin{minipage}[h]{.37\textwidth}
			{\includegraphics[width=\columnwidth]{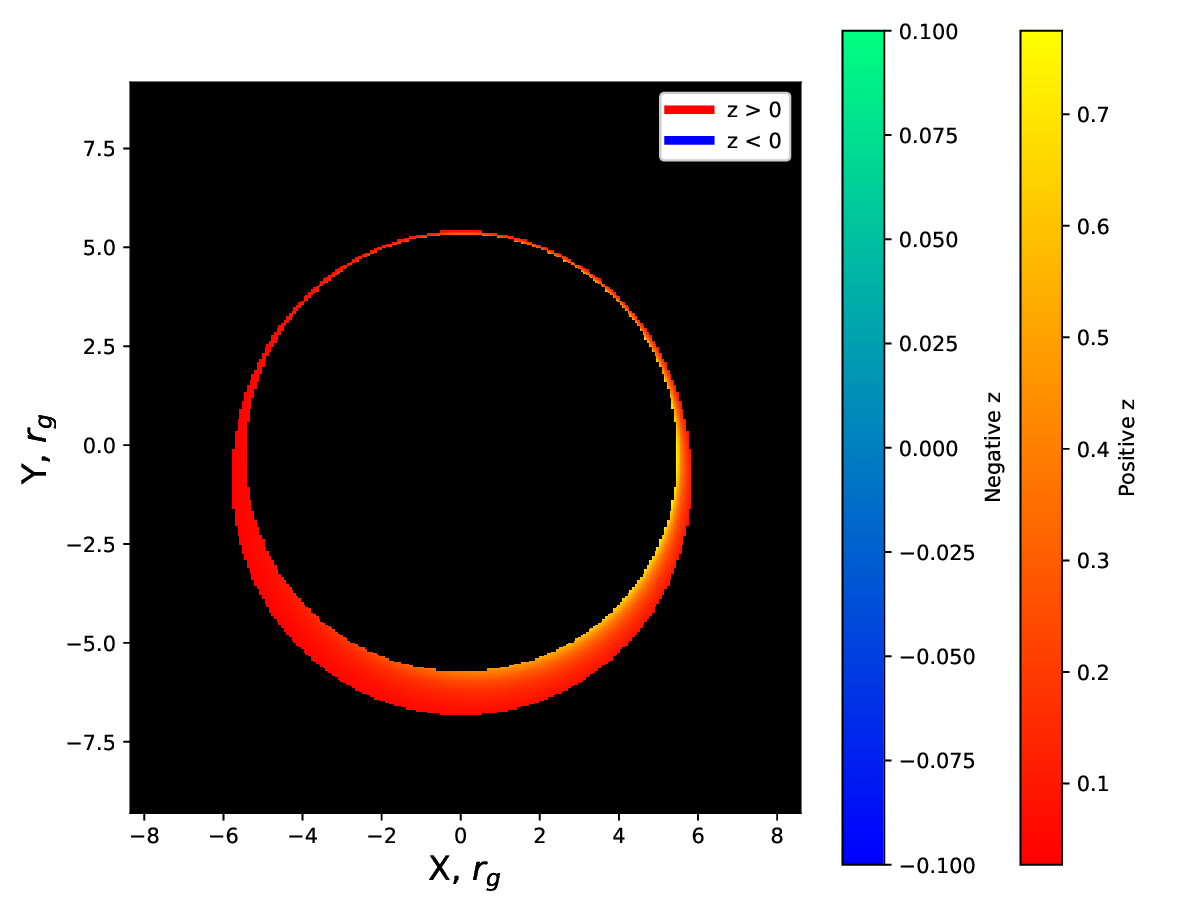}} \\ 

		\end{minipage}
\hfill
		\begin{minipage}[h]{.37\textwidth}
					
  {\small (III)}\\
  {\includegraphics[width=\columnwidth]{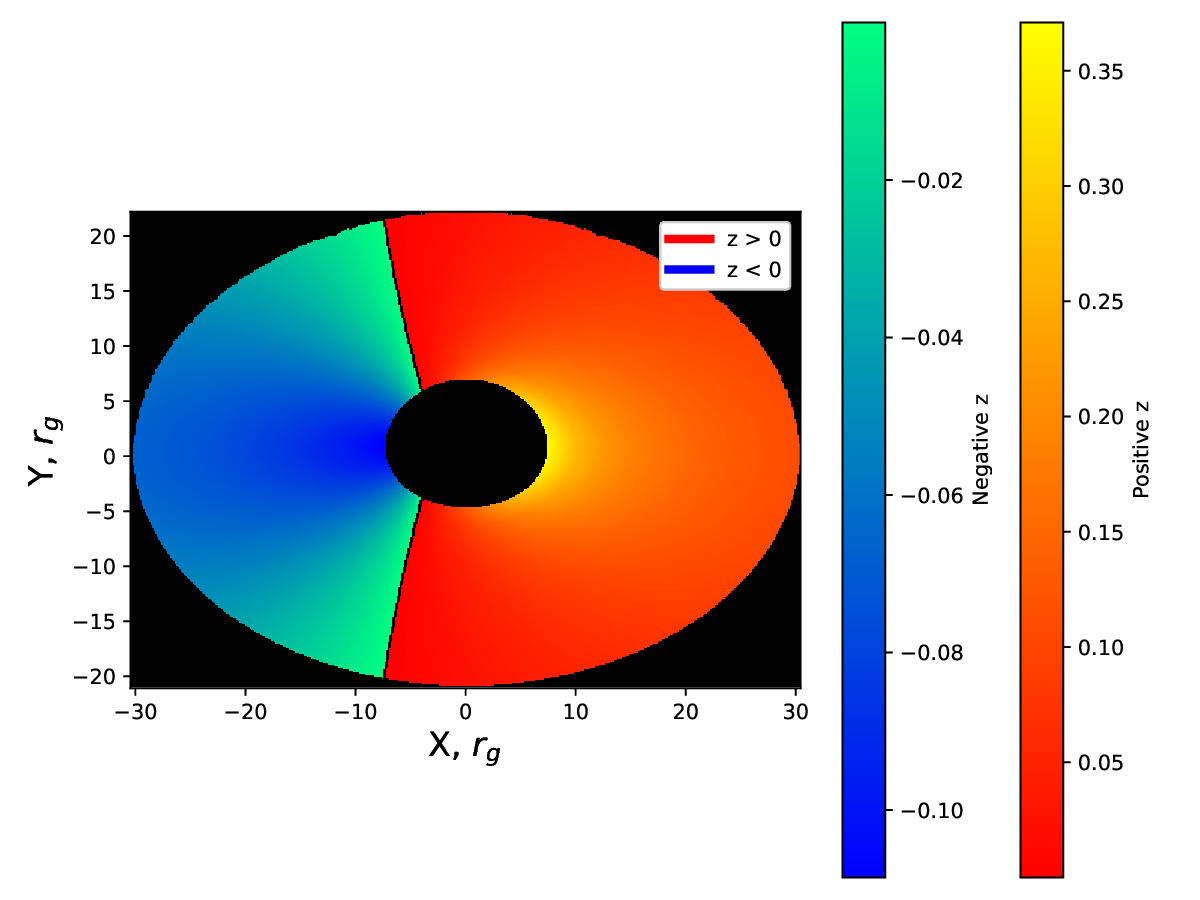}} \\ 
				
		\end{minipage}
\hfill
		\begin{minipage}[h]{.37\textwidth}
			
  {\small (IV)}\\
  {\includegraphics[width=\columnwidth]{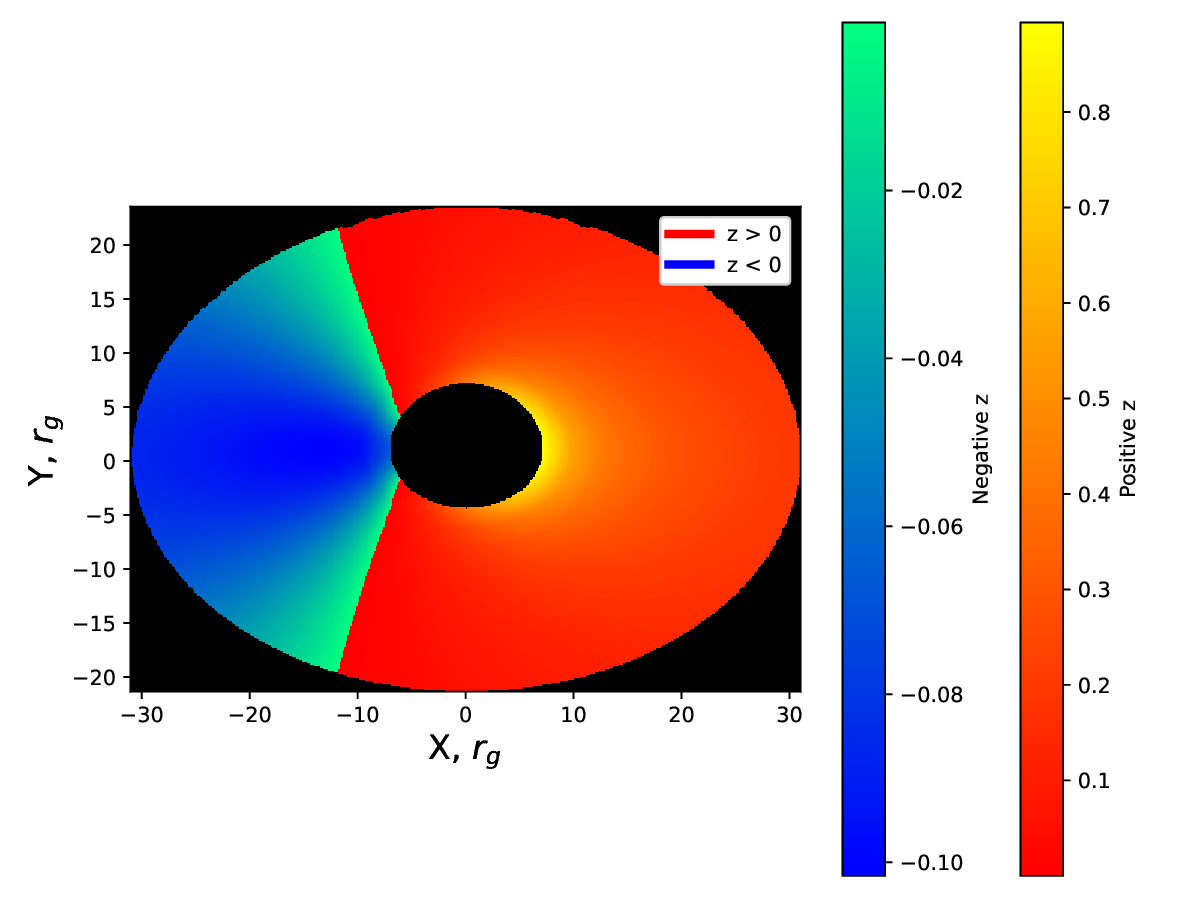}} \\ 
				
		\end{minipage}
\hfill
		\begin{minipage}[h]{.37\textwidth}
			{\includegraphics[width=\columnwidth]{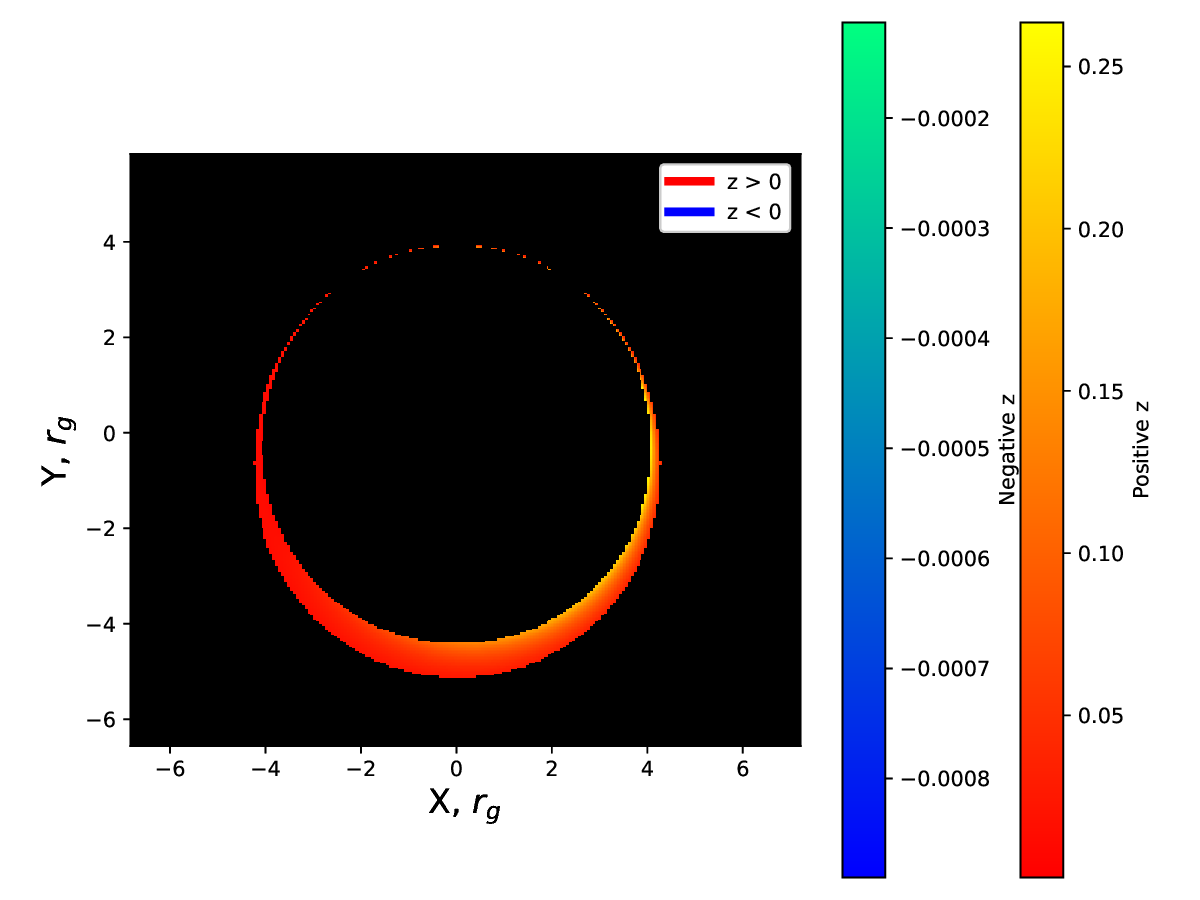}} \\ 
	
		\end{minipage}
\hfill
		\begin{minipage}[h]{.37\textwidth}
			{\includegraphics[width=\columnwidth]{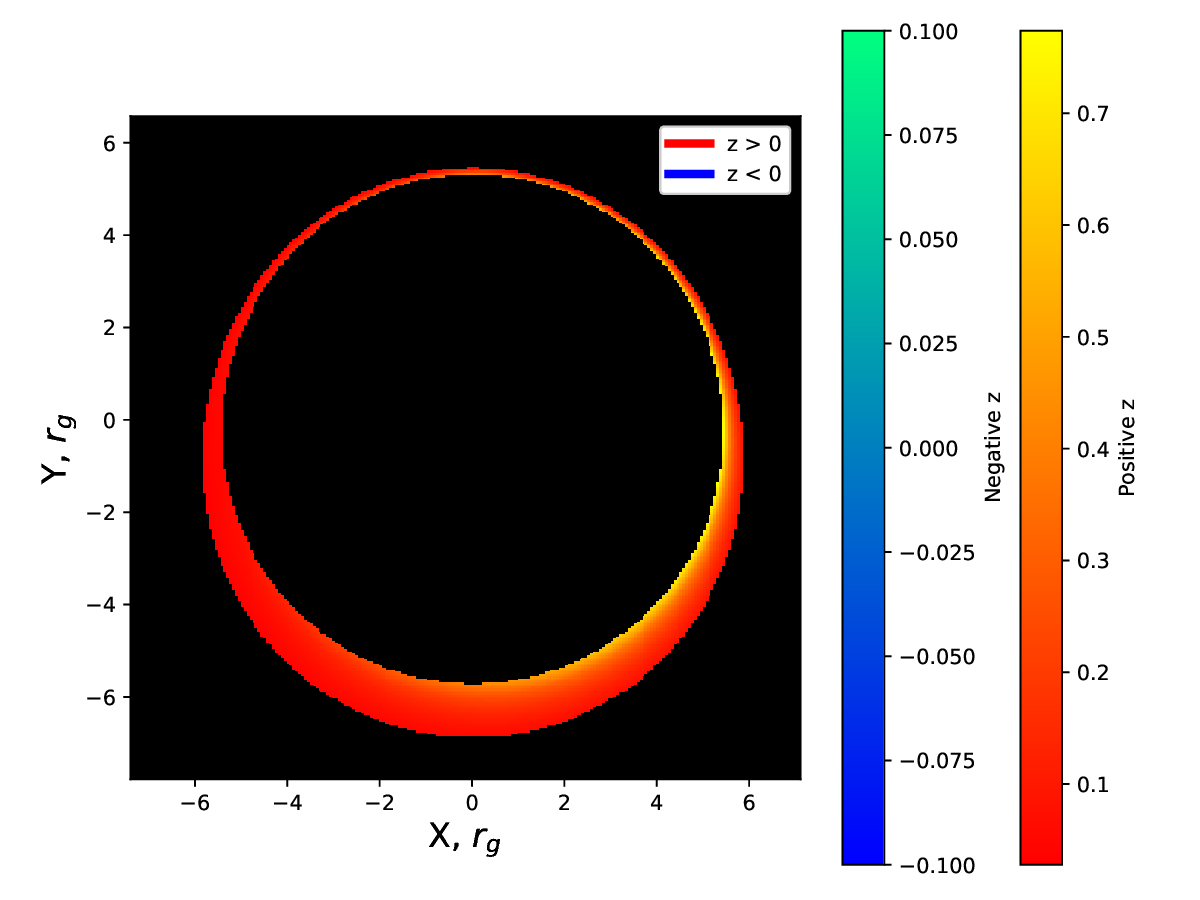}} \\ 

		\end{minipage}
	\caption{Continuous distributions of the redshift $z$ in the direct (odd rows) and secondary (even rows) images cast by the accretion disk for the configurations listed in Tab.\ref{tab:models}  at a fixed inclination angle of $i=45^\circ$. The inner and outer boundaries of the accretion disk correspond to stable circular orbits with radii $r=r_{\rm isco}$ and $r=30\,r_g$, respectively
}
\label{fig2b}	
\end{figure*}

\begin{figure*}
		\begin{minipage}[h]{.37\textwidth}
	{\small (I)}\\
			\includegraphics[width=\columnwidth]{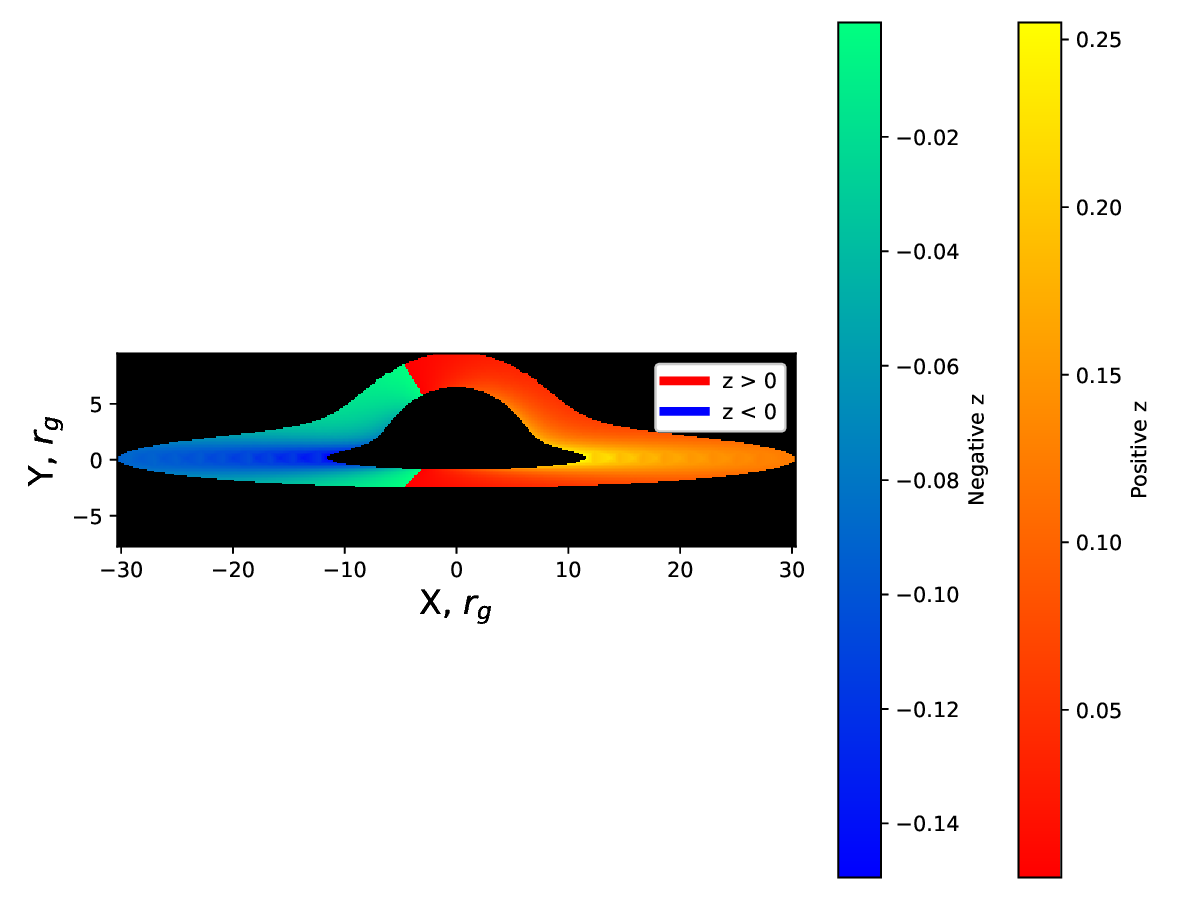} \\ 
					\end{minipage}
\hfill
		\begin{minipage}[h]{.37\textwidth}
	{\small (II)}\\
			\includegraphics[width=\columnwidth]{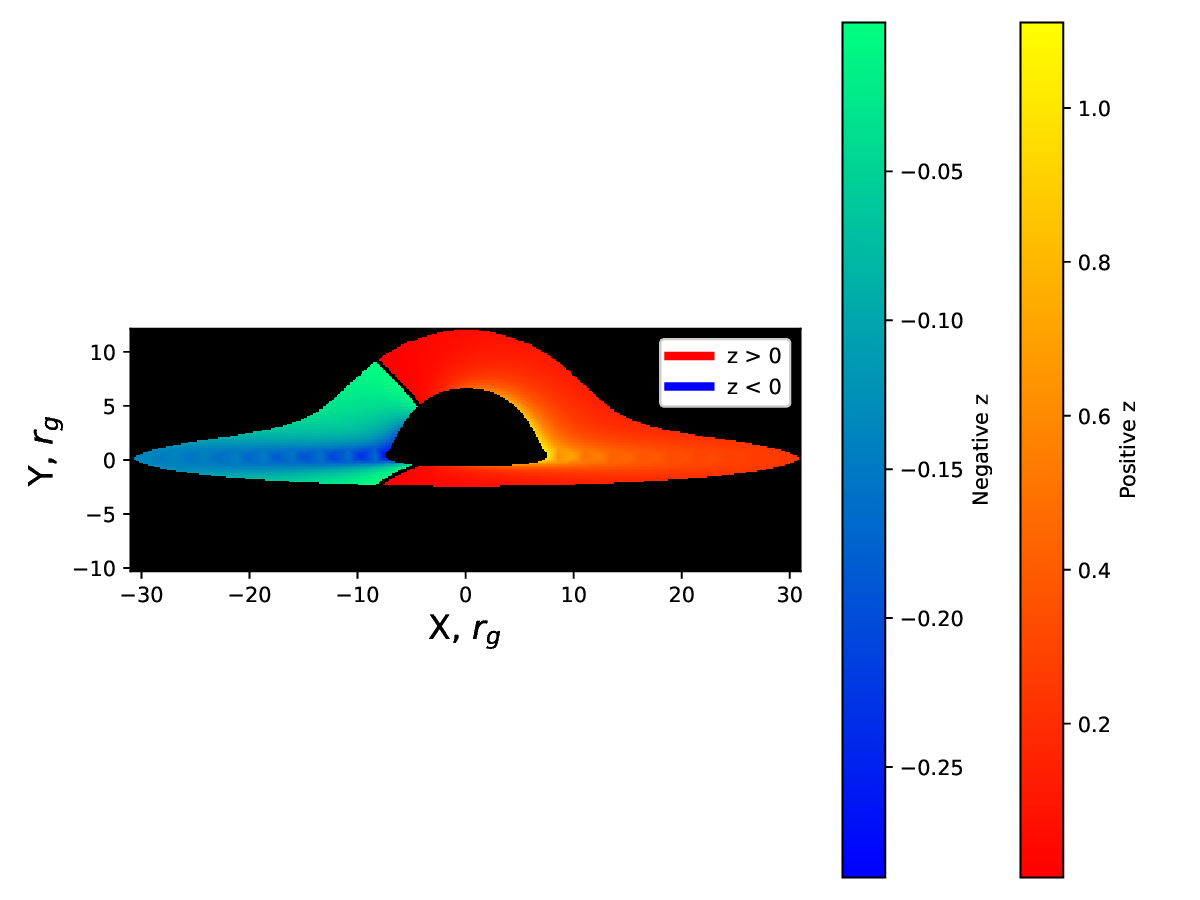} \\ 
					\end{minipage}
					\hfill
		\begin{minipage}[h]{.37\textwidth}
			\includegraphics[width=\columnwidth]{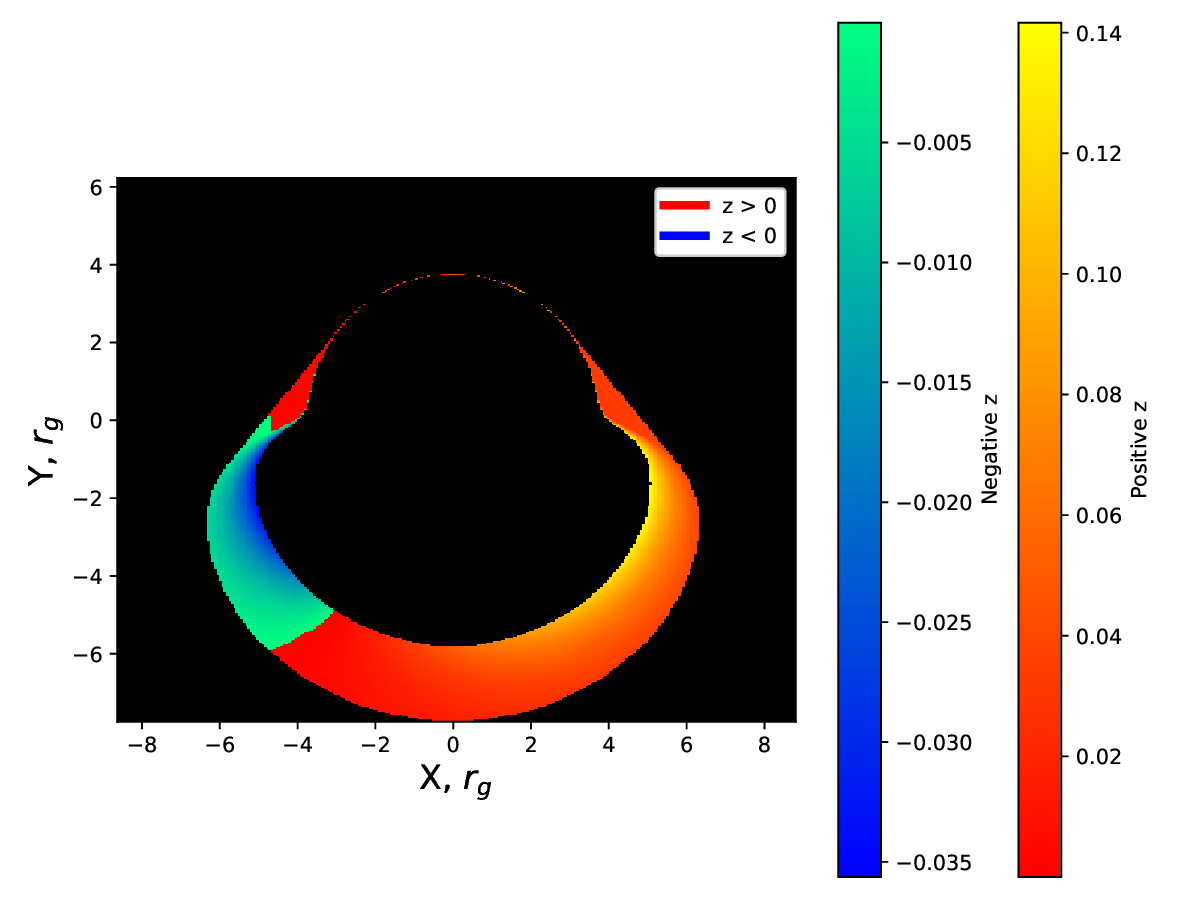} \\ 
					\end{minipage}
\hfill
		\begin{minipage}[h]{.37\textwidth}
			\includegraphics[width=\columnwidth]{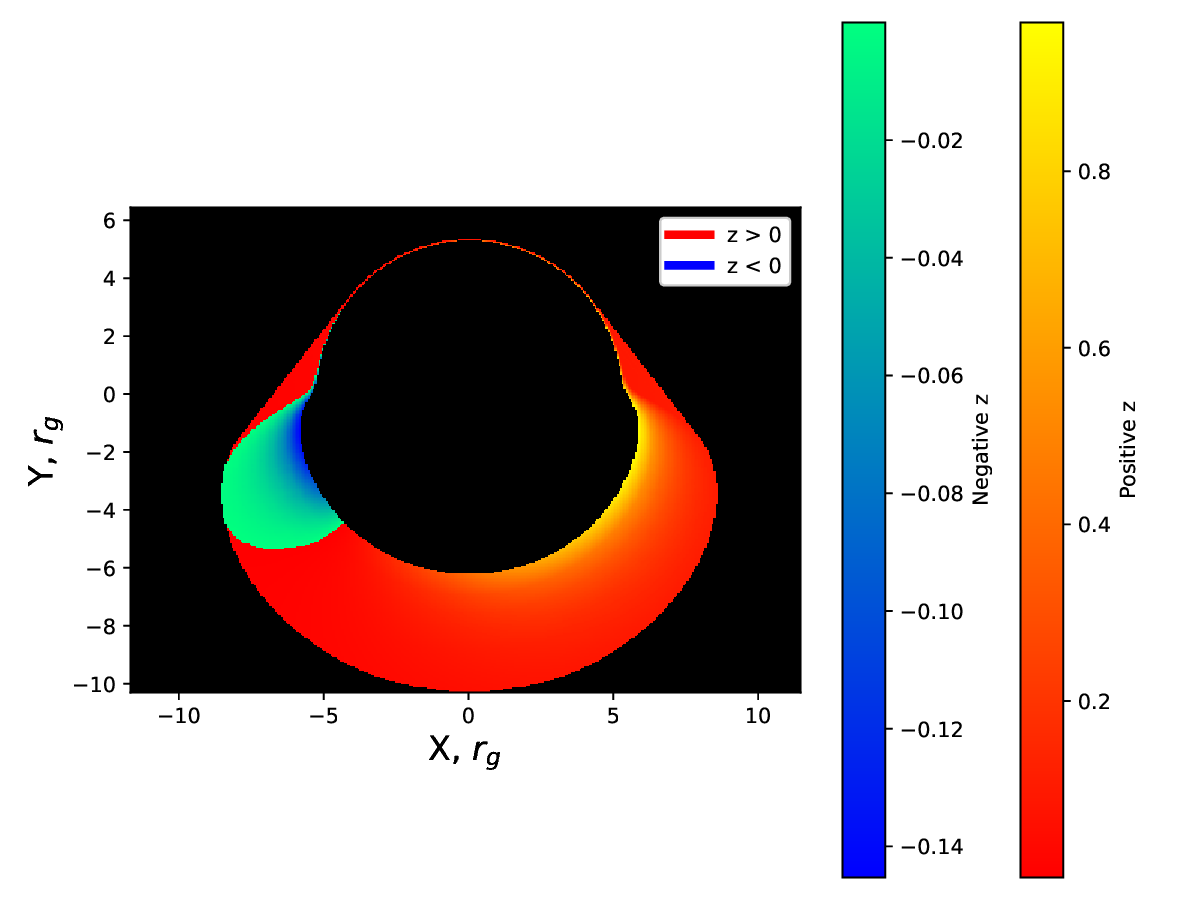} \\ 
					\end{minipage}
\hfill
		\begin{minipage}[h]{.37\textwidth}
		{\small (III)}\\
			\includegraphics[width=\columnwidth]{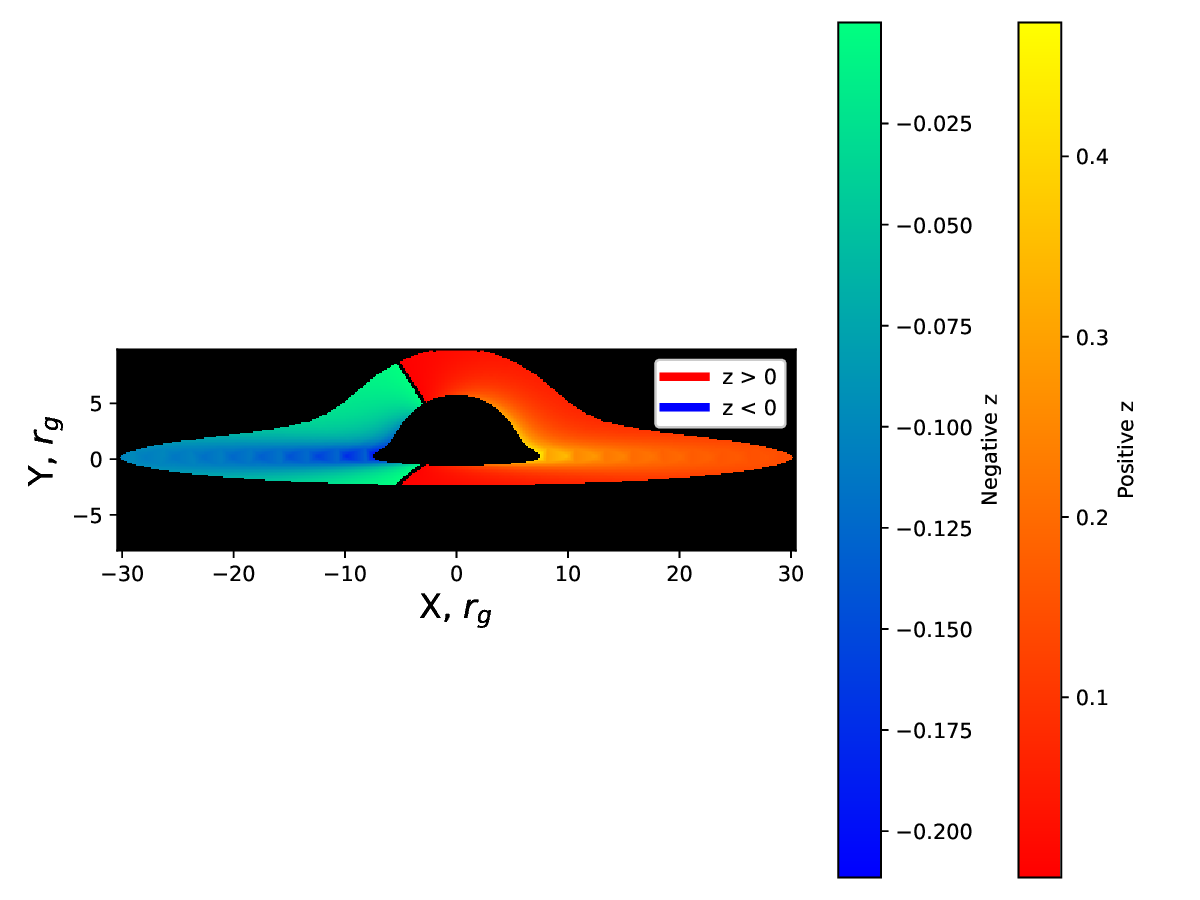} \\ 
					\end{minipage}
\hfill
		\begin{minipage}[h]{.37\textwidth}
		{\small (IV)}\\
			\includegraphics[width=\columnwidth]{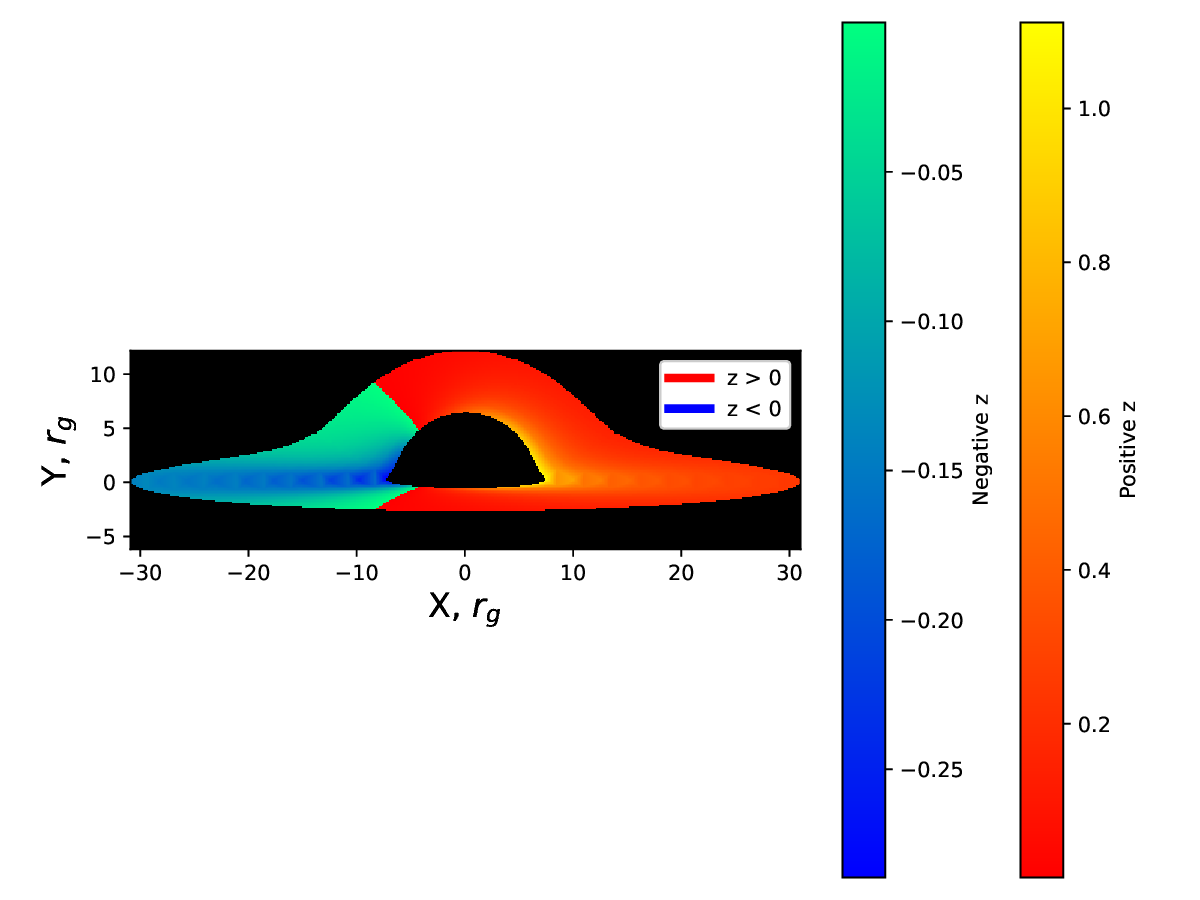} \\ 
					\end{minipage}
\hfill
		\begin{minipage}[h]{.37\textwidth}
			\includegraphics[width=\columnwidth]{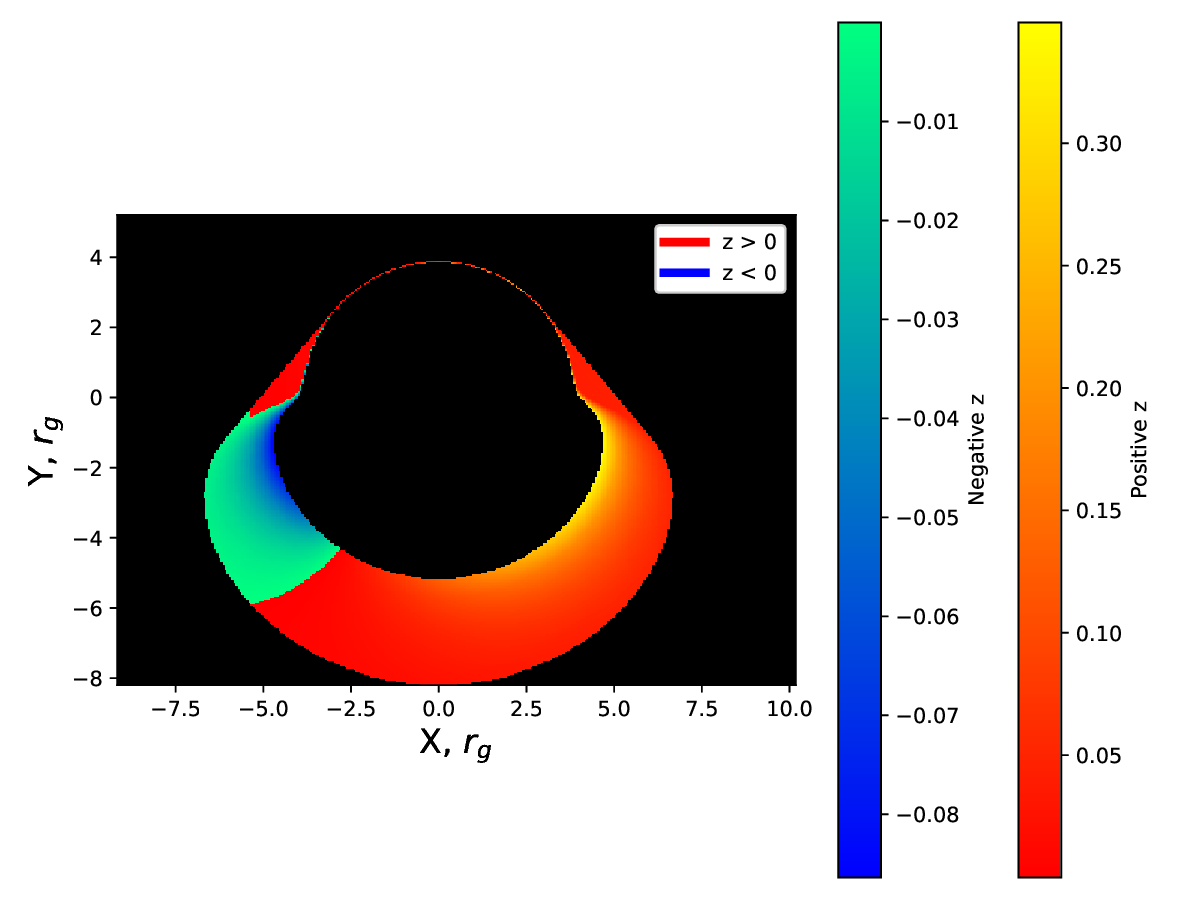} \\ 
					\end{minipage}
\hfill
		\begin{minipage}[h]{.37\textwidth}
			\includegraphics[width=\columnwidth]{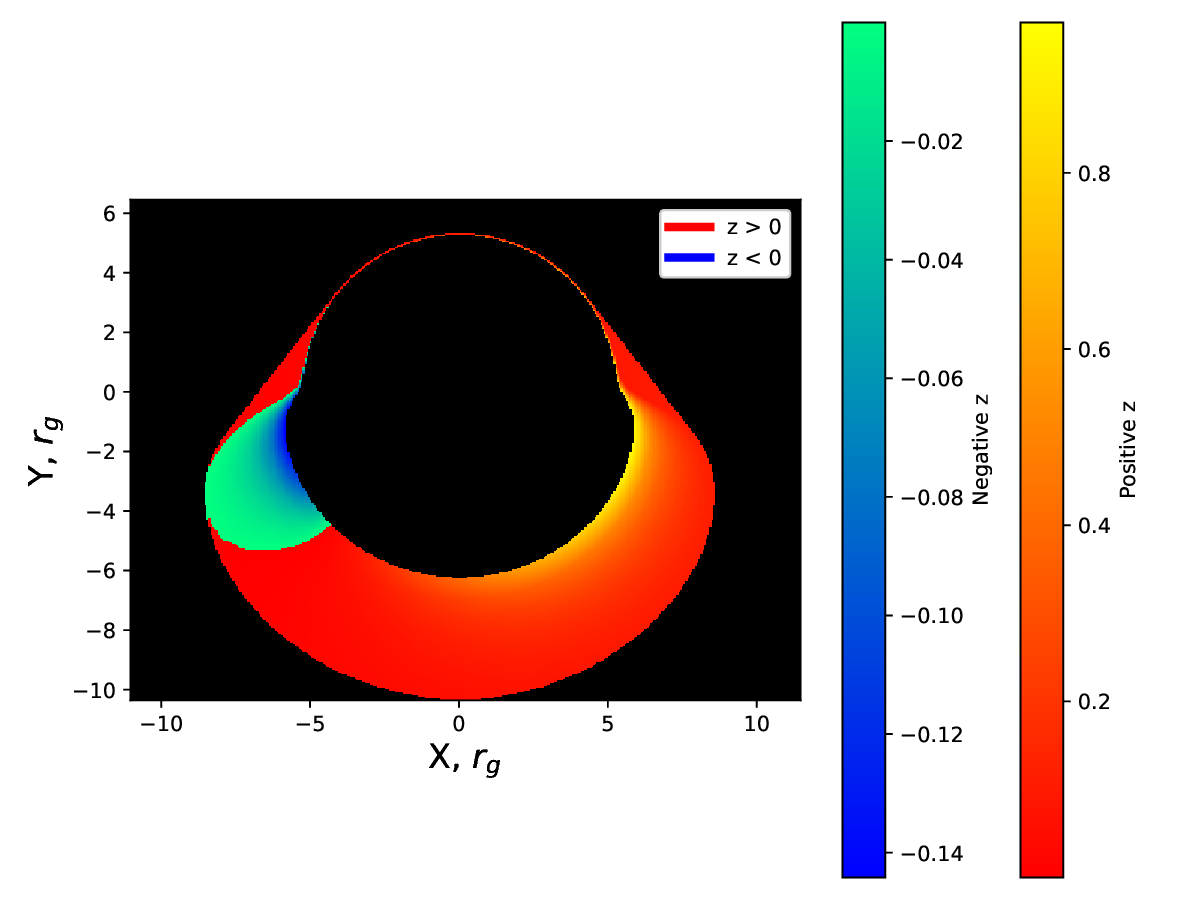} \\ 
					\end{minipage}
\caption{Continuous distributions of the redshift $z$ in the direct (odd rows) and secondary (even rows) images cast by the accretion disk for the configurations listed in Tab.\ref{tab:models}  at a fixed inclination angle of $i=85^\circ$. The inner and outer boundaries of the accretion disk correspond to stable circular orbits with radii $r=r_{\rm isco}$ and $r=30\,r_g$, respectively
}
\label{fig2c}	
		\end{figure*}

\begin{figure*}
		\begin{minipage}[h]{.5\textwidth}
		{\small (I)}\\
			\includegraphics[width=\columnwidth]{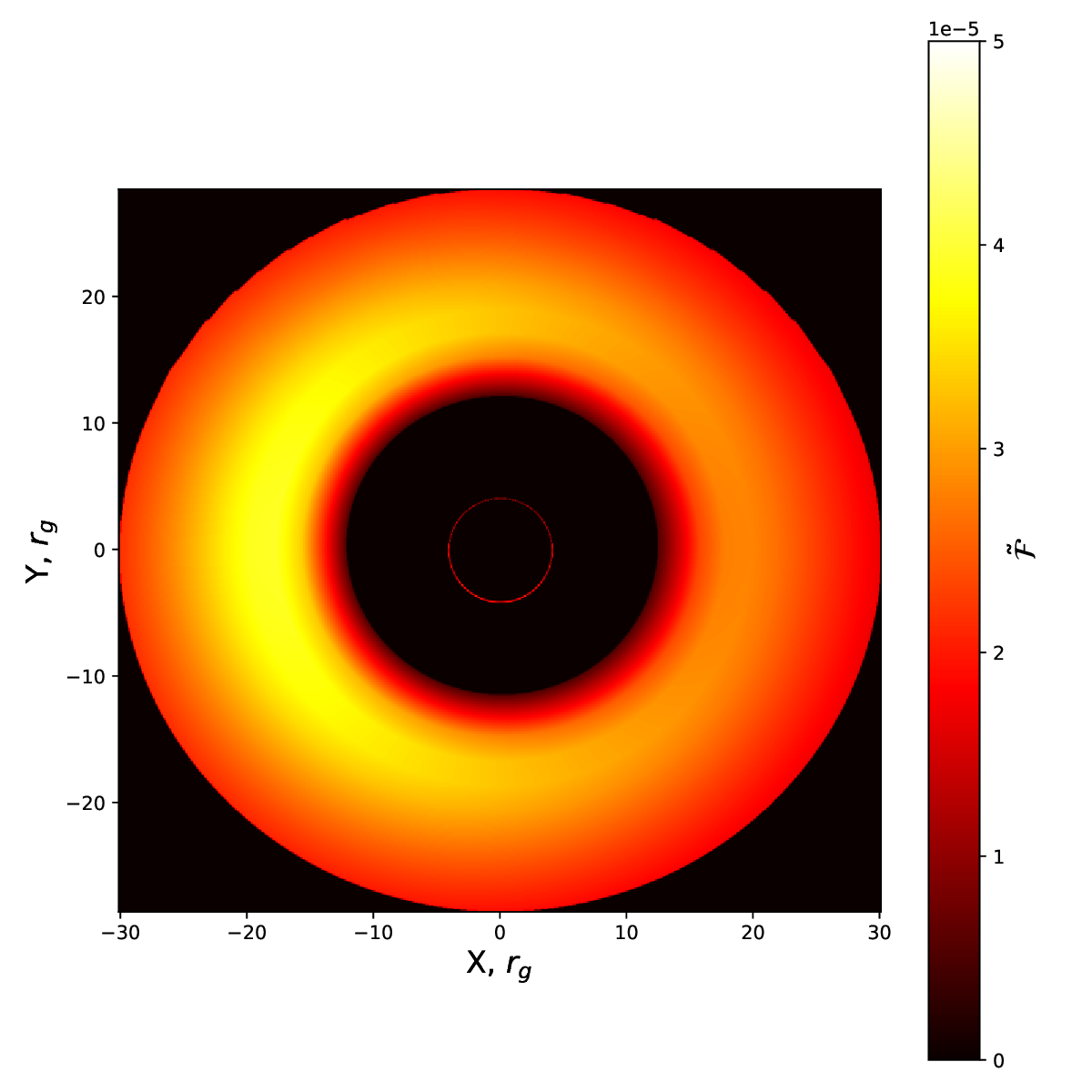} \\ 
					\end{minipage}
\hfill
		\begin{minipage}[h]{.5\textwidth}
	{\small (II)}\\		\includegraphics[width=\columnwidth]{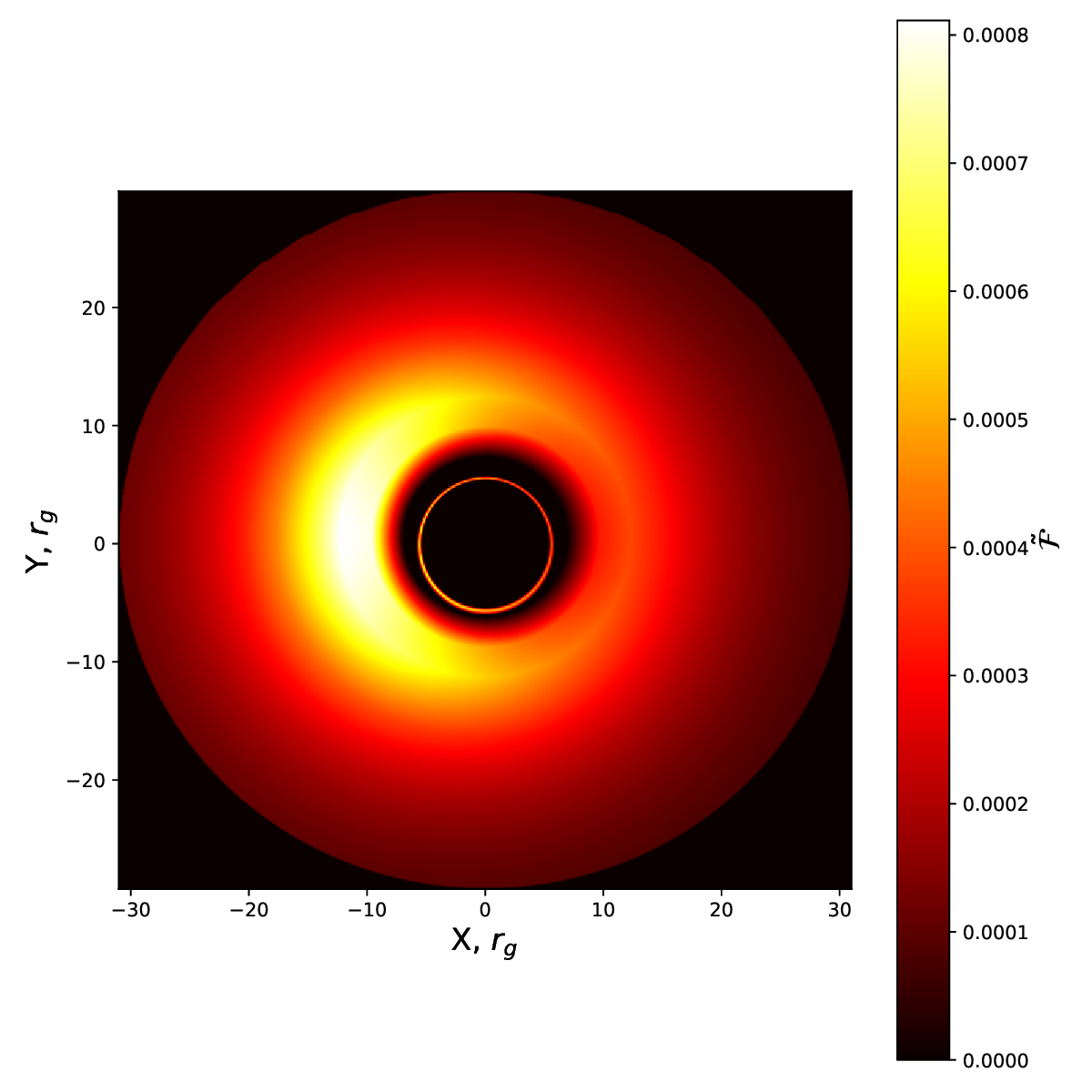} \\ 
					\end{minipage}
\hfill
\begin{minipage}[h]{.5\textwidth}
	{\small (III)}\\		\includegraphics[width=\columnwidth]{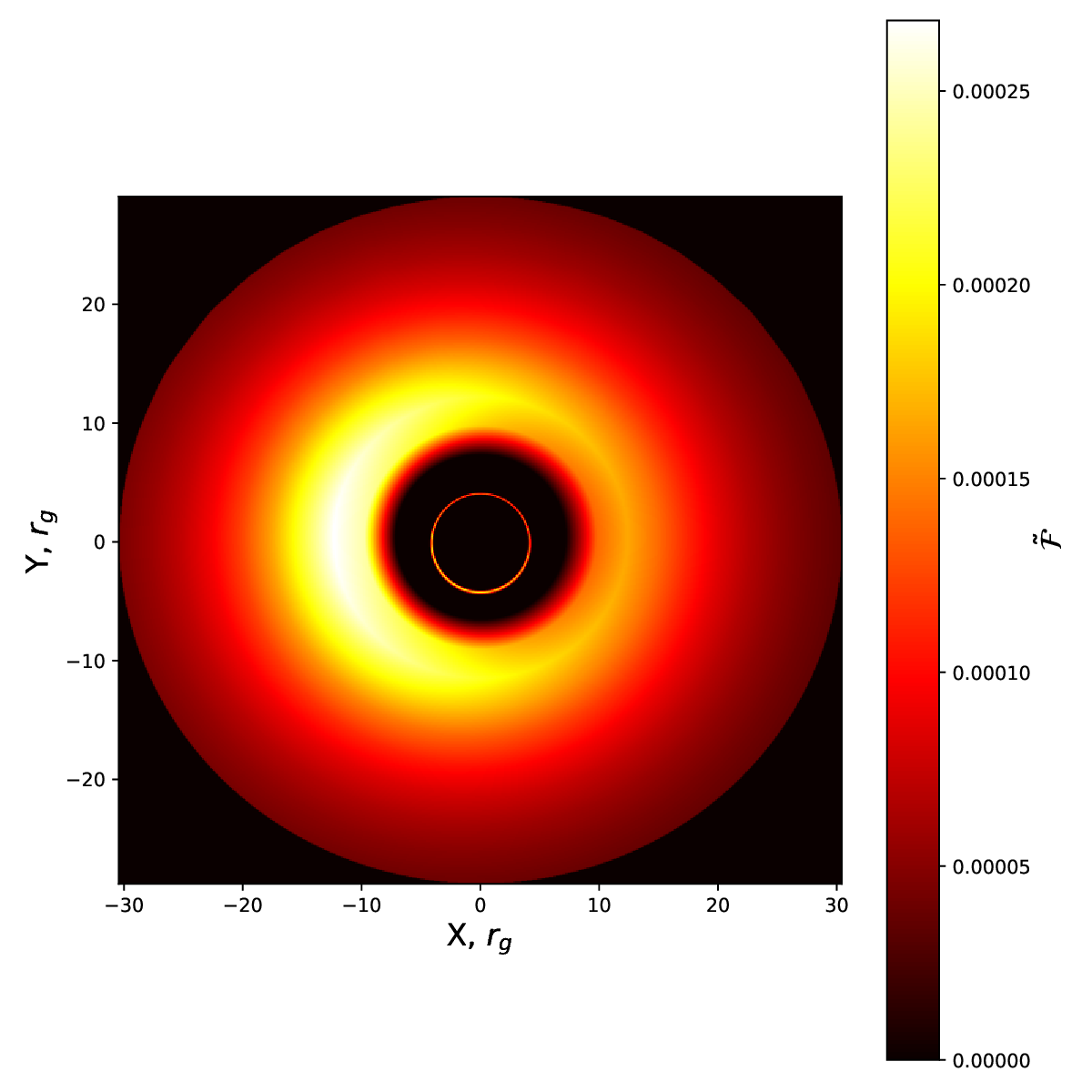} \\ 
					\end{minipage}
\hfill
					\begin{minipage}[h]{.5\textwidth}
		{\small (IV)}\\	\includegraphics[width=\columnwidth]{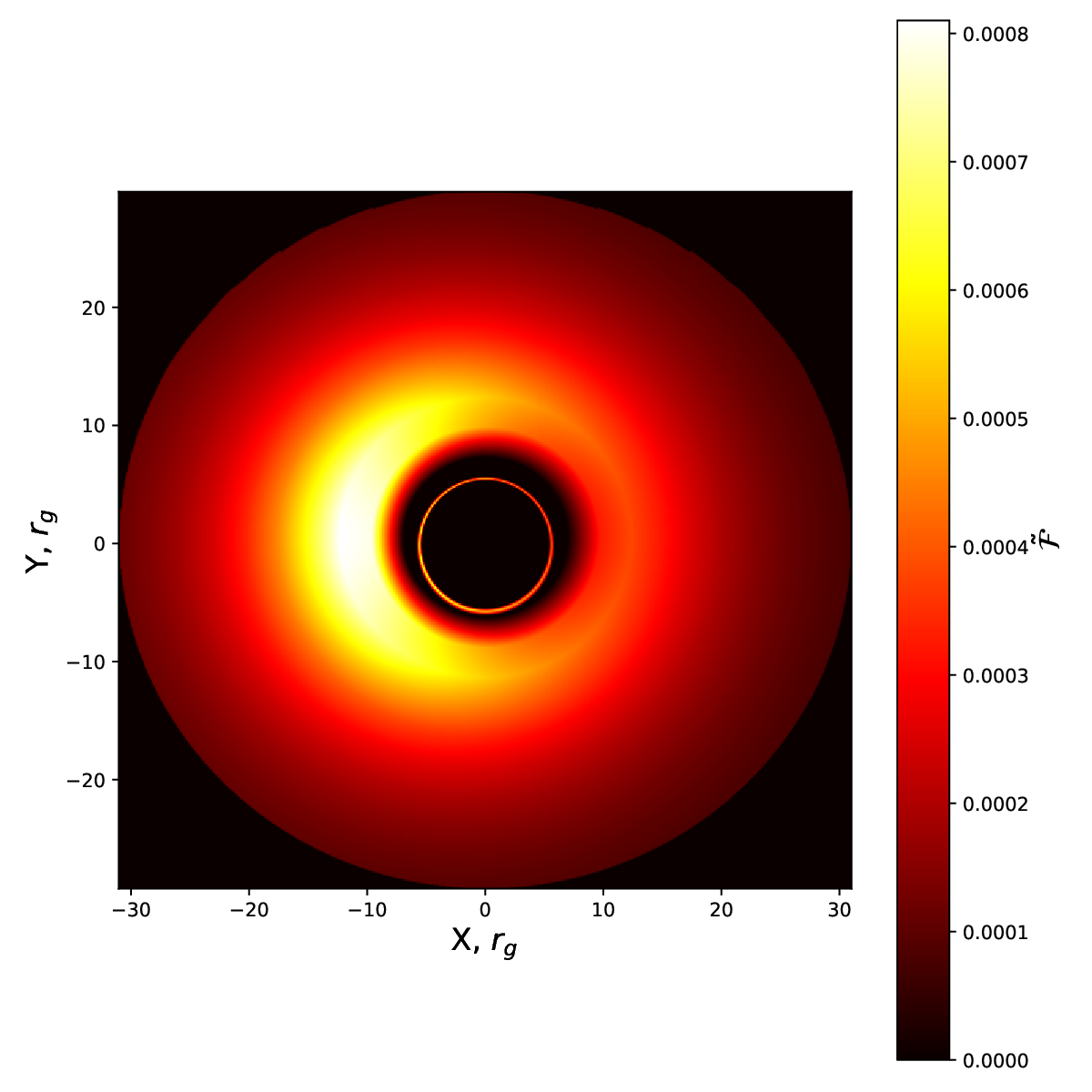} \\ 
					\end{minipage}
					\caption{The complete apparent images of the thin accretion disk for different scalar-field configurations (see Tab. \ref{tab:models}) at a fixed inclination angle of $i=17^\circ$. All panels are shown in generalized celestial coordinates on the observer's sky. The energy flux is normalized. The inner and outer boundaries of the accretion disk are stable circular orbits with radii $r=r_{\rm isco}$ and $r=30\,r_g$, respectively}
					\label{fig3a}	
				\end{figure*}

\begin{figure*}
		\begin{minipage}[h]{.5\textwidth}
		{\small (I)}\\	{\includegraphics[width=\columnwidth]{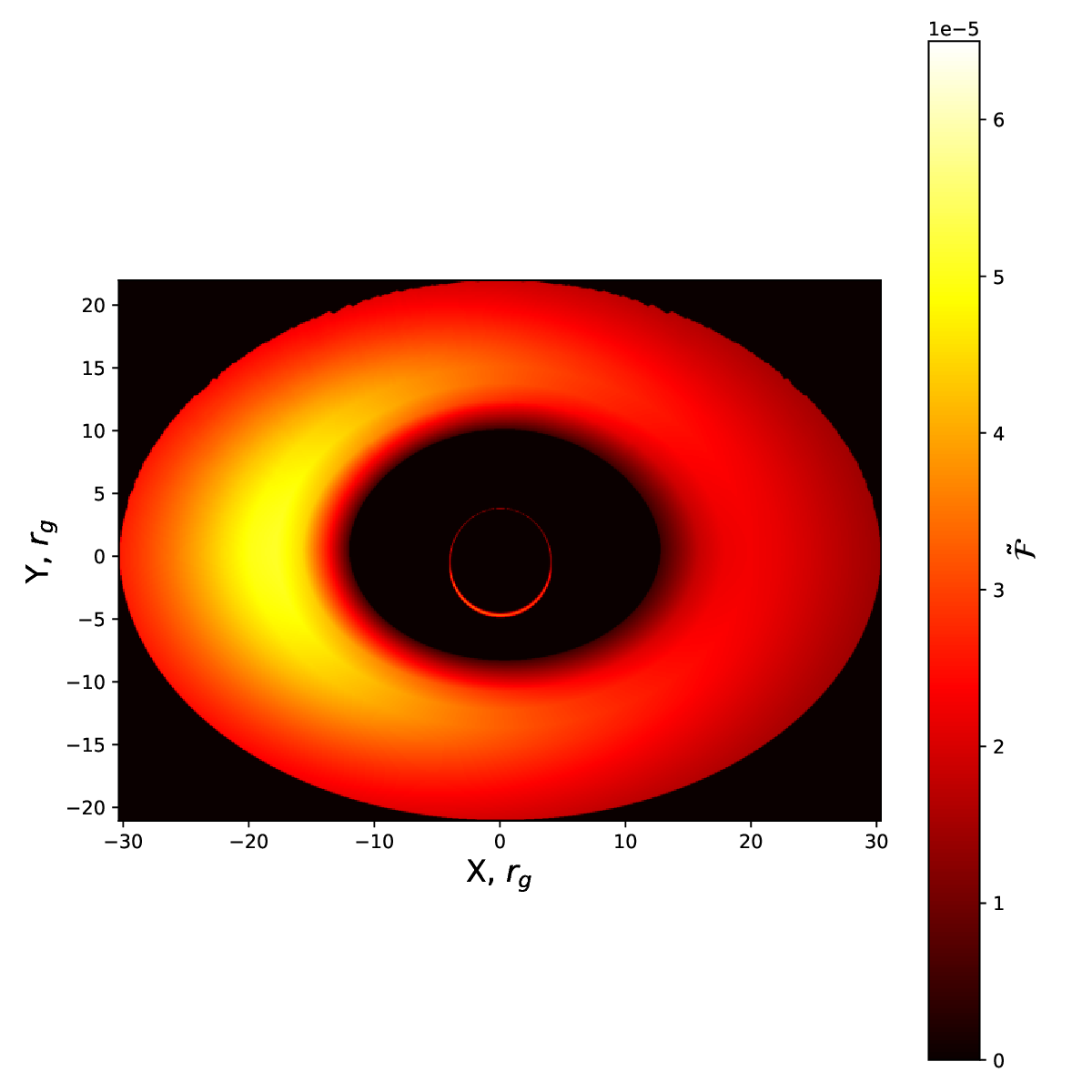}} \\ 
				
		\end{minipage}
\hfill
		\begin{minipage}[h]{.5\textwidth}
	{\small (II)}\\		{\includegraphics[width=\columnwidth]{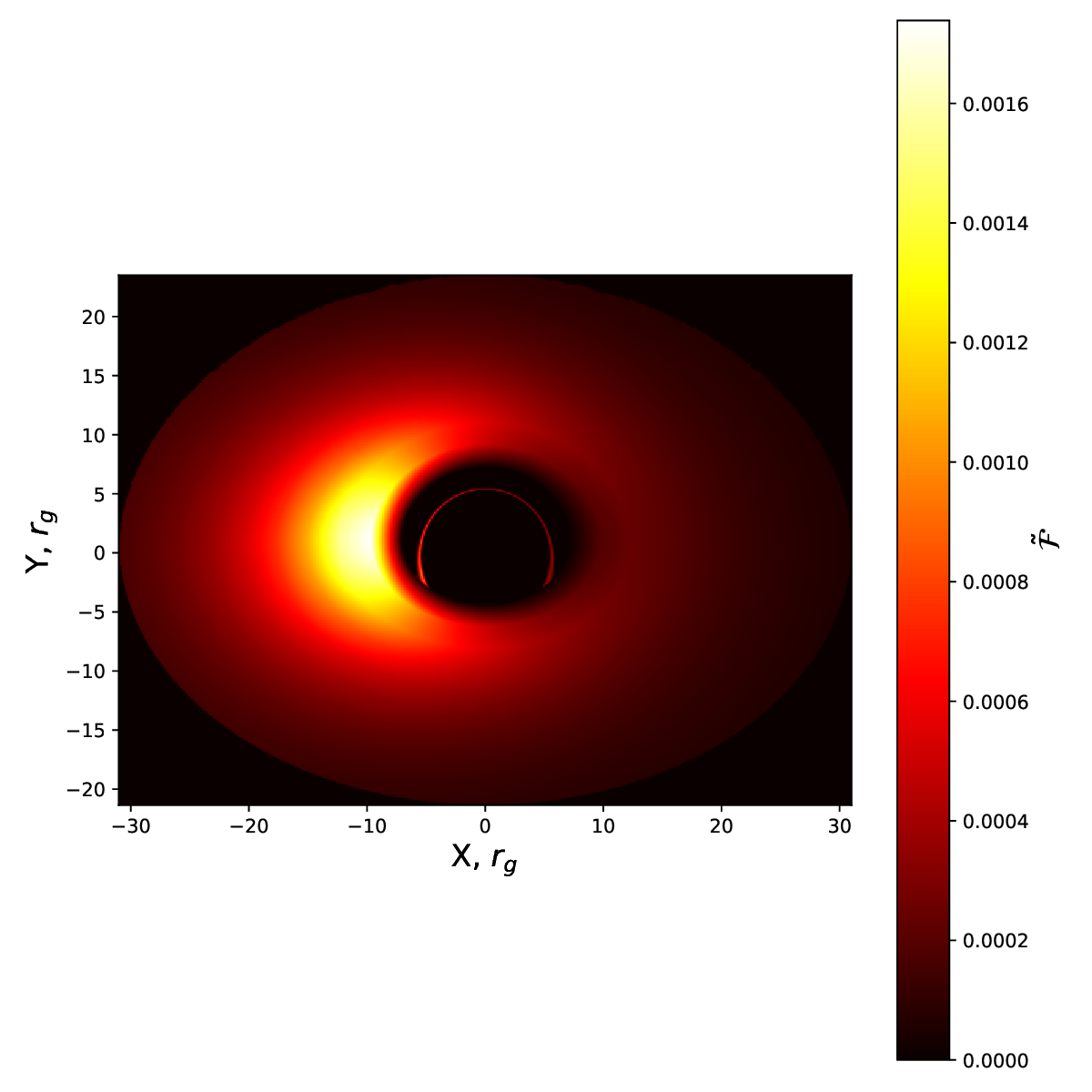}} \\ 
				
		\end{minipage}
\hfill
		\begin{minipage}[h]{.5\textwidth}
	{\small (III)}\\		\includegraphics[width=\columnwidth]{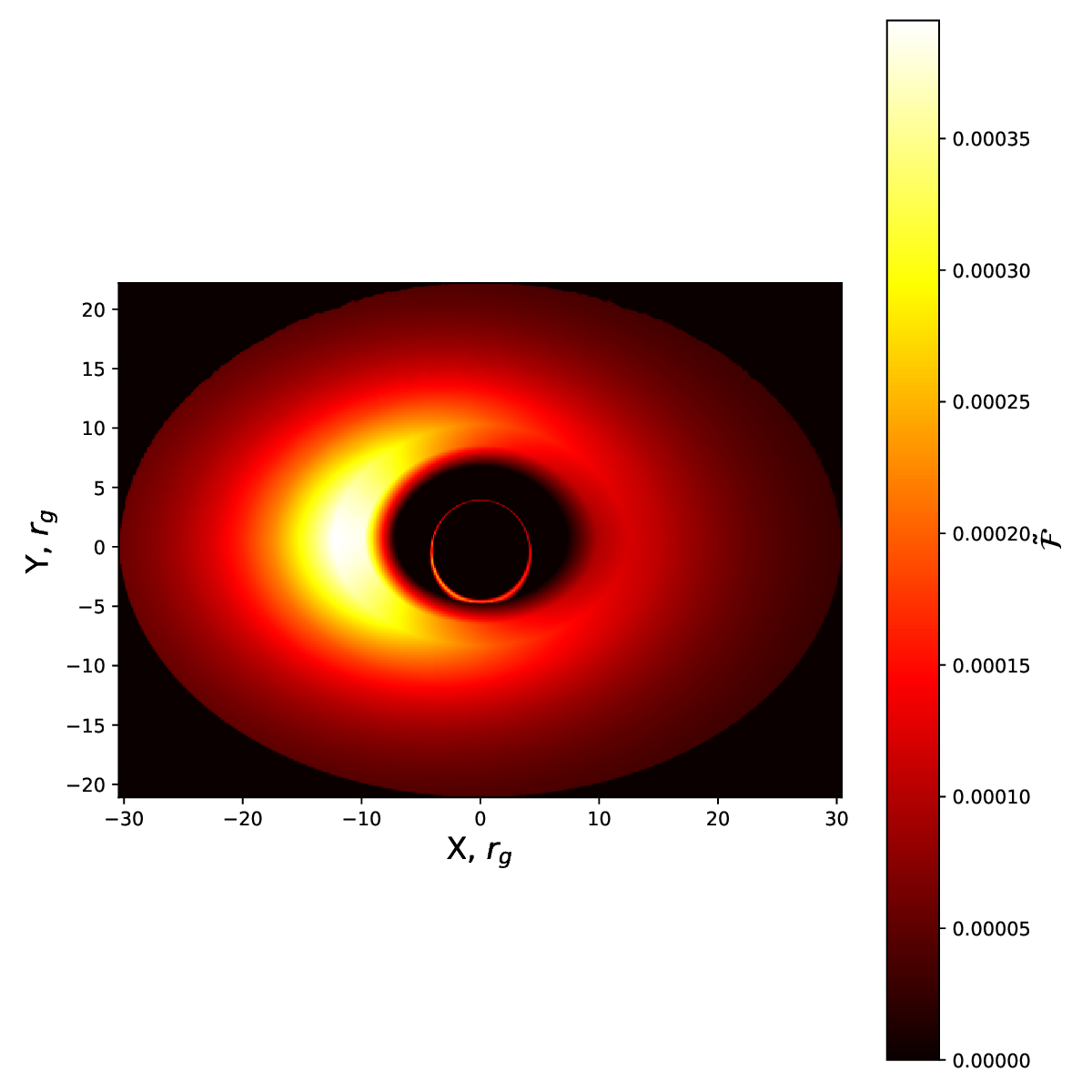} \\ 
					\end{minipage}
\hfill
		\begin{minipage}[h]{.5\textwidth}
		{\small (IV)}\\	\includegraphics[width=\columnwidth]{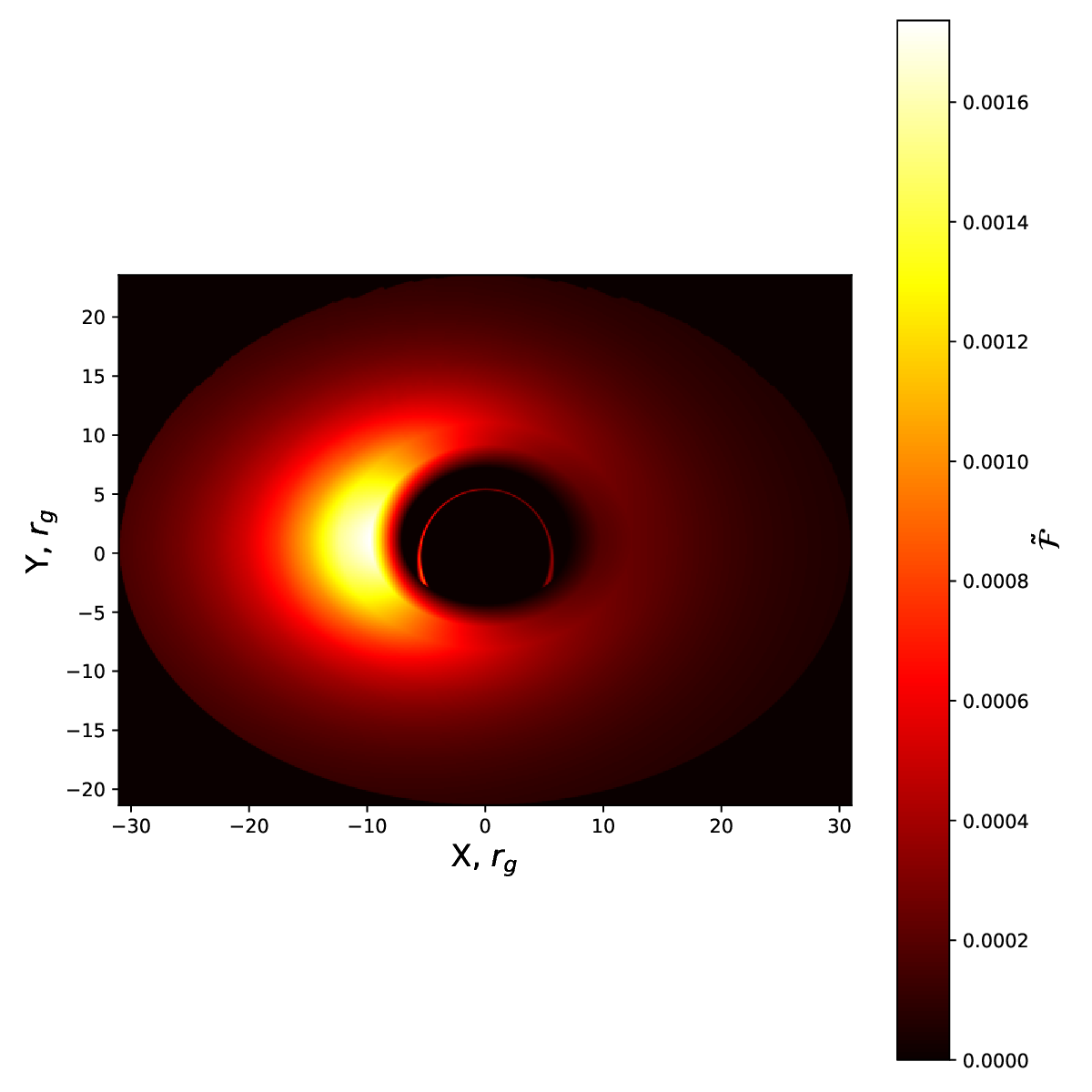} \\ 
					\end{minipage}
					\caption{	
					The complete apparent images of the thin accretion disk for different scalar-field configurations (see Tab. \ref{tab:models}) at a fixed inclination angle of $i=45^\circ$. All panels are shown in generalized celestial coordinates on the observer's sky. The energy flux is normalized. The inner and outer boundaries of the accretion disk are stable circular orbits with radii $r=r_{\rm isco}$ and $r=30\,r_g$, respectively}
\label{fig3b}	
\end{figure*}

\begin{figure*}
		\begin{minipage}[h]{.5\textwidth}
		{\small (I)}\\	\includegraphics[width=\columnwidth]{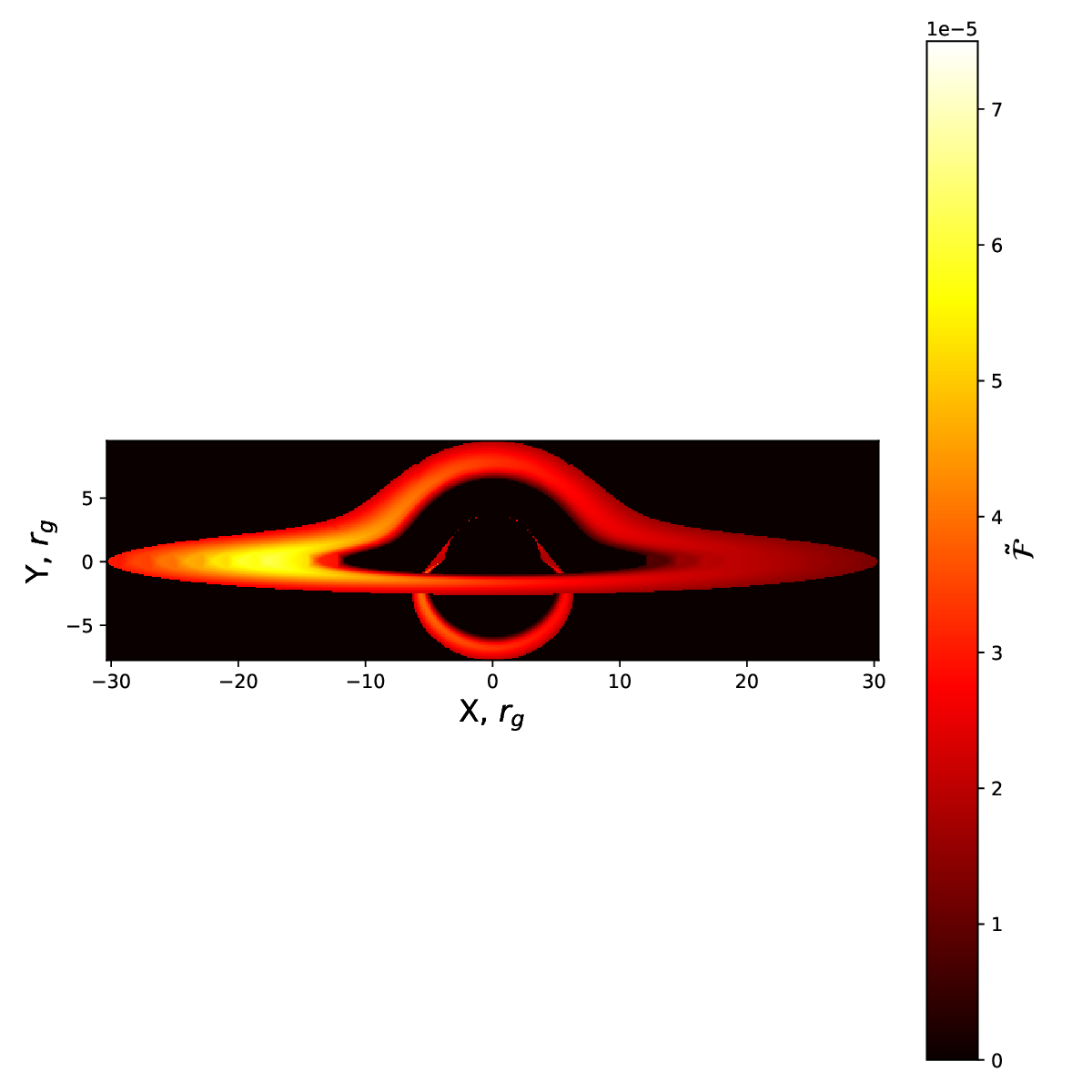} \\ 
					\end{minipage}
\hfill
		\begin{minipage}[h]{.5\textwidth}
	{\small (II)}\\		\includegraphics[width=\columnwidth]{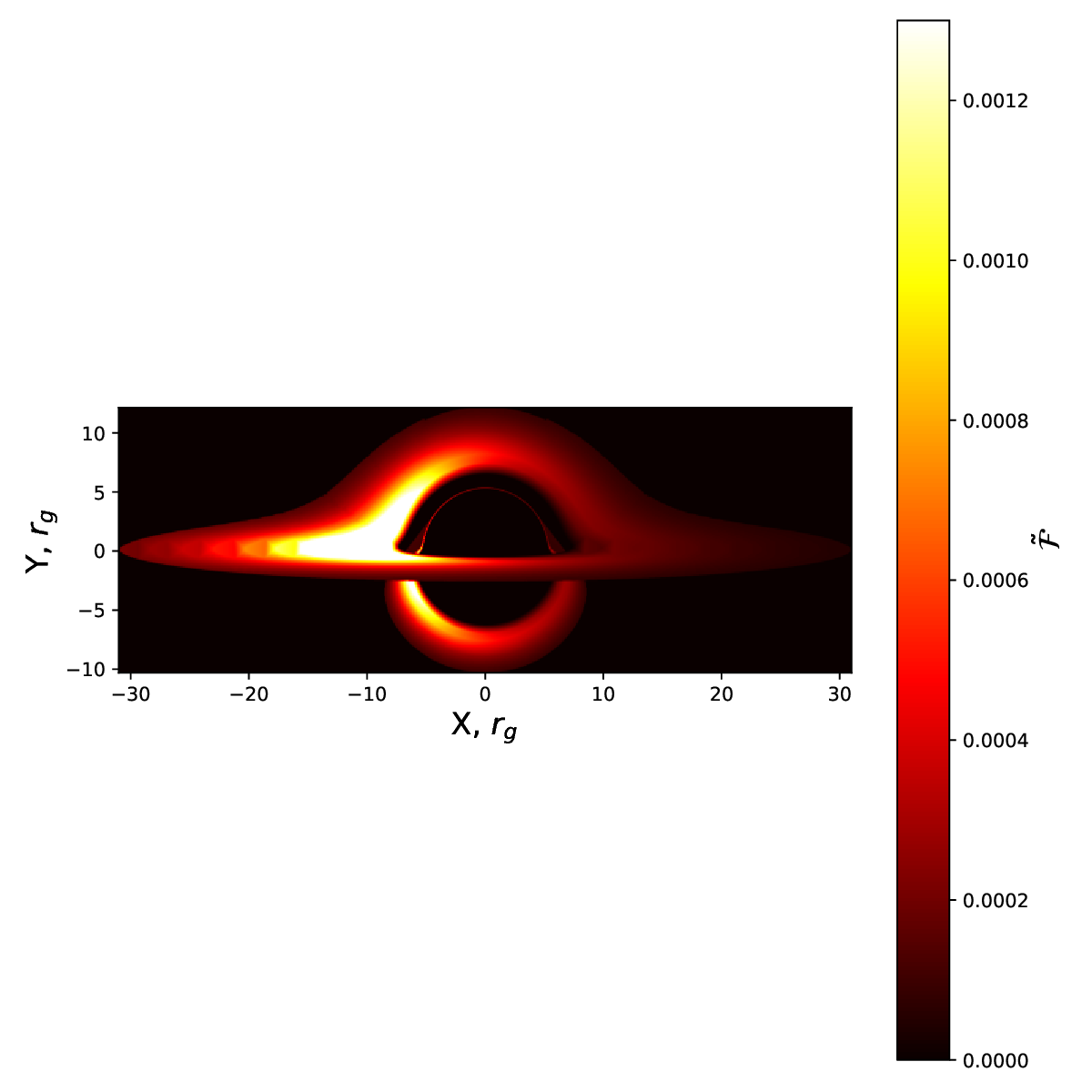} \\ 
					\end{minipage}		
\hfill
		\begin{minipage}[h]{.5\textwidth}
		{\small (III)}\\	{\includegraphics[width=\columnwidth]{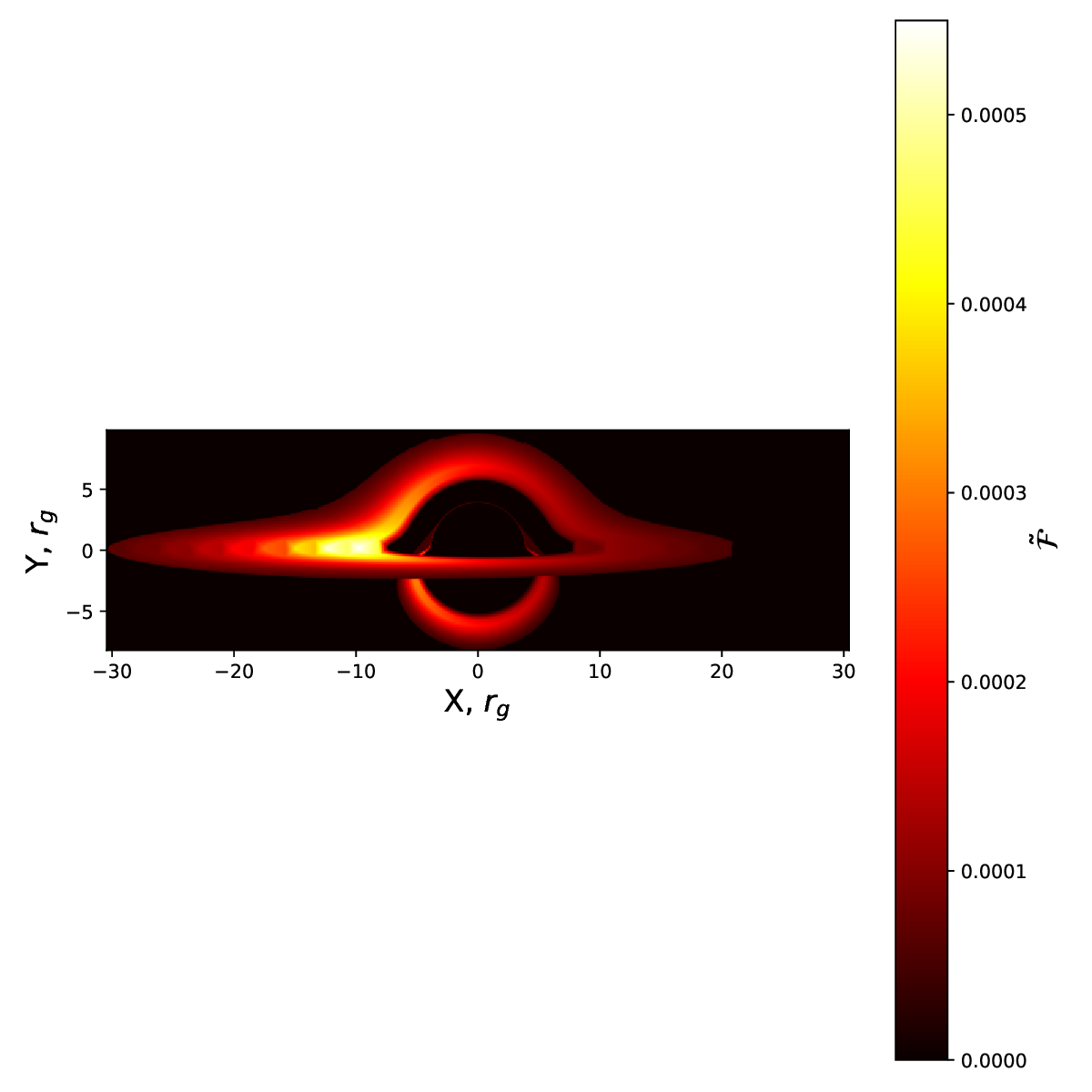}} 			
		\end{minipage}
\hfill
		\begin{minipage}[h]{.5\textwidth}
	{\small (IV)}\\		{\includegraphics[width=\columnwidth]{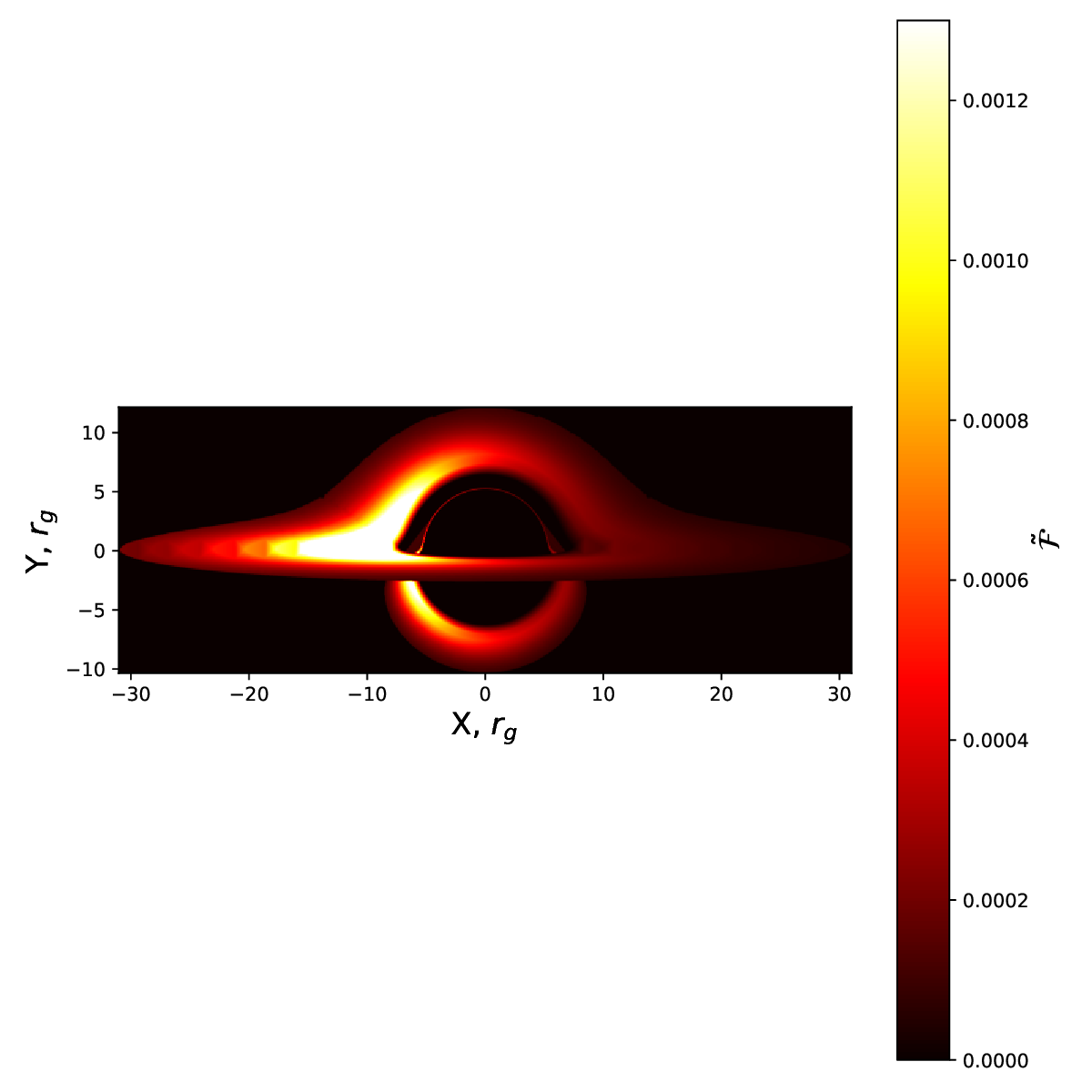}} 			
		\end{minipage}
\caption{The complete apparent images of the thin accretion disk for different scalar-field configurations (see Tab. \ref{tab:models}) at a fixed inclination angle of $i=85^\circ$. All panels are shown in generalized celestial coordinates on the observer's sky. The energy flux is normalized. The inner and outer boundaries of the accretion disk are stable circular orbits with radii $r=r_{\rm isco}$ and $r=30\,r_g$, respectively}

\label{fig3c}	
		\end{figure*}

	Further, we examine the redshift maps (contour plots of the redshift $z$) shown in Figs.~\ref{fig2a}--\ref{fig2c}. Then, we analyze the complete apparent images of the thin accretion disk presented in Figs.~\ref{fig3a}--\ref{fig3c}. The inclination angles are $i=17^\circ$, $45^\circ$, and $85^\circ$ in Figs.~\ref{fig2a}, \ref{fig2b}, and \ref{fig2c}, respectively; the same ordering is adopted for the disk images in Figs.~\ref{fig3a}, \ref{fig3b}, and \ref{fig3c}. Each figure contains four cases, as summarized in Tab. \ref{tab:models}. The disk spans $r\in[r_{\rm isco},30\,r_g]$, where $r_g=GM/c^2$ is the gravitational radius. 
All images are presented in generalized celestial coordinates $X$ and $Y$ in units of $r_g$, which makes them scale-free and avoids tying the results to the physical parameters of any particular source \cite{Perlick_2022}.

\begin{figure*}
		\begin{minipage}[h]{.45\textwidth}
		{\small (I)}\\	\includegraphics[width=\columnwidth]{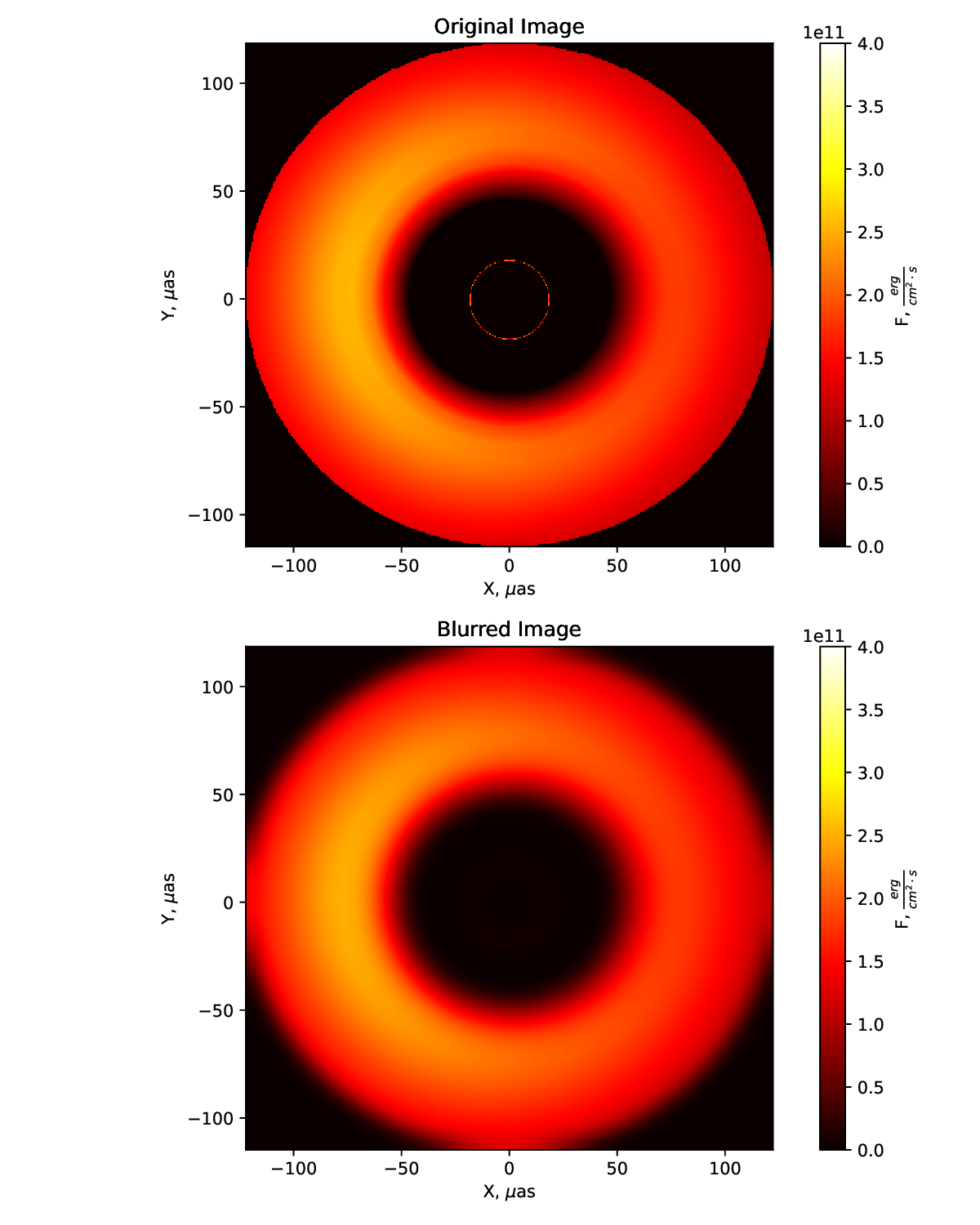} \\ 
					\end{minipage}
\hfill
		\begin{minipage}[h]{.45\textwidth}
		{\small (II)}\\	\includegraphics[width=\columnwidth]{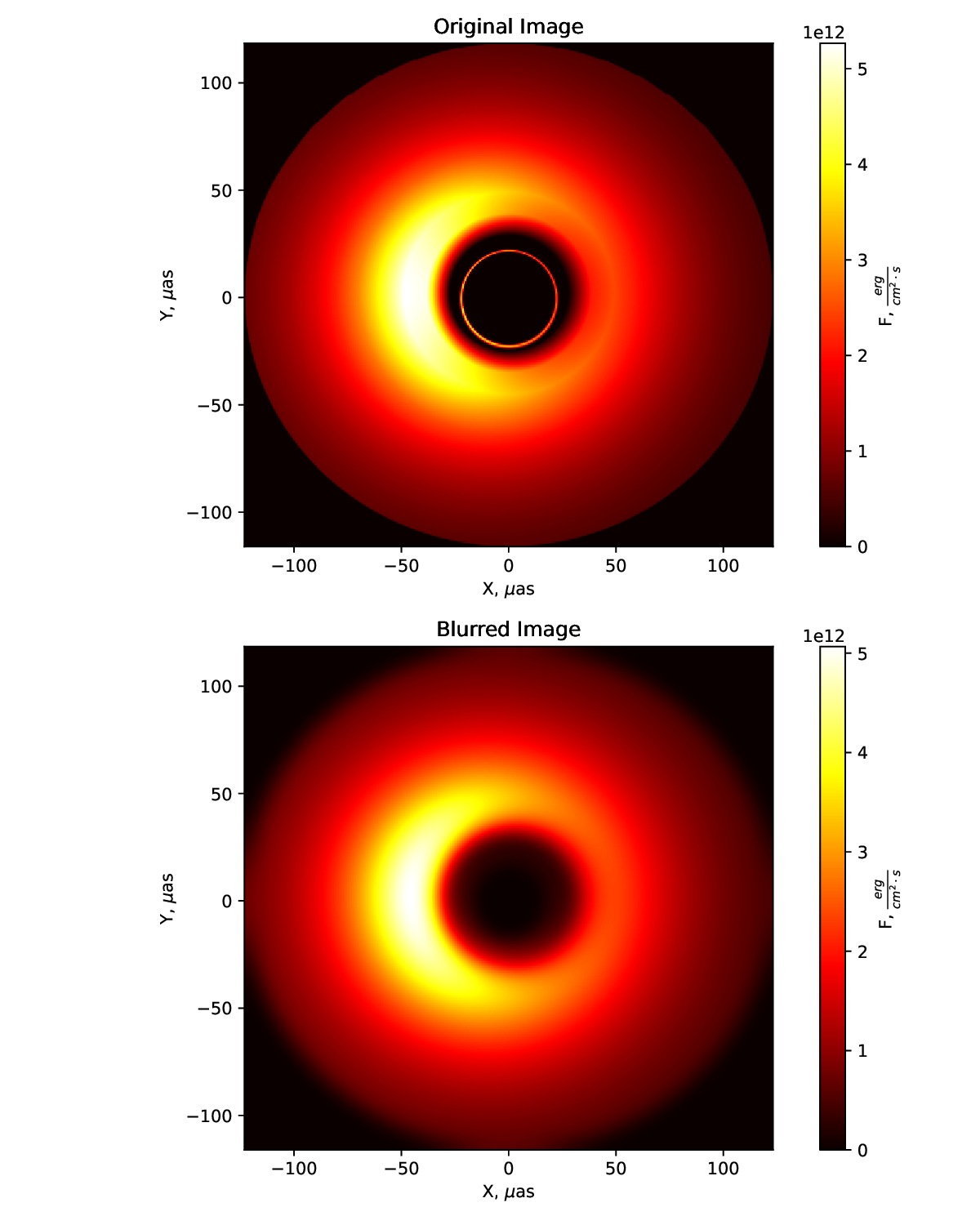} \\ 
					\end{minipage}					
					\hfill
		\begin{minipage}[h]{.45\textwidth}
		{\small (III)}\\	\includegraphics[width=\columnwidth]{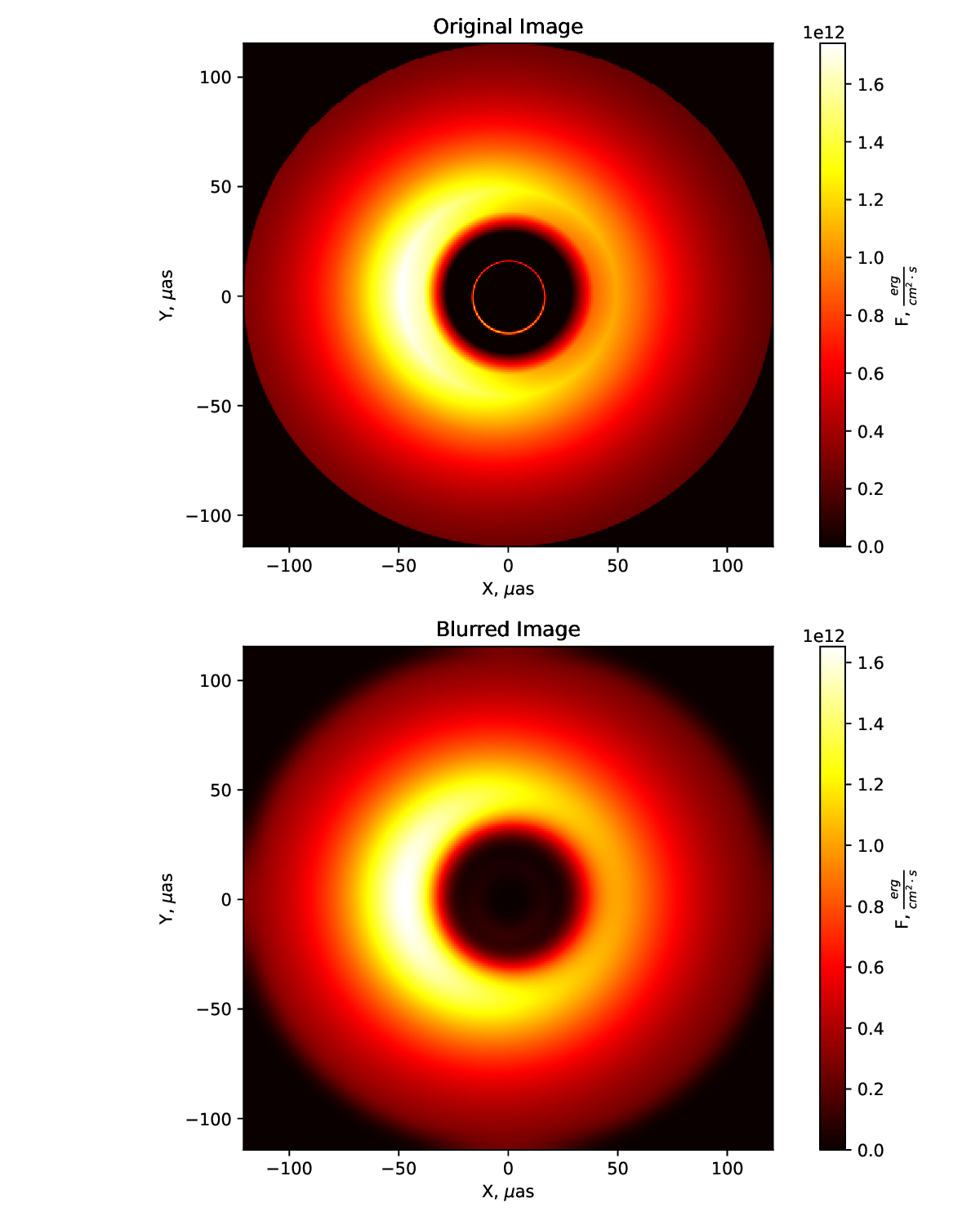} \\ 
					\end{minipage}						
						\hfill
		\begin{minipage}[h]{.45\textwidth}
		{\small (IV)}\\	\includegraphics[width=\columnwidth]{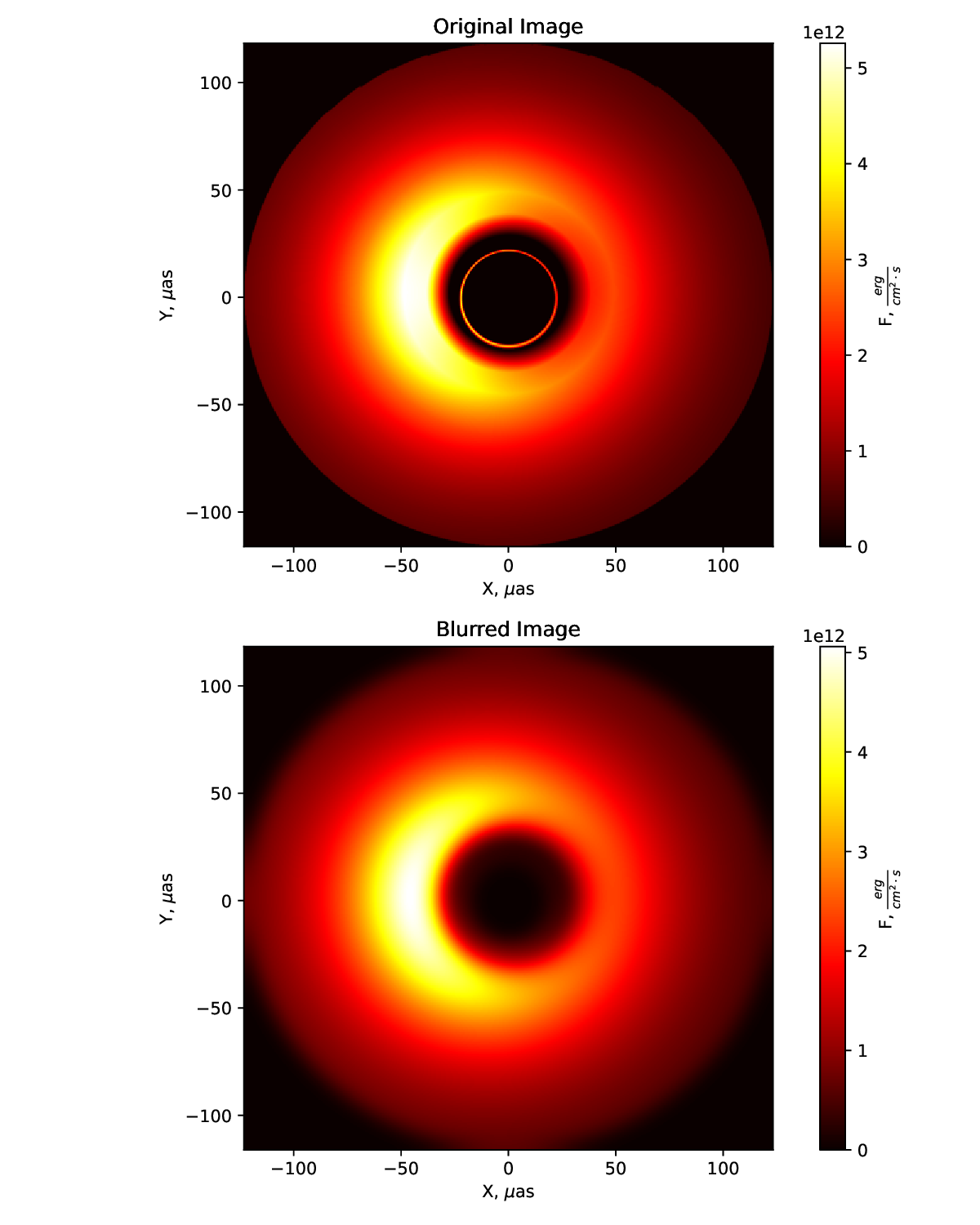} \\ 
					\end{minipage}
					
\caption{The complete apparent images of the thin accretion disk for different values of model's parameters (see Tab. \ref{tab:models}) at a fixed inclination angle of $i=17^\circ$. Even rows correspond to blurring images with a Gaussian filter in order to simulate the nominal resolution of the EHT. All images are constructed for a fiducial hypothetical BH with $M=4\times10^{9}\,M_\odot$, $D=10\,{\rm Mpc}$ and $\dot M=4\,M_\odot\,{\rm yr}^{-1}$. The inner and outer boundaries of the accretion disk are stable circular orbits with radii $r = r_{isco}$ and $r = 30 r_g$, respectively}
					
					\label{fig5}	
		\end{figure*}

\begin{figure*}
		
		\begin{minipage}[h]{.45\textwidth}
		{\small (I)}\\	\includegraphics[width=\columnwidth]{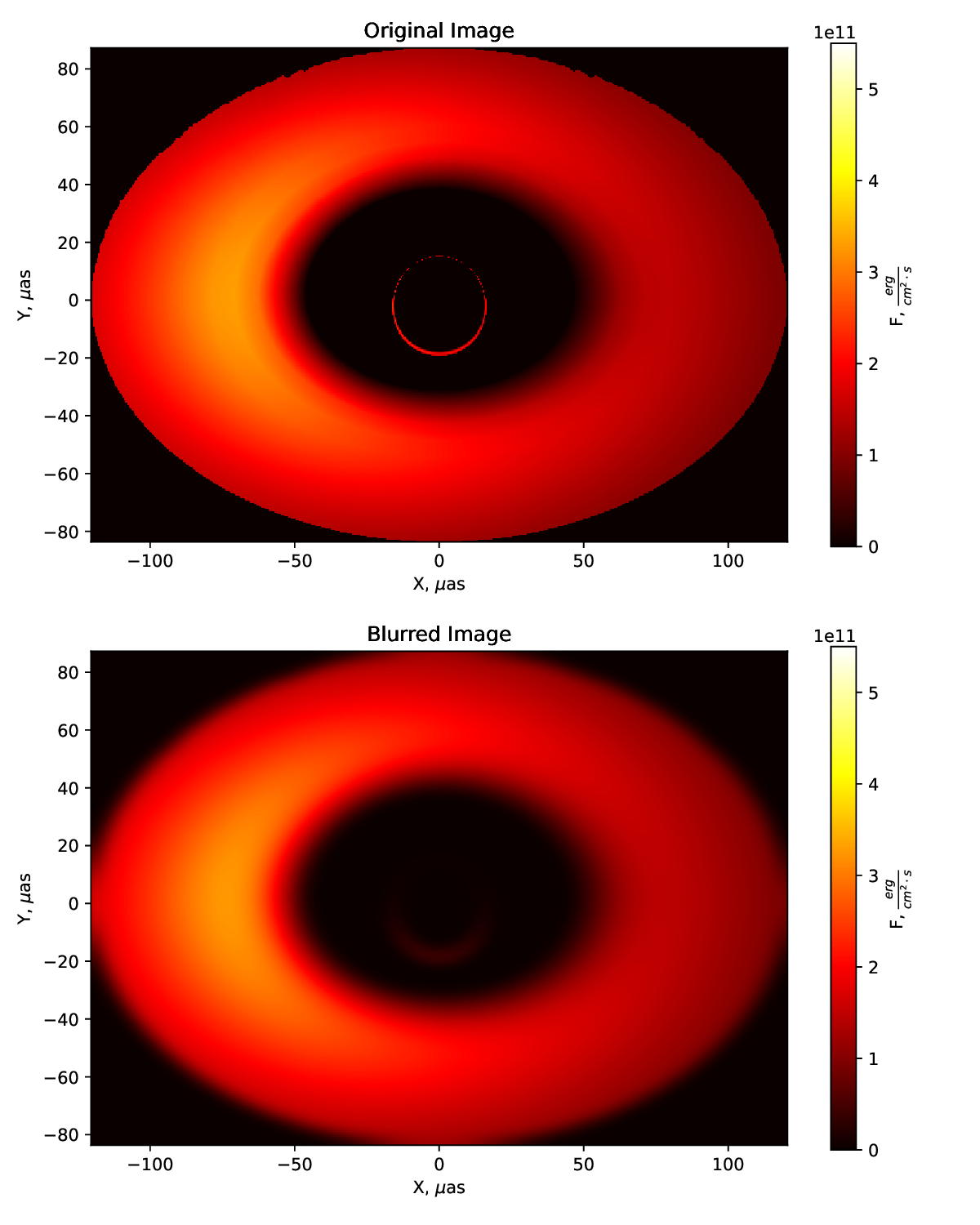} \\ 
					\end{minipage}
					\hfill		
		\begin{minipage}[h]{.45\textwidth}
		{\small (II)}\\	\includegraphics[width=\columnwidth]{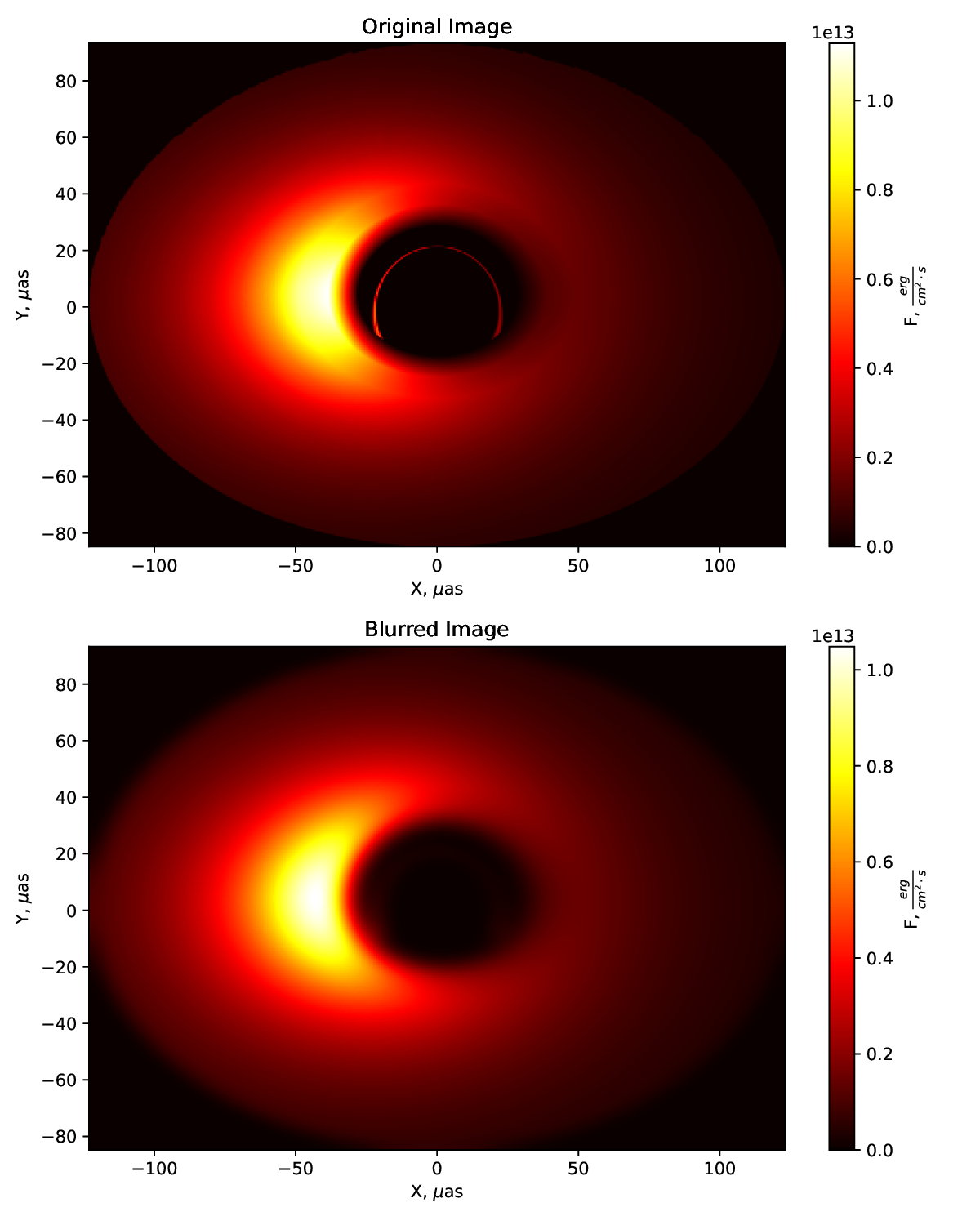} \\ 
					\end{minipage}					
						\hfill
		\begin{minipage}[h]{.45\textwidth}
		{\small (III)}\\	\includegraphics[width=\columnwidth]{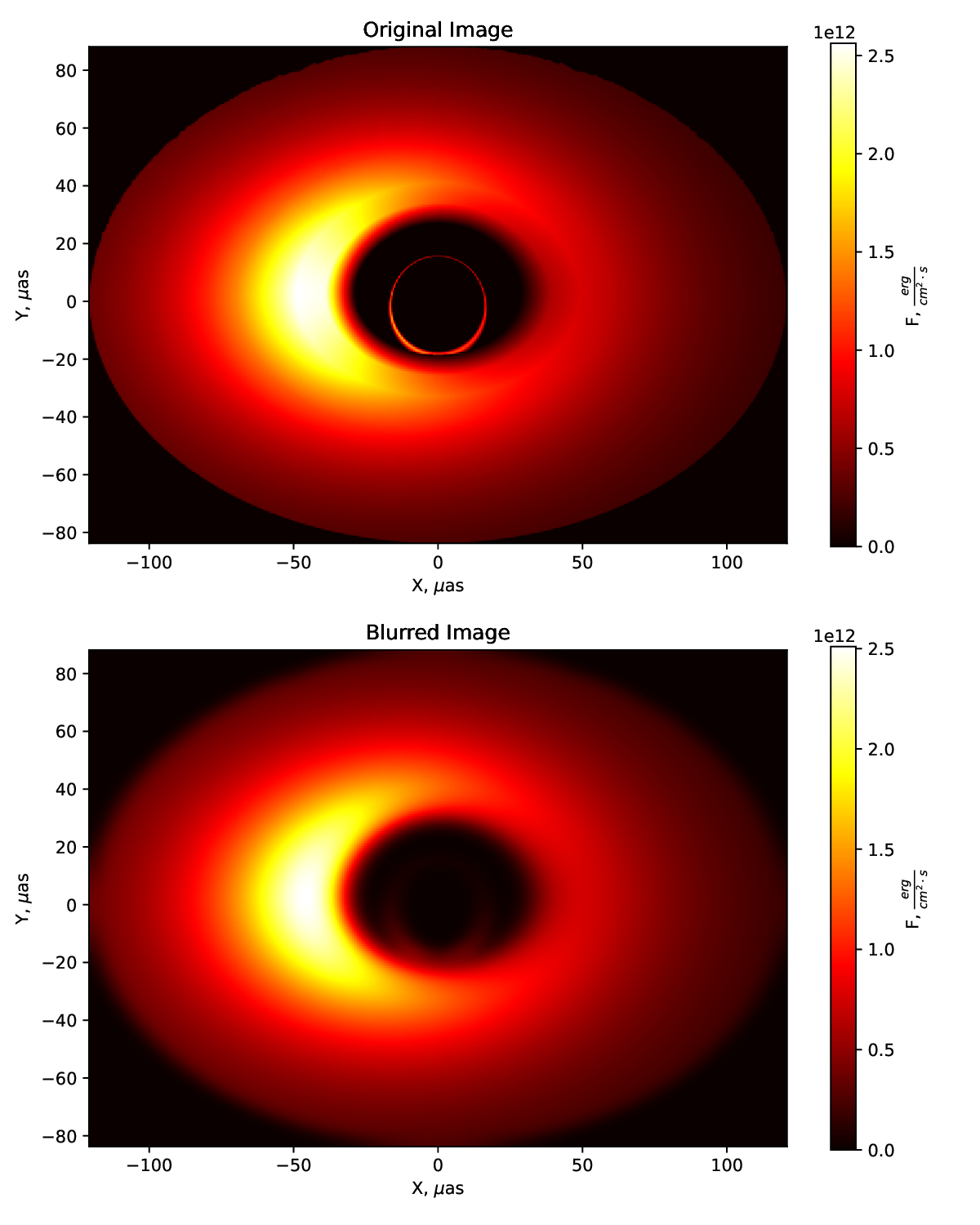} \\ 
					\end{minipage}						
					\hfill
		\begin{minipage}[h]{.45\textwidth}
		{\small (IV)}\\	\includegraphics[width=\columnwidth]{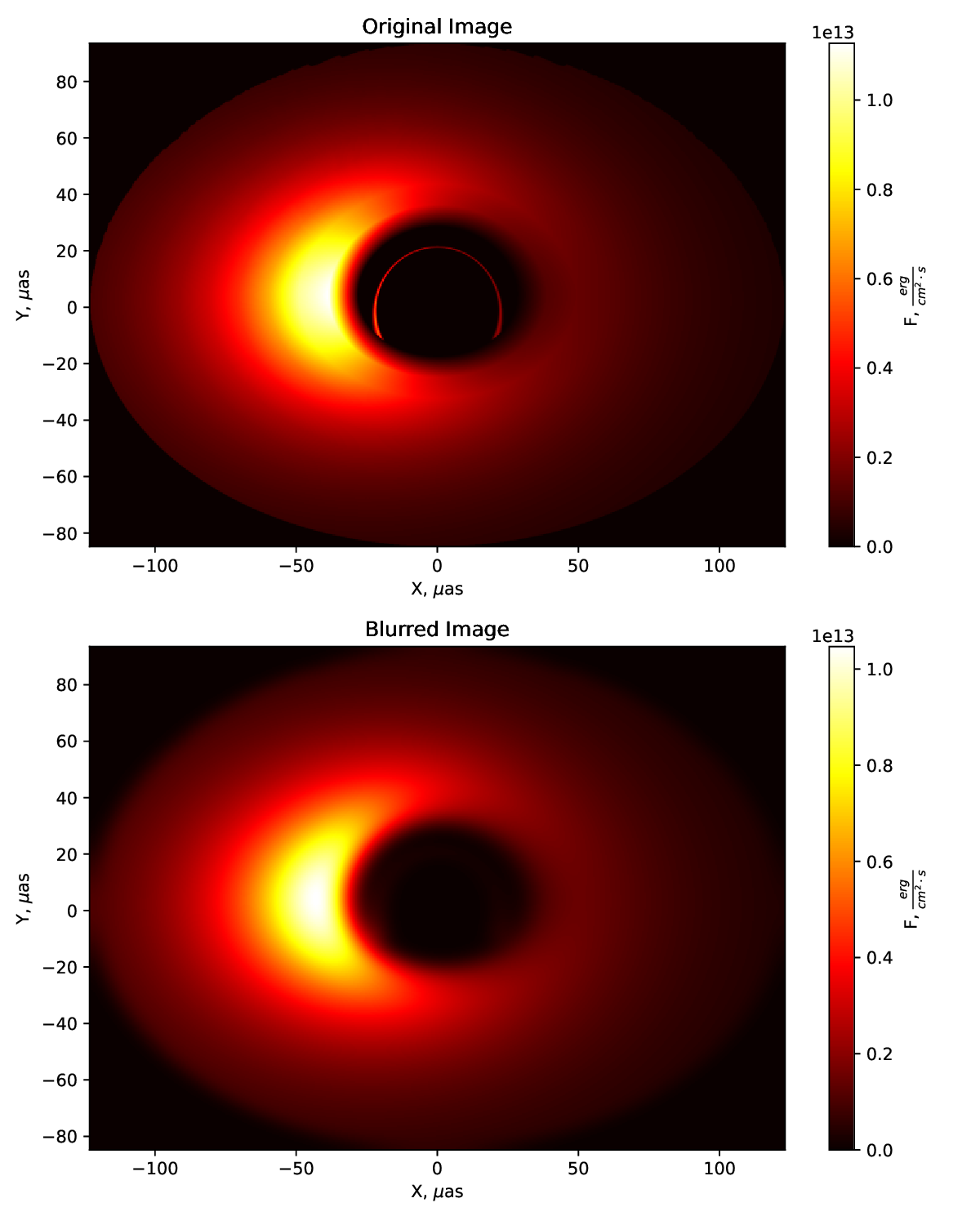} \\ 
					\end{minipage}
					
\caption{The complete apparent images of the thin accretion disk for different values of model's parameters (see Tab. \ref{tab:models}) at a fixed inclination angle of $i=45^\circ$. Even rows correspond to blurring images with a Gaussian filter in order to simulate the nominal resolution of the EHT. All images are constructed for a fiducial hypothetical BH with $M=4\times10^{9}\,M_\odot$, $D=10\,{\rm Mpc}$ and $\dot M=4\,M_\odot\,{\rm yr}^{-1}$. The inner and outer boundaries of the accretion disk are stable circular orbits with radii $r = r_{isco}$ and $r = 30 r_g$, respectively}
\label{fig6}	
		\end{figure*}

\begin{figure*}
			\begin{minipage}[h]{.45\textwidth}
		{\small (I)}\\	\includegraphics[width=\columnwidth]{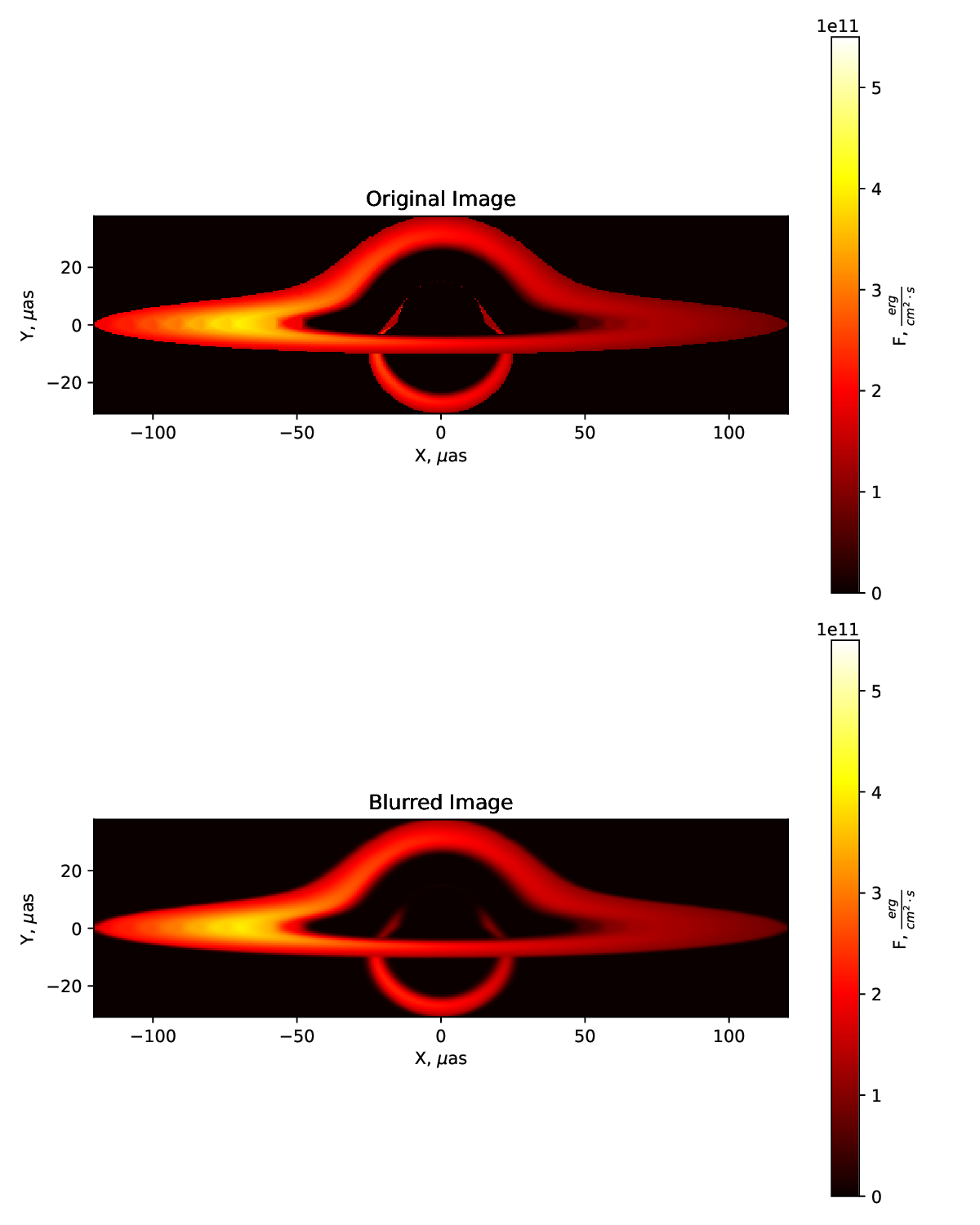} \\ 
					\end{minipage}
					\hfill		
		\begin{minipage}[h]{.45\textwidth}
		{\small (II)}\\	\includegraphics[width=\columnwidth]{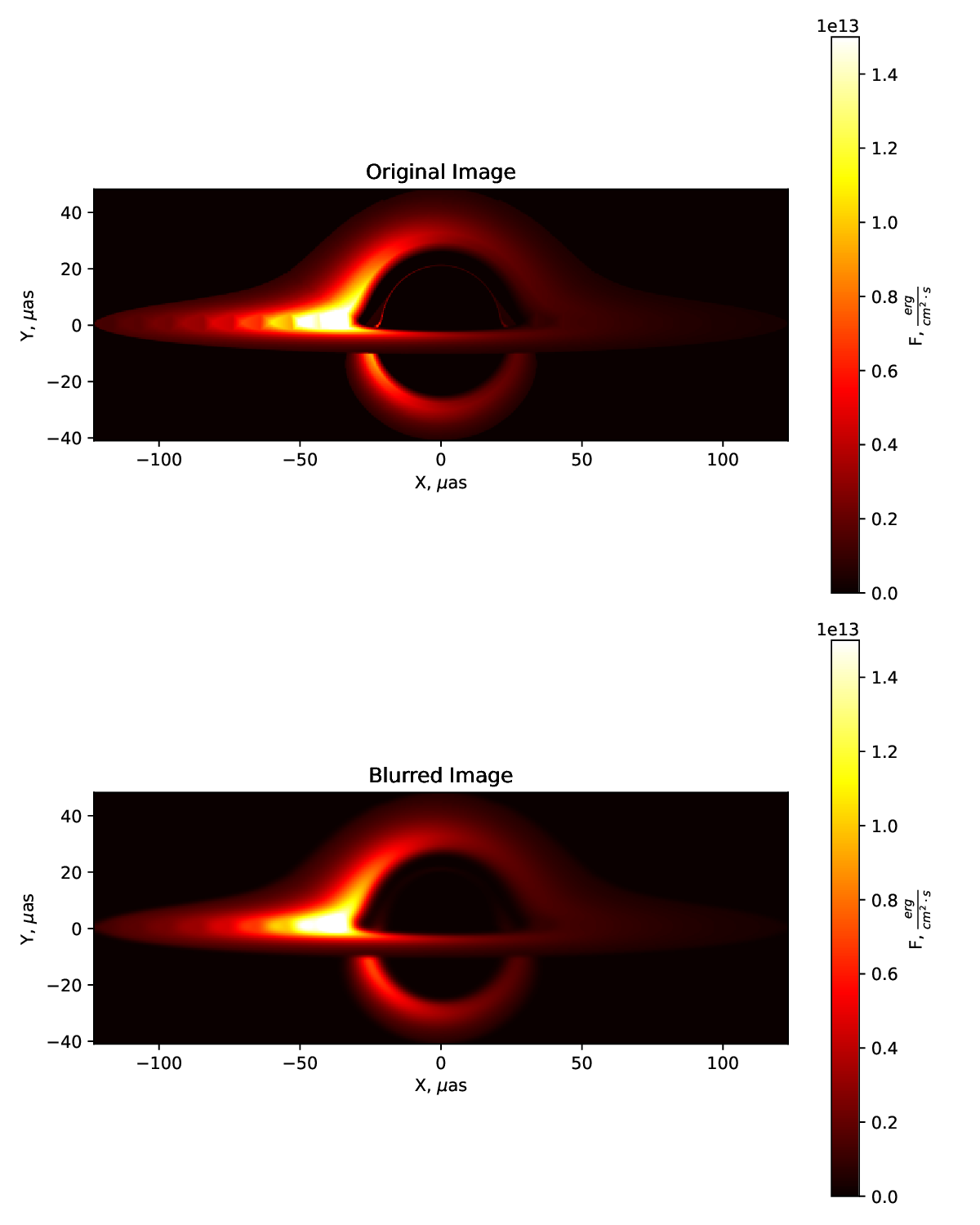} \\ 
					\end{minipage}
					\hfill		
		\begin{minipage}[h]{.45\textwidth}
		{\small (III)}\\	\includegraphics[width=\columnwidth]{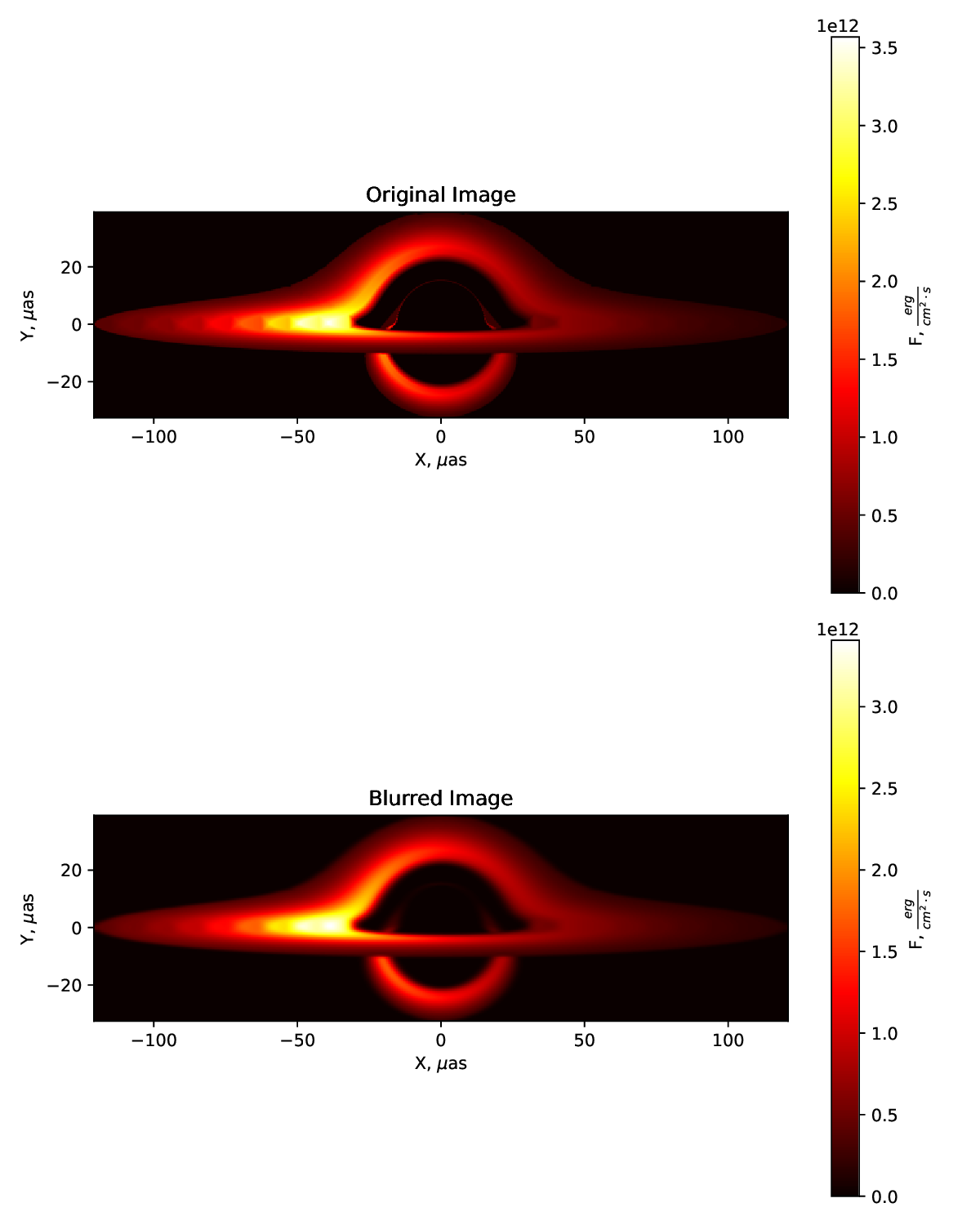} \\ 
					\end{minipage}
						\hfill	
		\begin{minipage}[h]{.45\textwidth}
		{\small (IV)}\\	\includegraphics[width=\columnwidth]{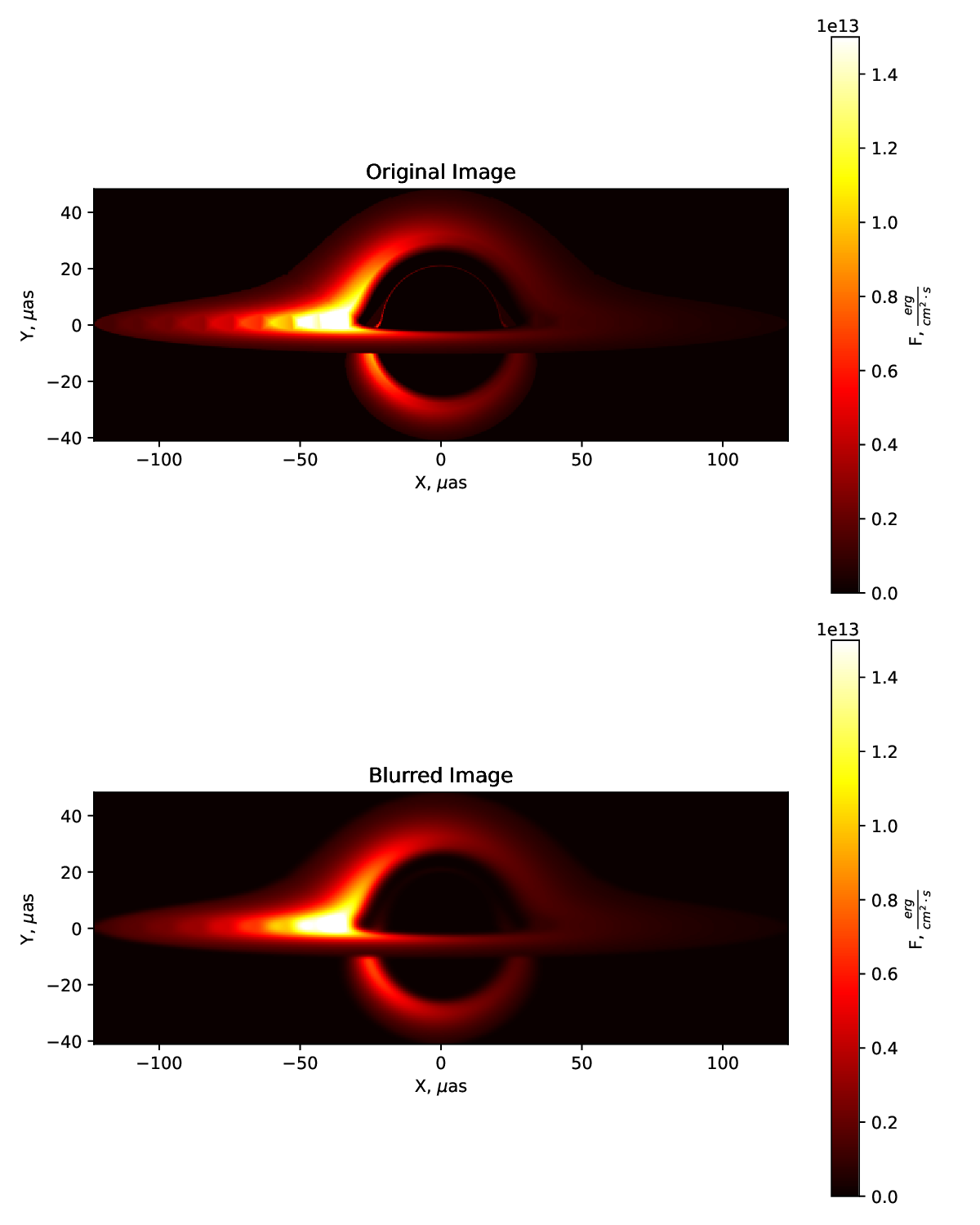} \\ 
					\end{minipage}
\caption{The complete apparent images of the thin accretion disk for different values of model's parameters (see Tab. \ref{tab:models}) at a fixed inclination angle of $i=85^\circ$. Even rows correspond to blurring images with a Gaussian filter in order to simulate the nominal resolution of the EHT. All images are constructed for a fiducial hypothetical BH with $M=4\times10^{9}\,M_\odot$, $D=10\,{\rm Mpc}$ and $\dot M=4\,M_\odot\,{\rm yr}^{-1}$. The inner and outer boundaries of the accretion disk are stable circular orbits with radii $r = r_{isco}$ and $r = 30 r_g$, respectively}
\label{fig7}	
		\end{figure*}

To illustrate how finite angular resolution affects the appearance of the same scale-free images, we additionally show the corresponding panels rescaled to angular units for a fiducial hypothetical black hole and convolved with a Gaussian kernel to mimic an EHT-like effective resolution (Figs.~\ref{fig5}--\ref{fig7}). The inclination angles are $i=17^\circ$, $45^\circ$, and $85^\circ$ in Figs.~\ref{fig5}, \ref{fig6}, and \ref{fig7}, respectively. This angular scaling is introduced solely for detectability estimates and does not represent a dedicated fit to any specific EHT source.




First, we analyze the redshift distribution (shown as contour maps of \(z\)) in the direct and secondary images at three inclination angles: \(i=17^\circ\) (Fig.~\ref{fig2a}), \(i=45^\circ\) (Fig.~\ref{fig2b}), and \(i=85^\circ\) (Fig.~\ref{fig2c}). Each odd row corresponds to direct images, while the even rows represent secondary imagies. Moreover, the pair of the upper direct and lower secondary images correspond to the same case. We can observe that the smallest spread in redshift values occurs for the case (I), while the largest spread is observed (II) and (IV) cases. This is due to the fact that, for (I), the ISCO lies significantly farther out ($\approx 11r_g$) compared to the other cases. As a result, the influence of the BH's gravitational field on the accretion disk is weaker than in the other scenarios. In all figures showing the direct image, the left-hand side contains a distinct region (shown in blue in the online version) corresponding to the blueshifted emission. The right-hand side corresponds to the redshifted emission (shown in red online).
For the secondary images, the blueshifted region is absent at low inclination angles (\( i = 17^\circ \), \( i = 45^\circ \)), while at the high inclination angle (\( i = 85^\circ \)) it appears on the left-hand side. Moreover, we observe that as the inclination angle decreases, the blueshift becomes weaker in all cases.  In addition, we observe that in the case (I), as well as in the presence of a potential (III), the regions of blueshift are more extended compared to both the case (II) and the Schwarzschild case (IV), for all inclination angles. Analyzing Fig. \ref{fig2a}--\ref{fig2c}, we find that  the case (II), is almost indistinguishable from the image of a Schwarzschild BH (IV). 

Let us consider Fig. \ref{fig3a}--\ref{fig3c}. In these images the energy flux intensity is depicted using a continuous color map, where white represents the maximum flux intensity, while dark (in online version red) indicates the minimum value. Analyzing Fig. \ref{fig3a}--\ref{fig3c}, we can also observe that in the case with a potential (III), the inner ring size is smaller than predicted by GR (IV) but still larger than in the potential-free case (I). Additionally, we find that the case (II), is almost indistinguishable from the image of a Schwarzschild BH (IV).

Besides, Fig. \ref{fig3a}--\ref{fig3c} shows that the shape of direct and secondary images strongly depends on the inclination angle. At an edge-on view, the direct image takes on a hat-like shape due to gravitational lensing, which bends light from obscured disk regions to the upper part of the BH. Additionally, as the inclination angle increases, some light rays with smaller deflection angles contribute to secondary images, leading to greater deviations from circularity.

Now we can consider the brightness distribution in the images. The brightness accumulates on the left side of each image. This effect becomes more noticeable with increasing inclination angle due to the Doppler boosting caused by the rotation of the accretion disk; it disappears for a face-on orientation of the disk. We see that case~(II) produces the brightest image, whereas case~(I) yields the dimmest one. The image corresponding to the Higgs-like potential (III) represents an intermediate case between these two. Furthermore, we can also observe that the Schwarzschild case~(IV) and case~(II) are practically indistinguishable from each other. This qualitative trend is quantified by the maximal observable flux values $\tilde{\mathcal{F}}_{\rm obs}^{\max}$ listed in Tab.~\ref{tab:Fmax}, and is fully consistent with the energy-flux maxima distribution shown in Fig.~\ref{fig:2} and with the results of \cite{meacc}.

\begin{table}[t]
\centering

\begin{tabular}{lccc}
\hline\\

 & \multicolumn{3}{c}{$\tilde{\mathcal{F}}_{\rm obs}^{\max}$ }\\
\cline{2-4}
Configuration & $i=17^\circ$ & $i=45^\circ$ & $i=85^\circ$ \\
\hline
(I) 
  &$3.917 \times10^{-5}$ &$5.110\times10^{-5}$  &$6.207\times10^{-5}$  \\
(II) 
  &$8.117\times10^{-4}$  &$1.741\times10^{-3}$ & $3.470\times10^{-3}$ \\
(III) 
  &$2.684\times10^{-4}$ &$3.949\times10^{-4}$  &$5.523\times10^{-4}$  \\
(IV) 
&$8.105\times10^{-4}$  & $1.738\times10^{-3}$ &$3.469\times10^{-3}$   \\
\hline
\end{tabular}
\caption{Maximal values of the observable flux $\tilde{\mathcal{F}}_{\rm obs}^{\max}$ for each black-hole configuration and inclination angle.  All configurations are defined in Tab.~\ref{tab:models}}
\label{tab:Fmax}
\end{table}

Next, we discuss the secondary image. In images with a $17^\circ$ inclination angle, it appears as a glowing circle inside the direct image. In the case of an $85^\circ$ inclination, it appears as a semicircle in the lower part of the disk. While in the case (II) the difference from the Schwarzschild BH (IV) is not noticeable, for (I) case, it becomes significantly apparent, with the angular size visibly reduced. In the case with a potential (III), we also observe a noticeable reduction in the angular size of the secondary image compared to GR (IV). While the effect is less extreme than in the previous case, it remains clearly visible. Thus the secondary images are closely related to the photon ring and can serve as the fingerprint of spacetime geometry. Therefore, we believe that it is possible to use this critical ring to identify the HMPG BHs.

Further, we should discuss the potential feasibility of verifying the constructed images. Currently, the best and only images of BHs have been obtained by EHT \cite{Event_Horizon_Telescope_Collaboration_2022a, Event_Horizon_Telescope_Collaboration_2022b, Event_Horizon_Telescope_Collaboration_2022c, Event_Horizon_Telescope_Collaboration_2022d, 2019a, 2019b, 2019c, 2019d, 2019e, 2019f}. This telescope has an angular resolution of approximately $20 \mu as$ at $1.3 mm$ and is limited by the diameter of the Earth. In the previous images, we used generalized celestial coordinates and normalized energy flux to avoid fitting our images to the physical parameters of any specific source. However, to illustrate how finite resolution affects the appearance of the images after converting to angular units, we consider a fiducial hypothetical BH with a mass $M = 4\times10^9 M_\odot$, located at a distance $D = 10 \ {\rm Mpc}$ with an accretion rate $\dot M=4 M_\odot/{\rm year}$. This fiducial choice is motivated by two requirements: (i) the corresponding angular size is in the range potentially resolvable by the EHT, and (ii) the adopted accretion rate places the system in the radiatively efficient regime where the geometrically thin disk approximation is applicable \cite{Yuan}. For all images, the energy flux distribution is computed using Eq.~\eqref{page_thorne_flux}. Also, on the x and y axes, we display the celestial coordinates, which can be obtained as follows \cite{2022hu, 2022hua}:
\begin{equation}
\cfrac{\delta}{\mu as}=\biggl(\cfrac{6.191165\times10^{-8}}{2\pi}\cfrac{\sigma}{D/Mpc}\biggr)\biggl(\cfrac{x}{M}\biggr),
\end{equation}
where $\sigma \equiv M/M_\odot$ and $D$ is the source distance.
To simulate the nominal resolution of the EHT, we apply a Gaussian blur to odd rows of Fig. \ref{fig5}--\ref{fig7} with a standard deviation equal to $1/12$ of the field of view \cite{Gralla_2019, Guo_2022, 2022hu}, obtaining the even rows of Fig. \ref{fig5}--\ref{fig7}.  It becomes evident that not only does the image's brightness decrease, but more importantly, the ring structure is significantly blurred. However, for large inclination angles, the size of the secondary image remains noticeably different from the predictions of GR in the case (I) and in the considered scenario with a Higgs-type potential (III). 

We stress that this Gaussian convolution is used only as a simplified proxy for finite angular resolution and is not intended as a direct comparison with current EHT data, which would require synthetic visibilities, realistic $(u,v)$ coverage, and a dedicated imaging pipeline. Moreover, our emission model assumes a geometrically thin, radiatively efficient disk and is therefore not aimed at modeling M87* and Sgr A*, whose millimeter emission is commonly described by ADAF/RIAF-type models. Consequently, in relation to the current EHT observations, the most robust observable that does not depend on the accretion model is the angular diameter of the photon ring, which is primarily determined by the spacetime geometry.

As a next step, we plan to extend our analysis beyond the Novikov-Thorne approximation toward more realistic accretion-disk models that incorporate physical processes such as disk thickness, magnetic fields, and radiative transfer. This will allow us to determine whether the geometric differences identified here persist under realistic astrophysical conditions. In parallel, future improvements in observational accuracy and the growing number of imaged black holes with diverse parameters together with the enhanced resolution of the next-generation EHT and the forthcoming Millimetron space observatory \cite{Likhachev_2022, 2021Andrianov, 2021Novikov} could make it possible to distinguish GR from the predictions of HMPG. Millimetron will overcome the limitations of Earth-based telescopes and achieve unprecedented angular resolution, offering new opportunities to probe the nature of gravity through black hole imaging.

\begin{figure}
\begin{center}
		\begin{minipage}[h]{.6\textwidth}
		\includegraphics[width=\columnwidth]{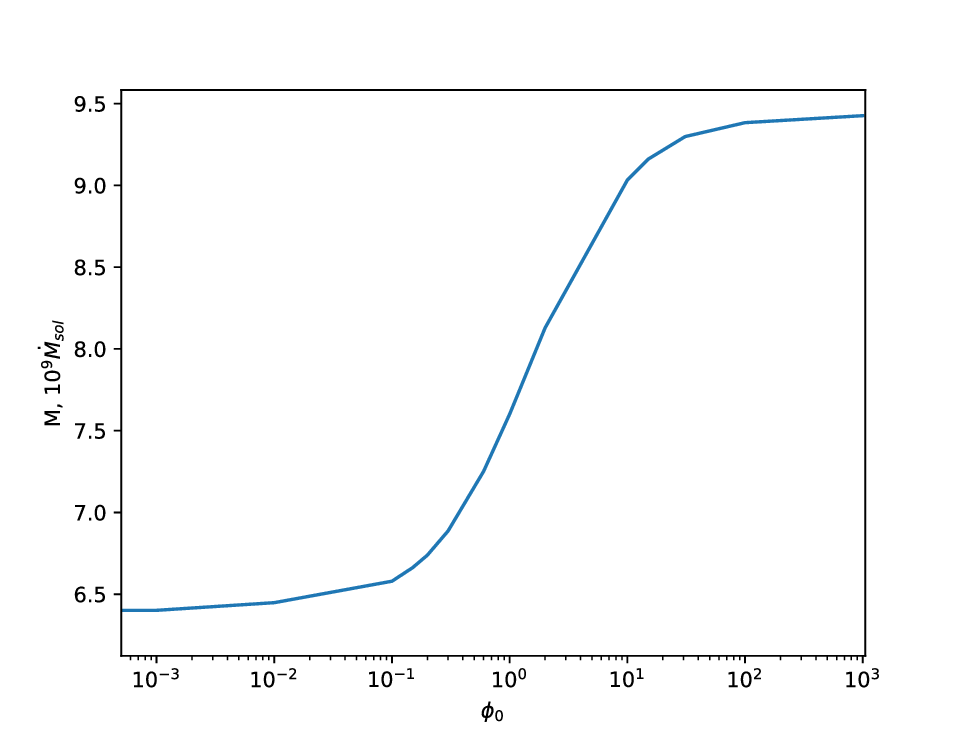} \\ 
					\end{minipage}
					
\caption{The critical ring diameter $S_d=42 \mu as$ of HMPG BHs in the two-dimensional space of parameter $\phi_0$ and mass M. Here the observational angle and BH distance are fixed at $17^\circ$ and 16.8 Mpc, respectively. The scale corresponding to $\phi_0$ is logarithmic for convenience}
\label{fig4}	
\end{center}
		\end{figure}

Now we can consider another important aspect of BH imaging. The key point is that two BHs can have the same angular size of the critical ring, yet, if they are described by different gravitational theories, the corresponding BH mass inferred for this angular scale may differ \cite{2022hu}. To illustrate this effect, we present Fig.~\ref{fig4}, where we analyze the case without a potential while assuming a post-Newtonian connection between \( \phi_0 \) and \( u_0 \) \eqref{u0}. In Fig.~\ref{fig4}, the BH mass is plotted on the vertical axis, while different values of the parameter \( \phi_0 \) are plotted on the horizontal axis, using a logarithmic scale. The curve represents the set of BH configurations that share a fixed (fiducial) angular size of the critical ring, $\theta_{\rm cr}=42~\mu$as. This observable is purely geometrical and does not depend on the details of the accretion flow. Therefore, in what follows we mention M87* only as a numerical reference for the geometric angular scale $\theta_{\rm cr}$ inferred from EHT analyses, without attempting any source-specific emission modeling.

 The Fig.~\ref{fig4} demonstrates that by varying \( \phi_0 \) and the BH mass, it is possible to obtain configurations with the same angular size of the critical ring as observed. Thus, if we have an independent estimate of the BH mass, we can impose constraints on the parameters of the model and break this degeneracy. For M87*, observational constraints on its mass exist. The EHT collaboration reports a mass estimate of \( M_{\text{shadow-method}} = (6.5 \pm 0.7) \times 10^9 M_\odot \) \cite{2019a, 2019b, 2019c, 2019d, 2019e, 2019f}. From Fig.~\ref{fig4}, we see that such a mass range is compatible with \(0< \phi_0 < 1 \) (in this paper, we do not consider negative values of the scalar field). However, it is important to note that this mass estimate is not completely theory-independent, meaning it cannot be used to impose strict bounds. Nevertheless, within HMPG the BH mass is not expected to deviate drastically from the GR-based estimate, which suggests that positive values of \( \phi_0 \) much larger than unity are disfavored. This qualitative conclusion is consistent with constraints obtained from other tests, such as observations in the Solar System \cite{Leanizbarrutia2017, Dyadina2019, Dyadina_2022} and binary pulsars \cite{Dyadina2018, Avdeev2020}. Thus, Fig.~\ref{fig4} provides an additional consistency check and illustrates how combining geometrical observables with independent mass information can help constrain the scalar-field sector of HMPG.

All these results give hope that, with future advancements in telescope resolution and space-based observatories, we will be able to distinguish black holes predicted by different gravitational theories with unprecedented precision. At the same time, progress in theoretical modeling within the framework of modified gravity, particularly in constructing more realistic accretion-disk simulations that include physical processes such as magnetic fields, radiative transfer, and disk thickness will complement observational developments. Together, these advances will not only refine our understanding of black hole properties but also impose stringent constraints on modified gravity models, bringing us closer to uncovering the true nature of gravity and the fundamental structure of our Universe.

\section{Conclusion}\label{conc}

This study focuses on the observational signatures of black holes predicted by hybrid metric-Palatini gravity  through the analysis of images of their surrounding accretion disks. We modeled the radiative appearance of geometrically thin, optically thick disks using semi-analytic ray-tracing techniques in curved spacetime. The Novikov-Thorne thin-disk model is adopted as a first step before addressing the more complex case of realistic accretion flows. This approach allows us to isolate and understand the geometric features of the disk without the additional complications that arise when modeling physically realistic disks. By systematically varying both the inclination angle and the parameters of the theory, we produced a set of images representing different physical scenarios, including configurations with a Higgs-type potential and those without a scalar potential. 

Our analysis reveals several notable features. First, the redshift distribution and brightness profiles of the disk images are sensitive to both the inclination angle and the scalar field configuration. In particular, we found that in HMPG, the inner disk regions are cooler and less luminous than in GR, which aligns with previous findings on energy flux profiles \cite{meacc}. While case (II) involving coupling between parameters $u_0$ and $\phi_0$ appear nearly indistinguishable from Schwarz\-schild scenarios, configuration without potential (I) or with Higgs-like potential (III) exhibit visible deviations in both direct and secondary images. The morphology and angular size of the secondary ring, in particular, provide a promising observational fingerprint of deviations from GR, as it is closely related to the photon ring.

We also simulated the effects of finite observational resolution by applying a Gaussian blur to our images, replicating the capabilities of EHT. The results indicate that, at current resolution limits, distinguishing between GR and HMPG BHs is challenging but may become feasible with future improvements. In particular, space-based missions such as Millimetron are expected to achieve significantly higher angular resolutions, potentially making such distinctions observationally accessible \cite{Likhachev_2022, 2021Andrianov, 2021Novikov}.

We also investigated the connection between the angular size of the BH’s photon ring and its mass within HMPG. Our analysis shows that, for a fixed angular diameter of the critical ring, different combinations of the scalar field parameter $\phi_0$ and BH mass can reproduce the same observational signature. We use M87* as an illustrative benchmark for the geometrical angular scale and the commonly quoted mass range, without attempting any source-specific emission modeling. Using this example, we demonstrated that only values of $0<\phi_0<1$ are consistent with the mass range inferred from the EHT shadow-based measurements, while the case of $\phi_0<0$ was not considered. Although these mass estimates are not entirely theory-independent, they still allow us to evaluate possible constraints on the model. This result aligns with previous bounds obtained from Solar System \cite{Leanizbarrutia2017, Dyadina2019, Dyadina_2022} and binary pulsar observations \cite{Dyadina2018, Avdeev2020, Dyadina2024}, further confirming the compatibility of HMPG with existing experimental data and reinforcing its viability as an alternative to GR.

In conclusion, this study shows that high-resolution imaging of accretion disks can serve as a powerful tool for testing gravitational theories beyond GR. Hybrid metric-Palatini gravity, as a well-motivated extension, yields distinctive yet potentially observable features in disk structure and emission. In this work, we employed the Novikov-Thorne thin-disk model as an initial approximation, representing a first step toward more realistic simulations that account for the physical conditions in actual accretion flows. With upcoming advancements in observational astronomy, including the next generation of very-long-baseline interferometry and space-based telescopes, as well as future theoretical developments incorporating realistic disk physics, we anticipate the emergence of strong constraints, and possibly observable signatures, of modified gravity theories in the strong-field regime.

\section*{Acknowledgements}
The authors thank Shiyang Hu for discussions and comments on the topics of this paper. P.I. Dyadina acknowledges support from  the Foundation for the Advancement of Theoretical Physics and Mathematics “BASIS”.

 \section*{Declarations}

\textbf{Conflict of interest} The authors declare that they have no known competing financial interests or personal relationships that could have appeared to influence the work reported in this paper.

\textbf{Data Availability Statement} All data that support the findings of this study are included within the article (and any supplementary information files).

\textbf{Code Availability Statement} The code used to generate the black hole shadow images in hybrid metric-Palatini gravity was developed by the authors specifically for this study. It is openly available on GitHub at \url{https://github.com/NikitaAvdeev/Shadow_BH}, and can be used to reproduce the results presented in this paper. The repository includes documentation and example scripts.
\bibliographystyle{JHEP}
\bibliography{imageshybrid2.bib}

\end{document}